\documentclass[aps,preprint,onecolumn,11pt,nofootinbib]{revtex4}
\usepackage{hyperref}
\usepackage{amsfonts}
\usepackage[utf8]{inputenc}
\usepackage{amsmath}

\usepackage{amssymb}
\usepackage{tensor}
\usepackage{graphicx}
\usepackage{bigints}
\usepackage{graphicx}
\usepackage{subfigure}
\usepackage{epsf}
\usepackage{bm}
\usepackage{amsmath}
\usepackage{amsfonts}
\usepackage{amssymb}
\usepackage{graphicx}
\usepackage{tabularx}
\usepackage{color}%
\providecommand{\U}[1]{\protect\rule{.1in}{.1in}}

\newcommand{\ie}{\begin{equation}}
\newcommand{\fe}{\end{equation}}

\newcommand{\mincir}{\raise
-3.truept\hbox{\rlap{\hbox{$\sim$}}\raise4.truept\hbox{$<$}\ }}
\newcommand{\magcir}{\raise
-3.truept\hbox{\rlap{\hbox{$\sim$}}\raise4.truept\hbox{$>$}\ }}

\providecommand{\U}[1]{\protect\rule{.1in}{.1in}}

\usepackage{tikz,xcolor,hyperref}

\definecolor{lime}{HTML}{A6CE39}
\DeclareRobustCommand{\orcidicon}{%
	\begin{tikzpicture}
	\draw[lime, fill=lime] (0,0) 
	circle [radius=0.16] 
	node[white] {{\fontfamily{qag}\selectfont \tiny ID}};
	\draw[white, fill=white] (-0.0625,0.095) 
	circle [radius=0.007];
	\end{tikzpicture}
	\hspace{-2mm}
}

\foreach \x in {A, ..., Z}{%
	\expandafter\xdef\csname orcid\x\endcsname{\noexpand\href{https://orcid.org/\csname orcidauthor\x\endcsname}{\noexpand\orcidicon}}
}

\begin{document}

\title{\Large{The impact of an antisymmetric tensor on charged black holes: evaporation process, geodesics, deflection angle,  scattering effects and quasinormal modes}}


\author{A. A. Ara\'{u}jo Filho\orcidB{}}
\email{dilto@fisica.ufc.br (The corresponding author)}


\affiliation{Departamento de Física, Universidade Federal da Paraíba, Caixa Postal 5008, 58051-970, João Pessoa, Paraíba, Brazil}

\author{N. Heidari\orcidD{}}
\email{heidari.n@gmail.com}

\affiliation{Center for Theoretical Physics, Khazar University, 41 Mehseti Street, Baku, AZ-1096, Azerbaijan.}

\author{J. A. A. S. Reis\orcidA{}}
\email{jalfieres@gmail.com}

\affiliation{Universidade Estadual do Sudoeste da Bahia (UESB), Departamento de Ciências Exatas e Naturais, Campus Juvino Oliveira, Itapetinga -- BA, 45700-00, Brazil}


\author{H. Hassanabadi\orcidC{}}
\email{hha1349@gmail.com}


\affiliation{Department of Physics, University of Hradec Kr$\acute{a}$lov$\acute{e}$, Rokitansk$\acute{e}$ho 62, 500 03 Hradec Kr$\acute{a}$lov$\acute{e}$, Czechia.}


\begin{abstract}

In this paper, we investigate the influence of anti--symmetric tensor effects, which trigger the Lorentz symmetry breaking, on charged spherically symmetric black holes. Initially, we address an overview of the model, laying the groundwork for deriving solutions to black holes. With this, we analyze the horizons, critical orbits, and geodesics. We compute quasinormal modes and the time--domain solution with a particular emphasis on vectorial perturbations. In addition, we derive the Hawking temperature to perform the calculation of the remnant mass. Additionally, we estimate the lifetime of the black holes until they reach their final stage after the evaporation process. Finally, we explore the emission rate, the deflection angle and, we investigate the correlation between quasinormal modes and shadows.

\end{abstract}
\maketitle

\tableofcontents

\pagebreak

\section{Introduction}

Lorentz symmetry, a cornerstone in contemporary physics, posits the consistent applicability of physical laws across all inertial reference frames. Although firmly ingrained as a fundamental principle, supported by extensive experimental and observational evidence, it has become apparent that Lorentz symmetry may deviate under specific energy conditions within various theoretical frameworks. These include string theory \cite{1}, loop quantum gravity \cite{2}, Horava--Lifshitz gravity \cite{3}, non--commutative field theory \cite{4}, Einstein--aether theory \cite{5}, massive gravity \cite{6}, $f(T)$ gravity \cite{7}, very special relativity \cite{8}, and others.

The breaking of Lorentz symmetry manifests itself in two discernible ways: explicit and spontaneous \cite{bluhm2006overview}. In the first scenario, there is a lack of Lorentz invariance in the Lagrangian density, leading to the formulation of distinct physical laws within certain reference frames. On the contrary, the second scenario unfolds when the Lagrangian density retains Lorentz invariance, but the ground state of a physical system does not display Lorentz symmetry \cite{bluhm2008spontaneous}.

The exploration of spontaneous Lorentz symmetry breaking \cite{9,10,11,12,13,KhodadiPoDU2023} is rooted in the Standard Model Extension. Within this conceptual framework, the simplest field theories are encapsulated by bumblebee models \cite{1,10,11,12,13, KhodadiEPJC2023,KhodadiEPJC20232,KhodadiPRD2022,CapozzielloJCAP2023}. In these models, a vector field, known as the bumblebee field, acquires a non--zero vacuum expectation value (VEV). This characteristic establishes a unique direction, resulting in the violation of local Lorentz invariance for particles, subsequently giving rise to notable consequences, such as impacts on thermodynamic properties \cite{petrov2021bouncing2,aaa2021thermodynamics,aa2021lorentz,araujo2021thermodynamic,aa2022particles,reis2021thermal,araujo2021higher,araujo2022thermal,araujo2022does,paperrainbow,anacleto2018lorentz,araujo2023thermodynamics}.

In the study presented by Ref. \cite{14}, an exact solution is detailed for a spacetime that is both static and spherically symmetric within the framework of bumblebee gravity. Similarly, a solution resembling Schwarzschild has been thoroughly scrutinized from various angles, encompassing aspects such as Hawking radiation \cite{16}, the intricacies of the accretion process \cite{17,18}, gravitational lensing phenomena \cite{15}, and the exploration of quasinormal modes \cite{19}.

Subsequent to these findings, Maluf et al. further advanced the research landscape by deducing a solution akin to (Anti--)de Sitter--Schwarzschild, relaxing the vacuum conditions \cite{20}. Additionally, Xu et al. introduced novel classifications of static spherical bumblebee black holes. They incorporated a background bumblebee field with a non--zero temporal component, exploring the thermodynamic properties and observational implications as expounded in Refs. \cite{21,22,23,24}.

Ding et al. explored the domain of rotating bumblebee black holes in their investigation detailed in Ref. \cite{25,26}. Their study covered a broad spectrum of topics, including accretion processes \cite{28}, shadows \cite{27}, quasi--periodic oscillations \cite{30}, and quasinormal modes \cite{29}. Additionally, an exact solution analogous to the rotating BTZ black hole was formulated in \cite{31}, and its quasinormal modes were analytically scrutinized in reference \cite{32}.

In an innovative development, a black hole resembling Schwarzschild, coupled with a global monopole, was introduced in \cite{33}. The quasinormal modes of this unique configuration underwent thorough analysis \cite{34,35}. Furthermore, diverse black hole solutions were explored within the conceptual framework of the Bumblebee gravity model in Refs. \cite{36,37,38}, along with investigations within the metric affine formalism \cite{nascimento2022vacuum,delhom2022radiative,delhom2022spontaneous,nascimento2024exact,nascimento2024gravitational}. Adding to this rich landscape, a recently proposed traversable bumblebee wormhole solution has captured attention in the literature \cite{39}, accompanied by subsequent studies exploring its corresponding gravitational waves \cite{40}.

Going beyond vector field theories, an alternative approach to explore Lorentz symmetry breaking (LSB) involves a rank--two antisymmetric tensor field known as the Kalb--Ramond field \cite{42,assunccao2019dynamical, maluf2019antisymmetric}. This field naturally appears in the bosonic string theory spectrum \cite{43}. When non--minimally coupled to gravity and acquiring a non--zero vacuum expectation value, it undergoes spontaneous Lorentz symmetry breaking. The work in Ref. \cite{44} presents an exact solution for a static and spherically symmetric configuration within this context. Following this breakthrough, Ref. \cite{45} conducts a comprehensive investigation into the dynamics of both massive and massless particles near this static spherical Kalb--Ramond black hole. Additionally, Ref. \cite{46} explores the gravitational deflection of light and the shadows cast by rotating black holes within this theoretical framework. Moreover, GUP--corrected approach and the gravitational parity violation for an antisymmetric tensor as well as the cosmological implications of Kalb--Ramond--like--particles were recently addressed in the literature \cite{baruah2023quasinormal,manton2024kalb,capanelli2023cosmological}.

In recent years, there has been a noteworthy emphasis on exploring gravitational waves and their spectra, as highlighted in contemporary studies \cite{bombacigno2023landau,aa2023analysis,boudet2022quasinormal,hassanabadi2023gravitational,amarilo2023gravitational}. This heightened interest finds its roots in substantial advancements in gravitational wave detection technology, exemplified notably by the VIRGO and LIGO detectors. These sophisticated instruments have played a pivotal role in providing profound insights into the intriguing realm of black hole physics \cite{abbott2016gw150914,abramovici1992ligo,grishchuk2001gravitational,vagnozzi2022horizon}. In the work presented by Ref. \cite{yang2023static}, novel exact solutions for static and spherically symmetric spacetime are introduced, both in the presence and absence of the cosmological constant. These solutions are derived within the context of a non--zero vacuum expectation value background of the Kalb--Ramond field. Adding to these advancements, the authors recently proposed, within the same framework, a charged black hole \cite{duan2023electrically}.

Based on the black hole exploration, a notable area of focus centers around the study of quasinormal modes. These intricate oscillation frequencies manifest as a result of black holes relating to their initial perturbations. The derivation of these frequencies necessitates the imposition of specific boundary conditions \cite{berti2009quasinormal,konoplya2011quasinormal}. The scrutiny of gravitational waves and their spectra in recent years \cite{bombacigno2023landau,aa2023analysis,boudet2022quasinormal,hassanabadi2023gravitational,amarilo2023gravitational} has been particularly pronounced, paralleling advancements in gravitational wave detectors such as VIRGO and LIGO. These instruments have significantly enriched our understanding of black hole physics \cite{abbott2016gw150914,abramovici1992ligo,grishchuk2001gravitational,vagnozzi2022horizon}.

In this study, we explore the influence of anti--symmetric tensor effects, which induce Lorentz symmetry breaking, on charged spherically symmetric black holes. Initially, we provide an overview of the model, laying the foundation for deriving the black hole solutions. Following this, we conduct a thorough analysis covering the characteristics of horizons, critical orbits, and geodesics. In order to get more information about the stability of the theory, we calculate the quasinormal modes and the time--domain solution, focusing on vectorial perturbations. In parallel, we determine the Hawking temperature to probe the calculation of the remnant mass. Additionally, we estimate the lifetime for the final stage of the black hole when the evaporation process ends. Furthermore, we investigate the emission rate, also known as Hawking radiation, and the deflection angle. Finally, we examine the correspondence between quasinormal modes and shadows.


\section{The antisymmetric black hole solution}

The Kalb--Ramond field, denoted as $B_{\mu\nu}$, manifests as a rank--two antisymmetric tensor obeying the condition $B_{\mu\nu} = -B_{\nu\mu}$. Its field strength is defined as the 3--form $H_{\mu\nu\rho} \equiv \partial_{[\mu}B_{\nu\rho]}$. Under the gauge transformation $B_{\mu\nu} \rightarrow B_{\mu\nu} + \partial_{[\mu}\Gamma_{\nu]}$, the field strength remains invariant. To facilitate our analysis, we decompose the KR field as $B_{\mu\nu} = \tilde{E}_{[\mu}v_{\nu]} + \epsilon_{\mu\nu\alpha\beta}v^{\alpha}\tilde{B}^{\beta}$, where $\tilde{E}_{\mu}v^{\mu} = \tilde{B}_{\mu}v^{\mu} = 0$, and $v^{\alpha}$ represents a timelike 4--vector \cite{44,45}. Consequently, the spacelike pseudo--vector fields $\tilde{E}_{\mu}$ and $\tilde{B}_{\mu}$ can be seen as the pseudo--electric and pseudo--magnetic fields, respectively, drawing an direct analogy with Maxwell electrodynamics.

We initiate our analysis by examining the Einstein--Hilbert action, non--minimally coupled to a self--interacting Kalb--Ramond field, as presented in the form detailed by the expression below \cite{altschul2010lorentz}:
\ie
\begin{split}
\label{action}
S = & \frac{1}{2 \kappa^{2}} \int \mathrm{d}^{4}x \sqrt{-g} \left[ R - 2\Lambda - \frac{1}{6}H^{\mu\nu\rho}H_{\mu\nu\rho} - V(B_{\mu\nu}B^{\mu\nu}) + \xi_{2} B^{\rho\mu}B\indices{^\nu_\mu} R_{\rho\nu}  + \xi_{3} B^{\mu\nu}B_{\mu\nu} R        \right] \\
& + \int \mathrm{d}^{4}x\sqrt{-g} \mathcal{L}_{m}.
\end{split}
\fe
Here, $\xi_{2}$ and $\xi_{3}$ denote the coupling constants between the Ricci tensor and the Kalb--Ramond field, $\kappa = 8 \pi G$, $\Lambda$ represents the cosmological constant, and the field strength is defined as $H_{\mu\nu\rho} \equiv \partial_{[\mu}B_{\nu\rho]}$. The potential $V(B_{\mu\nu}B^{\mu\nu})$ plays a crucial role in inducing spontaneous Lorentz symmetry breaking. This aspect allows us to preserve the invariance of the theory under local Lorentz transformations. In pursuit of charged solutions, we adopt the matter Lagrangian $\mathcal{L}_{m}$ to describe the electromagnetic field, expressed as $\mathcal{L}_{m} = -\frac{1}{4}F_{\mu\nu}F^{\mu\nu} + \mathcal{L}_{\text{int}}$. Notice that, $F_{\mu\nu} = \partial_\mu A_\nu - \partial_\nu A_\mu$ denotes the field strength of the electromagnetic field, and $\mathcal{L}_{\text{int}}$ characterizes the interaction between the Kalb--Ramond and the electromagnetic fields.

The potential $V(B_{\mu\nu}B^{\mu\nu} \pm b^2)$ is intricately designed to hinge on $B_{\mu\nu}B^{\mu\nu}$, ensuring the theory's invariance when one considers local Lorentz transformations. While the cosmological constant $\Lambda$ is treated independently, the potential assumes a zero value at its minimum. This minimal state is dictated by the condition $B_{\mu\nu}B^{\mu\nu} = \mp b^2$, with the choice of $\pm$ rendering $b^2$ a positive constant. Consequently, the Kalb--Ramond field takes on a nonzero vacuum expectation value, denoted as $\langle B_{\mu\nu} \rangle = b_{\mu\nu}$. Due to the non--minimal coupling of the KR field to gravity, this nonzero VEV background, $b_{\mu\nu}$, spontaneously breaks local Lorentz invariance at the particle level. In the vacuum configuration, the interaction term $\xi_3B_{\mu\nu}B^{\mu\nu}R = \mp \xi_3b^2R$ in Eq. (\ref{action}) can be seamlessly absorbed into the Einstein--Hilbert terms through variable redefinition.

Furthermore, we posit the assumption that the vacuum state of the Kalb--Ramond field adopts a pseudo--electric configuration. In this arrangement, the sole non--vanishing components are defined by $b_{10} = -b_{01} = \tilde{E}(r)$, where the pseudo--electric field $\tilde{E}(r)$ adheres to the constant norm condition $b_{\mu\nu} b^{\mu\nu} = \mp b^2$ \cite{45}. Consequently, this specific configuration automatically renders the KR field strength null, i.e., $H_{\lambda\mu\nu} = 0$.

To describe electrically charged black hole solutions, we employ the conventional approach of introducing an electrostatic vector potential $A_{\mu} = -\Phi(r)\delta^{t}_{\mu}$. It is essential to recognize that a consistent representation of charged black holes necessitates more than just a standalone electromagnetic field. Hence, we must incorporate the interaction between the electromagnetic field and the Kalb--Ramond field. One approach is to modify the KR field strength $H_{\mu\nu\rho}$ by incorporating a $U(1)$ electromagnetic Chern--Simons three--form, expressed as $\tilde{H}_{\mu\nu\rho} = H_{\mu\nu\rho} + A_{[\mu}F_{\nu\rho]}$ \cite{majumdar1999parity,duan2023electrically}.

Nevertheless, when examining the vacuum KR configuration and the electrostatic vector potential, it is evident that all interactions in the modified kinetic term $\tilde{H}_{\mu\nu\rho}\tilde{H}^{\mu\nu\rho} = H_{\mu\nu\rho}H^{\mu\nu\rho} + H_{\mu\nu\rho}A_{[\mu}F_{\nu\rho]} + A^{[\mu}F^{\nu\rho]}A_{[\mu}F_{\nu\rho]}$ still vanish. To introduce nontrivial aspects concerning the dynamics of the spacetime, we therefore consider an interaction term as:
\ie
\mathcal{L}_{\text{int}} = -\eta B^{\alpha\beta}B^{\gamma\rho}F_{\alpha\beta}F_{\gamma\rho},
\fe
in which $\eta$ represents the coupling constant. It is worth mentioning that such an interaction term is fundamental in enabling the emergence of electrically charged black hole solutions.

To derive the gravitational field equations, we vary Eq. (\ref{action}) with respect to the metric tensor $g^{\mu\nu}$, yielding:
\ie
R_{\mu\nu} - \frac{1}{2}g_{\mu\nu}R + \Lambda g_{\mu\nu} = T^{m}_{\mu\nu} + T^{KR}_{\mu\nu} = \mathcal{T}_{\mu\nu}.
\fe
Here, $\mathcal{T}_{\mu\nu}$ stands for the total stress--energy tensor, with $T^{m}_{\mu\nu}$ and $T^{KR}_{\mu\nu}$ representing the stress--energy of the electromagnetic field
\ie
T^{m}_{\mu\nu} = \alpha\left( 4 F_{\mu\alpha}F_{\nu}^{\phantom{\nu}\alpha} - g_{\mu\nu}F_{\alpha\beta}F^{\alpha\beta}\right) + \eta \left(8B^{\alpha\beta}B\indices{_\nu^\gamma}F_{\alpha\beta}F_{\mu\gamma} - g_{\mu\nu}B^{\alpha\beta}B^{\gamma\rho}F_{\alpha\beta}F_{\gamma\rho} \right),
\fe
and the Kalb--Ramond field respectively
\ie
\begin{split}
T^{KR}_{\mu\nu} = & \frac{1}{2}H_{\mu\alpha\beta}H\indices{_\nu^\alpha
^\beta} - \frac{1}{12}g_{\mu\nu}H^{\alpha\beta\rho}H_{\alpha\beta\rho} + 2 V^{\prime}B_{\alpha\mu}B\indices{^\alpha_\nu} - g_{\mu\nu}V \\
& \xi_{2} \left\{ \frac{1}{2}g_{\mu\nu}B^{\alpha\gamma}B\indices{^\beta_\gamma}R_{\alpha\beta} - B\indices{^\alpha_\mu}B\indices{^\beta_\nu} R_{\alpha\beta} - B^{\alpha\beta}B_{\nu\beta}R_{\mu\alpha} - B^{\alpha\beta}B_{\mu\beta}R_{\nu\alpha} \right. \\
& \left.  \frac{1}{2} \nabla_{\alpha}\nabla_{\mu}(B^{\alpha\beta}B_{\nu\beta})  + \frac{1}{2} \nabla_{\alpha}\nabla_{\nu}(B^{\alpha\beta}B_{\mu\beta}) - \frac{1}{2} \nabla^{\alpha}\nabla_{\alpha}B\indices{_\mu^\gamma}B_{\nu\gamma}  \right. \\
& \left. -\frac{1}{2}g_{\mu\nu} \nabla_{\alpha}\nabla_{\beta}(B^{\alpha\gamma}B\indices{^\beta_\gamma})          \right\}.
\end{split}
\fe

In this context, the prime symbol indicates differentiation with respect to the respective function's argument. Employing the Bianchi identity, we can confirm the conservation of $\mathcal{T}_{\mu\nu}$. Also, the modified Maxwell are written as 
\ie
\label{mes}
\nabla^{\nu}(F_{\mu\nu} + 2 \eta B_{\mu\nu} B^{\alpha\beta}F_{\alpha\beta}) = 0,
\fe
where, naturally, if we consider $\eta \rightarrow 0$, we recover the well--known Maxwell equations.


\section{Charged black hole solutions}

In this study, our emphasis lies in the exploration of a static and spherically symmetric spacetime, featuring a non--zero vacuum expectation value (VEV) for the KR field. The metric tensor delineating this spacetime is articulated by the following line element:
\ie
\mathrm{d}s^2 = -F(r) \mathrm{d}t^2 + G(r) \mathrm{d}r^2 + r^2 \mathrm{d}\theta^2 + r^2 \sin^2 \theta \mathrm{d}\phi^2.
\fe
In this investigation, \(F(r)\) and \(G(r)\) denote functions determined by the radial coordinate \(r\) within the dynamics of the system, incorporating the influence of the KR field. Consequently, the pseudo--electric field \( \tilde{E}(r) \) can be reformulated as $
\tilde{E}(r) = \left| b \right| \sqrt{ \frac{F(r) G(r)}{2} },
$
ensuring adherence to the constant norm condition \( b_{\mu\nu} b^{\mu\nu} = -b^2 \). Also, it is appropriate to express the field equation in terms of $b$ as given below:
\ie
\begin{split}
R_{\mu\nu} =& \,\, T^{m}_{\mu\nu} - \frac{1}{2}g_{\mu\nu}T^{m} + \Lambda g_{\mu\nu} \\
& + V^{\prime}( b_{\mu\alpha} b\indices{_\nu^\alpha} + b^{2} g_{\mu\nu}) + \xi_{2} \left[ g_{\mu\nu} b^{\alpha\gamma} b^\beta{}_\gamma R_{\alpha\beta} - b\indices{^\alpha_\mu} b\indices{^\beta_\nu} R_{\alpha\beta} - b^{\alpha\beta}b_{\mu\beta}R_{\nu\alpha}   \right. \\ 
& \left. - b^{\alpha\beta} b_{\nu\beta} R_{\mu\alpha}
 + \frac{1}{2} \nabla_\alpha \nabla_\mu (b^{\alpha\beta} b_{\nu\beta}) + \frac{1}{2} \nabla_\alpha \nabla_\nu (b^{\alpha\beta} b_{\mu\beta}) - \frac{1}{2} \nabla^\alpha \nabla_\alpha (b\indices{_\mu^\gamma} b_{\nu\gamma}) \right], \label{modifiedfieldeq}
\end{split}
\fe
where $T^{m} \equiv T^{m}_{\alpha
\beta}g^{\alpha\beta}$.

With the prescribed metric ansatz, we can explicitly reformulate the field equations denoted in Eq. (\ref{modifiedfieldeq}) as follows \cite{duan2023electrically}:
\begin{subequations}
    \begin{align} \label{9a}
& \frac{2 F''}{F} - \frac{F'G'}{FG} - \frac{F'^{2}}{F^{2}} + \frac{4 F^{\prime}}{r F} + \frac{4 \Lambda G}{1-l} - \frac{4(1-2\eta b^{2})\Phi'^{2}}{(1-l)F} = 0, \\ \label{9b}
& \frac{2F''}{F} - \frac{F'G'}{FG} - \frac{F'^{2}}{F^{2}} - \frac{4 G'}{r G} + \frac{4 \Lambda G}{1-l} - \frac{4(1-2\eta b^{2})\Phi'^{2}}{(1-l)F} = 0,\\ \label{9c}
& \frac{2 F''}{F} - \frac{F'G'}{FG} - \frac{F'^{2}}{F^{2}} + \frac{1+l}{l r}\left( \frac{F'}{F} - \frac{G'}{G} \right) - (1-\Lambda r^{2} - b^{2} r^{2} V')\frac{2G}{l r^{2}}\\ \nonumber
& +  \frac{2(1-l)}{l r^{2}} - \frac{2(1-6\eta b^{2})\Phi'^{2}}{l F} = 0,
 \end{align}
\end{subequations}
where $ l \equiv \frac{\xi_2 b^2}{2}$. Notice that by subtracting \ref{9a} and \ref{9b}, we obtain 
\ie
\frac{4}{r} \left( \frac{F^{'}}{F} + \frac{G^{'}}{G}   \right) = 0, \quad \text{or} \quad \frac{F^{'}G + F G^{'}}{FG} = 0, \quad \text{so that} \quad \frac{\mathrm{d}(F G)}{FG} = 0, \quad \text{which leads to} \quad F G = 1.
\fe

Furthermore, the modified Maxwell equation (\ref{mes}) is properly rewritten as: 
\ie
\label{mmes}
(1-2\eta b^{2}) \left[ \Phi'' + \frac{\Phi'}{2} \left(  \frac{4}{r} - \frac{F'}{F} - \frac{G'}{G}   \right)   \right] = 0.
\fe


\section{The case where $\Lambda$ vanishes}

\begin{figure}
    \centering
    \includegraphics[scale=0.42]{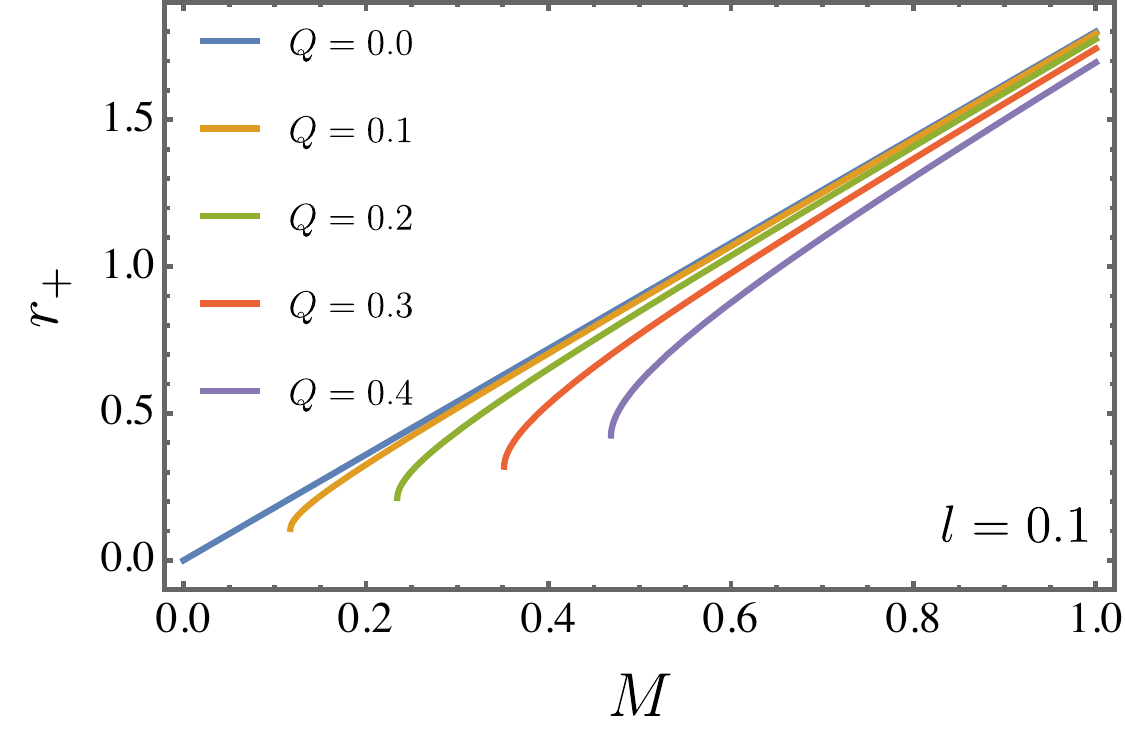}
    \includegraphics[scale=0.42]{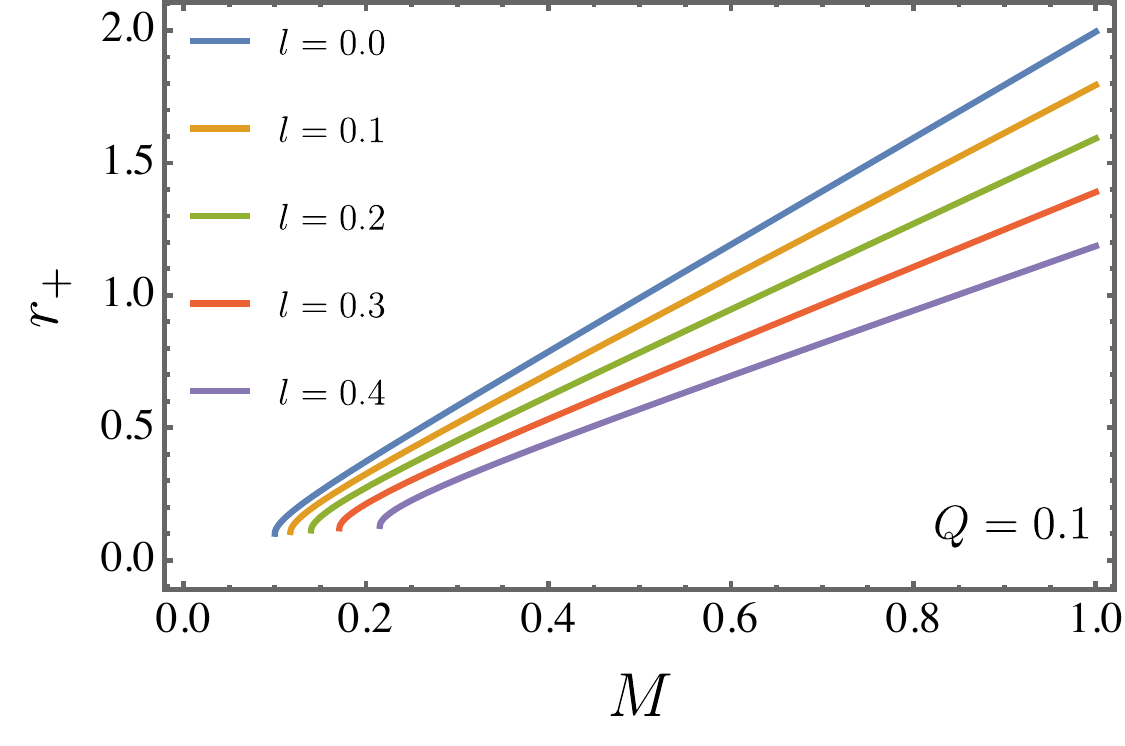}
    \caption{The event horizon is displayed, denoted as $r_{+}$, for two distinct configurations: one where the electric charge increases (on the left), and another where the parameter $l$ is varied (on the right).}
    \label{events}
\end{figure}

\begin{figure}
    \centering
    \includegraphics[scale=0.42]{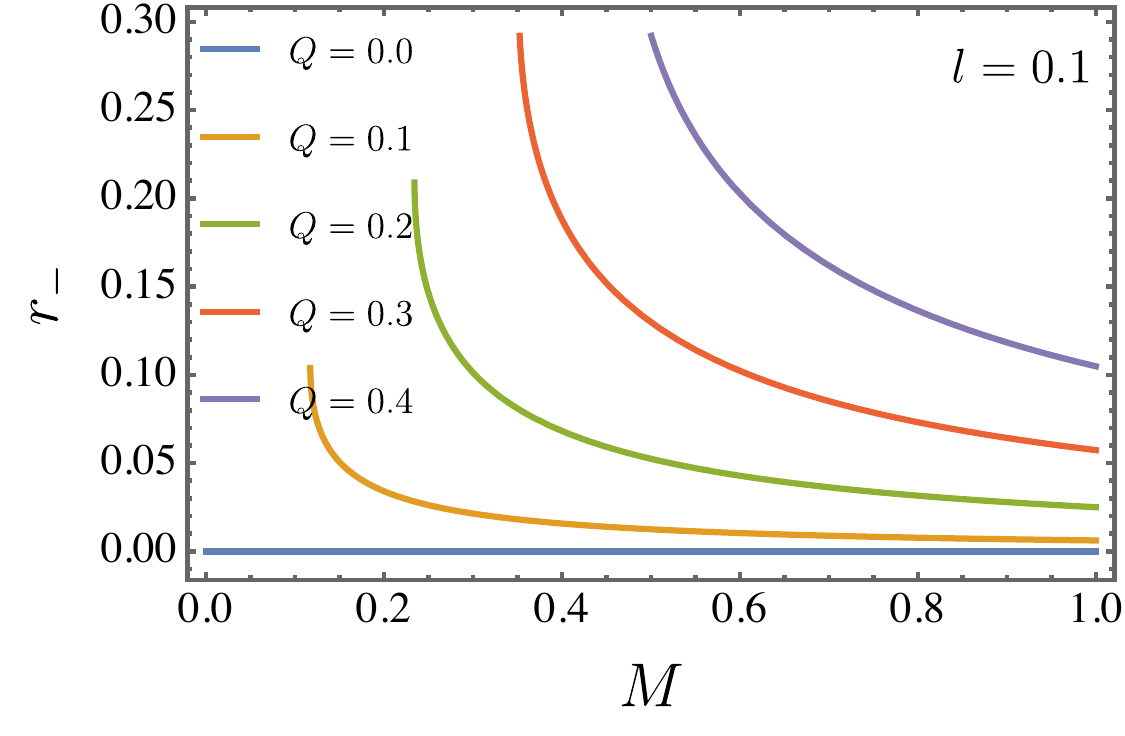}
    \includegraphics[scale=0.42]{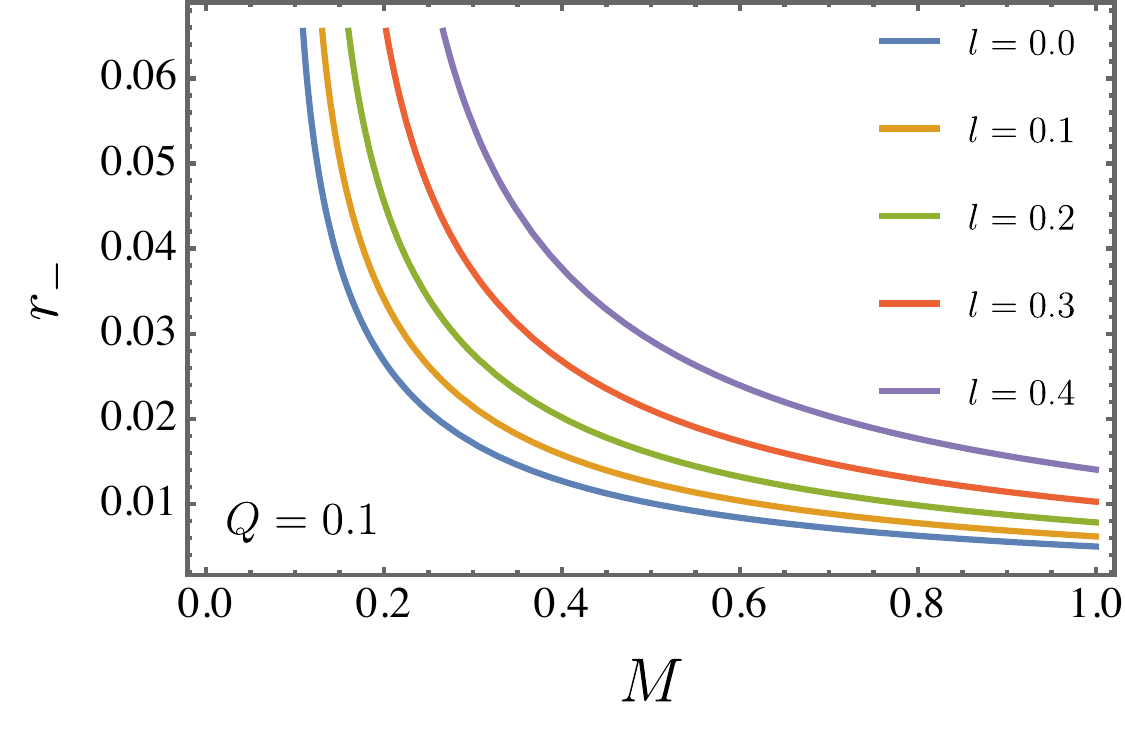}
    \caption{The behavior of the Cauchy horizon is shown, identified as $r_{-}$, under different values of $Q$ (on the left panel) and varying $l$ (on the right panel).}
    \label{events2}
\end{figure}

Upon substitution into the modified Maxwell equation (\ref{mmes}), we derive the expression for the electrostatic potential as \cite{duan2023electrically}:
\ie
\Phi(r) = \frac{c_{1}}{r} + c_{2},
\fe
setting the integration constant \(c_2\) to zero aligns with fixing the zero point of the potential at infinity. Nevertheless, this adjustment is made while considering the modification of the conserved current to \(J_\mu = \nabla_\nu (F^{\mu\nu} + 2\eta B^{\mu\nu} B^{\alpha\beta} F_{\alpha\beta}\)), the integration constant \(c_2\) can be determined through the application of Stokes's theorem
\ie
\begin{split}
Q = & - \frac{1}{4} \int_{\Sigma} \mathrm{d}^{3} x \sqrt{\gamma^{(3)}} n_{\mu}J^{\mu}\\
= & - \frac{1}{4 \pi} \int_{\partial\Sigma}\mathrm{d}\theta \mathrm{d}\phi \sqrt{\gamma^{(2)}}n_{\mu}\sigma_{\nu}(F^{\mu\nu} + 2 \eta B^{\mu\nu} B^{\alpha\beta}F_{\alpha\beta})\\
& = (1-2b^{2}\eta)c_{1}.
\end{split}
\fe
In the given context, let $\Sigma$ represent a three--dimensional spacelike region characterized by the induced metric $\gamma^{(3)}$. The boundary of $\Sigma$, denoted as $\partial\Sigma$, manifests as a two--sphere situated at spatial infinity, with the induced metric $\gamma^{(2)}_{ij} = \frac{1}{r^2} d\theta^2 + \sin^2 \theta d\phi^2$. In this framework, the vectors $n^\mu = (1, 0, 0, 0)$ and $\sigma^\mu = (0, 1, 0, 0)$ correspond to the unit normal vectors associated with $\Sigma$ and $\partial\Sigma$, respectively. Consequently, the electrostatic potential is expressed as:
\ie
\Phi(r) = \frac{Q}{(1-2\eta b^{2})r},
\fe
and, after some algebraic manipulations, $F(r)$ can be written as \cite{duan2023electrically}: 
\ie
F(r) = \frac{1}{1-l} - \frac{2M}{r} + \frac{1 + l -2(3-l)\eta b^{2} Q^{2}}{(1-l)^{2}(1-2\eta b^{2})^{2}r^{2}}.
\fe
Here, the integration constant has been precisely determined, allowing for the recovery of the Schwarzschild--like solution as presented in Ref. \cite{yang2023static}, particularly in the scenario where the electric charge $Q$ becomes negligible. Moreover, upon substituting the derived results into all the field equations, it becomes evident that the solutions exhibit consistency only if
\ie
\eta = \frac{l}{2b^{2}}.
\fe
Therefore, it is clear that, in the presence of Lorentz violation, the inclusion of the interaction term $\mathcal{L}_{\text{int}}$ is indispensable for obtaining a proper solution depicting a charged black hole. As a consequence, the charged metric with Lorentz--violating effects is given by
\ie
\mathrm{d}s^{2} = - \left( \frac{1}{1-l} - \frac{2M}{r} + \frac{Q^{2}}{(1-l)^{2}r^{2}}    \right) \mathrm{d}t^{2} + \frac{\mathrm{d}r^{2}}{\left( \frac{1}{1-l} - \frac{2M}{r} + \frac{Q^{2}}{(1-l)^{2}r^{2}}    \right)} + r^{2} \mathrm{d} \theta^{2} + r^{2}\sin^{2}\theta \mathrm{d}\phi^{2}.
\fe
It is worth mentioning that the manifestation of Lorentz violation, induced by the non--zero vacuum expectation value of the Kalb--Ramond field, is expressed by the dimensionless parameter \(l\). This parameter's small value is intricately constrained through classical gravitational experiments conducted within the Solar System \cite{yang2023static}. As the Lorentz--violating parameter \(l \to 0\), we recover the well--established result encountered in the literature: Reissner--Nordström case. Above metric, gives rise to the following horizons
\ie
r_{\pm} = (1-l)\left( M \pm \sqrt{M^{2} - \frac{Q^{2}}{(1-l)^{3}}} \right).
\fe
Actually, if we regard $Q \to 0$ in the horizons, we obtain the results established in Refs. \cite{yang2023static,reis2023exploring}. To gain a deeper understanding of the horizon, we showcase Figs. \ref{events} and \ref{events2}. Initially, we illustrate the event horizon, denoted as $r_{+}$, for two distinct configurations: one where the electric charge increases (on the left), and another where the parameter $l$ is varied (on the right). These features are visually represented in Fig. \ref{events}. It is noteworthy that, as the electric charge $Q$ approaches zero, the behavior aligns with what was observed in Ref. \cite{yang2023static}.

Moreover, in Fig. \ref{events2}, we explore the behavior of the Cauchy horizon, identified as $r_{-}$, under different values of $Q$ (on the left panel) and varying $l$ (on the right panel). Furthermore, we present Tabs. \ref{horizon1} and \ref{horizon2} to provide a quantitative analysis that complements the previously displayed plots.

\begin{table}[!h]
\begin{center}
\caption{\label{horizon1} The event horizon, represented by $r_{+}$, is depicted across different values of the mass $M$, charge $Q$, and parameter $l$.}
\begin{tabular}{c c c c ||| c c c c  ||| c c c c c } 
 \hline\hline
 $M$ & $l$ & $Q$ & $r_{+}$ & $M$ & $l$ & $Q$ & $r_{+}$ & $M$ & $l$ & $Q$ & $r_{+}$ &  \\ [0.2ex] 
 \hline 
 0.00 & 0.10 & 0.10 & -------- & 1.00 & 0.00 & 0.10 & 1.99499 & 1.00 & 0.10 & 0.00 & 1.80000 &  \\ 

 1.00 & 0.10 & 0.10 & 1.79381 & 1.00 & 0.05 & 0.10 & 1.89444 & 1.00 & 0.10 & 0.05 & 1.79846 &  \\
 
 2.00 & 0.10 & 0.10 & 3.59691 & 1.00 & 0.10 & 0.10 & 1.79381 & 1.00 & 0.10 & 0.10 & 1.79381 & \\
 
 3.00 & 0.10 & 0.10 & 5.39794 & 1.00 & 0.20 & 0.10 & 1.59215 & 1.00 & 0.10 & 0.20 & 1.77496 &   \\
 
 4.00 & 0.10 & 0.10 & 7.19846 & 1.00 & 0.30 & 0.10 & 1.38972 & 1.00 & 0.10 & 0.30 & 1.74261 &  \\
 
 5.00 & 0.10 & 0.10 & 8.99877 & 1.00 & 0.40 & 0.10 & 1.18595 & 1.00 & 0.10 & 0.40 & 1.69512 & \\
 
 6.00 & 0.10 & 0.10 & 10.7990 & 1.00 & 0.50 & 0.10 & 0.97958 & 1.00 & 0.10 & 0.50 & 1.62954 &  \\
 
 7.00 & 0.10 & 0.10 & 12.5991 & 1.00 & 0.60 & 0.10 & 0.76742 & 1.00 & 0.10 & 0.60 & 1.54031 &   \\
 
 8.00 & 0.10 & 0.10 & 14.3992 & 1.00 & 0.70 & 0.10 & 0.53804 & 1.00 & 0.10 & 0.70 & 1.41532 &   \\
 
9.00 & 0.10 & 0.10 & 16.1993 & 1.00 & 0.80 & 0.10 & -------- & 1.00 & 0.10 & 0.80 & 1.21447 &  \\
 
 10.0 & 0.10 & 0.10 & 17.9994 & 1.00 & 0.90& 0.10 & -------- &  1.00 & 0.10 & 0.90 & -------- &   \\[0.2ex] 
 \hline \hline
\end{tabular}
\end{center}
\end{table}

\begin{table}[!h]
\begin{center}
\caption{\label{horizon2}  The Cauchy horizon, represented by $r_{-}$, is shown across various values of the mass $M$, charge $Q$, and parameter $l$.}
\begin{tabular}{c c c c ||| c c c c  ||| c c c c c } 
 \hline\hline
 $M$ & $l$ & $Q$ & $r_{-}$ & $M$ & $l$ & $Q$ & $r_{-}$ & $M$ & $l$ & $Q$ & $r_{-}$ &  \\ [0.2ex] 
 \hline 
 0.00 & 0.10 & 0.10 & ----------- & 1.00 & 0.00 & 0.10 & 0.0050125 & 1.00 & 0.10 & 0.00 & ----------- &  \\ 

 1.00 & 0.10 & 0.10 & 0.0061941 & 1.00 & 0.05 & 0.10 & 0.0055564 & 1.00 & 0.10 & 0.05 & 0.0015445 &  \\
 
 2.00 & 0.10 & 0.10 & 0.0030890 & 1.00 & 0.10 & 0.10 & 0.0061941 & 1.00 & 0.10 & 0.10 & 0.0061941 & \\
 
 3.00 & 0.10 & 0.10 & 0.0020584 & 1.00 & 0.20 & 0.10 & 0.0078510 & 1.00 & 0.10 & 0.20 & 0.0250397 &   \\
 
 4.00 & 0.10 & 0.10 & 0.0015435 & 1.00 & 0.30 & 0.10 & 0.0102796 & 1.00 & 0.10 & 0.30 & 0.0573850 &  \\
 
 5.00 & 0.10 & 0.10 & 0.0012347 & 1.00 & 0.40 & 0.10 & 0.0140535 & 1.00 & 0.10 & 0.40 & 0.1048760 & \\
 
 6.00 & 0.10 & 0.10 & 0.0010289 & 1.00 & 0.50 & 0.10 & 0.0204168 & 1.00 & 0.10 & 0.50 & 0.1704640 &  \\
 
 7.00 & 0.10 & 0.10 & 0.0008818 & 1.00 & 0.60 & 0.10 & 0.0325765 & 1.00 & 0.10 & 0.60 & 0.2596880 &   \\
 
 8.00 & 0.10 & 0.10 & 0.0007716 & 1.00 & 0.70 & 0.10 & 0.0619524 & 1.00 & 0.10 & 0.70 & 0.3846790 &   \\
 
9.00 & 0.10 & 0.10 & 0.0006859 & 1.00 & 0.80 & 0.10 & ----------- & 1.00 & 0.10 & 0.80 & 0.5855340 &  \\
 
 10.0 & 0.10 & 0.10 & 0.0006170 & 1.00 & 0.90& 0.10 & ----------- &  1.00 & 0.10 & 0.90 & ----------- &   \\[0.2ex] 
 \hline \hline
\end{tabular}
\end{center}
\end{table}

\bigskip
\bigskip
\subsection{Critical orbits and geodesics}

For a thorough grasp of particle dynamics and the intricate patterns of light rays near black holes, a profound understanding of photon spheres is imperative. These spherical regions play a important role in unraveling the shadows cast by black holes and elucidating the impact of the electric charge and the KR field on the spacetime under examination. Several estimations exist for the bounds on the parameter $l$. For instance, constraints derived from Mercury's precession ($-3.7\times10^{-12} < l < +1.9\times10^{-11}$), light deflection ($-1.1\times10^{-10} < l < +5.4\times10^{-10}$), and the Shapiro time delay ($-6.1\times10^{-13} < l < +2.8\times10^{-14}$) have been recently reported in the literature \cite{yang2023static}.

Within the context of black hole dynamics, the photon sphere, often denoted as the critical orbit, assumes paramount importance. Employing the Lagrangian method, particularly in the computation of null geodesics, becomes essential to investigate the features of such regions. The primary objective of this study is to investigate the influence of specific parameters, namely $M$, $Q$, and $l$, on the photon sphere. To provide further clarity, we assert:
\ie
\mathcal{L} = \frac{1}{2} g_{\mu\nu}\Dot{x}^{\mu}\Dot{x}^{\nu}.
\fe
When the angle is set to \(\theta = \pi/2\), the previous equation turns out to:
\ie
g_{00}^{-1} E^{2} + g_{11}^{-1} \Dot{r}^{2} + g_{33}^{-1}L^{2} = 0.\label{sep123a}
\fe
Here, \(L\) denotes the angular momentum, and \(E\) represents the energy \cite{HouPRD2022}. As a consequence, Eq. (\ref{sep123a}) can be articulated as:
\ie
\Dot{r}^{2} = E^{2} - \left( \frac{1}{1-l} - \frac{2M}{r} + \frac{Q^{2}}{(1-l)^{2}r^{2}}    \right)\left(  \frac{L^{2}}{r^{2}} \right).
\fe
Notice that the effective potential is represented as $\mathcal{V} \equiv \left( \frac{1}{1-l} - \frac{2M}{r} + \frac{Q^{2}}{(1-l)^{2}r^{2}}    \right)\left(  \frac{L^{2}}{r^{2}} \right)$. To determine the critical radius, one needs to solve the equation \(\partial \mathcal{V}/\partial r = 0\). Remarkably, it turns out that two solutions give rise to. In other words, we have two photon spheres (critical orbits) given as follows:
\ie
r_{c_{out}} = -\frac{\sqrt{9 (l-1)^4 M^2+8 (l-1) Q^2}+3 (l-1)^2 M}{2 (l-1)},
\fe
and,
\ie
r_{c_{in}} = \frac{\sqrt{9 (l-1)^4 M^2+8 (l-1) Q^2}-3 (l-1)^2 M}{2 (l-1)}.
\fe

In this context, let $r_{c_{in}}$ and $r_{c_{out}}$ denote the radii of the critical orbits, referring to the inner and outer orbits, respectively. Analogously to the Schwarzschild and Reissner--Nordström black holes, the photon spheres of black hole under consideration is unstable. This instability arises from the nature of the effective potential for photon motion, which has a maximum at $r_{c_{out}}$ and $r_{c_{in}}$. Any small perturbation (inward or outward) causes photons to either spiral into the black hole or escape to infinity.  Notably, as $l \to 0$, we recover the standard case. Moreover, the introduction of Lorentz symmetry breaking imposes a constraint on the parameter $l$, resulting in a singularity when $l = 1$ is considered. It is crucial to emphasize that dual photon spheres have been observed in the context of the Simpson--Visser and Hayward--like solutions \cite{tsukamoto2021gravitational, tsukamoto2022retrolensing, guerrero2022multiring, aa2023implications, aa2023analysis}. In Tab. \ref{p1h2o3t4onspheres} and \ref{p1nspheres2}, we present the characteristics of the outer and the inner photon spheres, denoted as $r_{c_{out}}$ and $r_{c_{in}}$, respectively. Specifically, we observe that as the mass increases while keeping the parameters $l$ and $Q$ constant, there is a corresponding expansion in the radius of the associated critical orbits. Conversely, when the angular momentum $l$ and the charge $Q$ increase (while maintaining fixed values for the mass $M$), we observe a contraction in the size of the critical orbit.

In addition, it is worth noting that, in agreement with Tabs. \ref{horizon1} and \ref{horizon2}, which illustrate the quantitative behavior of the event and Cauchy horizons respectively, only the outer photon sphere, $ r_{c_{\text{out}}}$, lies outside the event horizon. As such, it is the only one relevant for physical purposes in this scenario. More so, recent studies in the literature have explored the geodesic structure within the context of Kalb--Ramond gravity, including the consideration of two photon spheres as well \cite{duan2023electrically}.

Furthermore, photon sphere stability near black holes is determined by the geometric and topological properties of optical spacetime, with conjugate points playing an important role. Perturbations to photon paths reveal different behaviors depending on the stability of the photon sphere. For unstable photon spheres, small deviations cause photons to either escape to infinity or be captured by the black hole. In contrast, stable photon spheres allow photons to remain confined to nearby orbits \cite{qiao2022curvatures,qiao2022geometric}.

This behavior is linked to the presence or absence of conjugate points in the spacetime manifold. Stable photon spheres exhibit conjugate points, while unstable ones do not. The Cartan--Hadamard theorem, which relates Gaussian curvature, $\mathcal{K}(r)$, to them, offers a framework for determining the stability of critical orbits \cite{qiao2024existence}. In this manner, we write the null geodesics, which satisfy $\mathrm{d}s^2=0$, yielding \cite{araujo2024effects,heidari2024absorption}:
\ie
\mathrm{d}t^2=\gamma_{ij}\mathrm{d}x^i \mathrm{d}x^j=\frac{1}{f(r)^2}\mathrm{d}r^2+\frac{r^2}{f(r)}\mathrm{d}\Omega^2,
\fe
in which, $i$ and $j$ lies on $1$ to $3$, and $\gamma_{ij}$ account for the optical metric. Also, the Gaussian curvature reads 
\ie
\label{dffdsf}
\mathcal{K}(r)=\frac{R}{2}= \frac{f(r)}{2} \frac{\mathrm{d}^{2}}{\mathrm{d} r^{2}}f(r) -\frac{\left(\frac{\mathrm{d}}{\mathrm{d} r}f(r)\right)^{2}}{4} = -\frac{6 M Q^2}{(1-l)^2 r^5}-\frac{2 M}{(1-l) r^3}+\frac{2 Q^4}{(1-l)^4 r^6}+\frac{3 Q^2}{(1-l)^3 r^4}+\frac{3 M^2}{r^4}.
\fe

\begin{figure}
    \centering
    \includegraphics[scale=0.67]{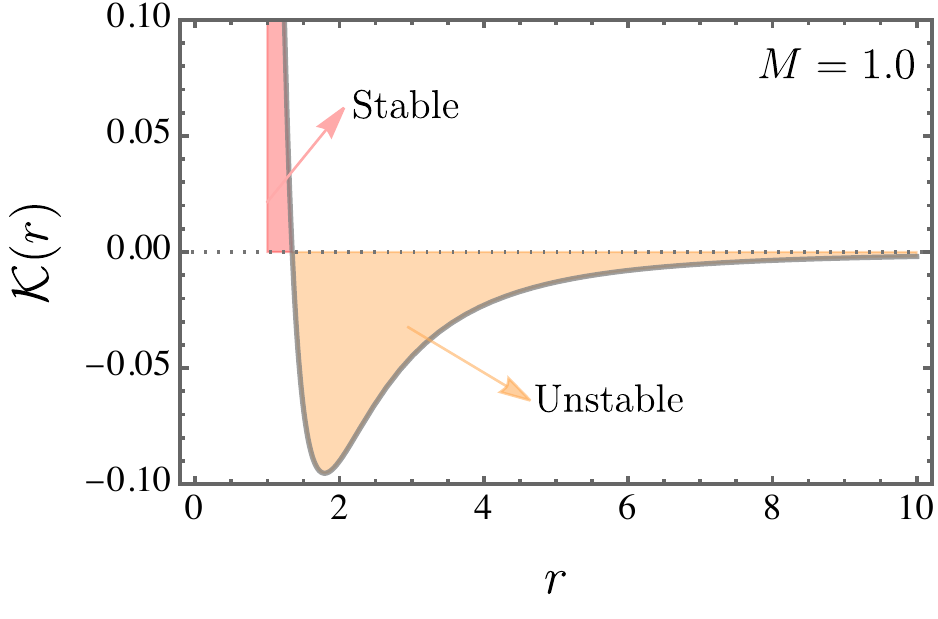}
    \caption{ The Gaussian curvature $\mathcal{K}(r)$ is plotted as a function of $r$, highlighting two distinct regions: stable and unstable. For this analysis, we set the parameters as $M = 1$, $Q = 0.1$, and $l = 0.1$.}
    \label{Gaussiancurvature}
\end{figure}

As discussed in Refs. \cite{qiao2022curvatures, qiao2022geometric,qiao2024existence}, the conditions $\mathcal{K}(r) < 0$ and $\mathcal{K}(r) > 0$ correspond to unstable and stable photon spheres, respectively. To illustrate the positivity of Eq. (\ref{dffdsf}), we present Fig. \ref{Gaussiancurvature}, where the Gaussian curvature $\mathcal{K}(r)$ is depicted as a function of $r$, emphasizing two distinct regions corresponding to stability and instability. The parameters used for accomplishing these analyses are $M = 1$, $Q = 0.1$, and $l = 0.1$.

The stability of $r_{c_{in}}$ and $r_{c_{out}}$ can be interpreted based on Fig. \ref{Gaussiancurvature} and Tabs. \ref{p1h2o3t4onspheres} and \ref{p1nspheres2}. In other words, $r_{c_{in}}$ corresponds to a region where the Gaussian curvature $\mathcal{K}(r)$ is positive, indicating a stable critical orbit, while $r_{c_{out}}$ lies in a region where $\mathcal{K}(r)$ is negative, signifying an unstable photon sphere.

\begin{table}[!h]
\begin{center}
\caption{\label{p1h2o3t4onspheres} The critical orbit, denoted by \(r_{c_{out}}\), is illustrated for a range of values pertaining to mass \(M\), $Q$, and parameter \(l\).}
\begin{tabular}{c c c c ||| c c c c  ||| c c c c c } 
 \hline\hline
 $M$ & $l$ & $Q$ & $r_{c_{out}}$ & $M$ & $l$ & $Q$ & $r_{c_{out}}$ & $M$ & $l$ & $Q$ & $r_{c_{out}}$ &  \\ [0.2ex] 
 \hline 
 0.00 & 0.10 & 0.10 & -------- & 1.00 & 0.00 & 0.10 & 2.99332 & 1.00 & 0.10 & 0.00 & 2.70000 &  \\ 

 1.00 & 0.10 & 0.10 & 2.69174 & 1.00 & 0.05 & 0.10 & 2.84259 & 1.00 & 0.10 & 0.05 & 2.69794 &  \\
 
 2.00 & 0.10 & 0.10 & 5.39588 & 1.00 & 0.10 & 0.10 & 2.69174 & 1.00 & 0.10 & 0.10 & 2.69174 & \\
 
 3.00 & 0.10 & 0.10 & 8.09726 & 1.00 & 0.20 & 0.10 & 2.38954 & 1.00 & 0.10 & 0.20 & 2.66667 &   \\
 
 4.00 & 0.10 & 0.10 & 10.7979 & 1.00 & 0.30 & 0.10 & 2.08631 & 1.00 & 0.10 & 0.30 & 2.62377 &  \\
 
 5.00 & 0.10 & 0.10 & 13.4984 & 1.00 & 0.40 & 0.10 & 1.78129 & 1.00 & 0.10 & 0.40 & 2.56117 & \\
 
 6.00 & 0.10 & 0.10 & 16.1986 & 1.00 & 0.50 & 0.10 & 1.47284 & 1.00 & 0.10 & 0.50 & 2.47559 &  \\
 
 7.00 & 0.10 & 0.10 & 18.8988 & 1.00 & 0.60 & 0.10 & 1.15678 & 1.00 & 0.10 & 0.60 & 2.36119 &   \\
 
 8.00 & 0.10 & 0.10 & 21.5990 & 1.00 & 0.70 & 0.10 & 0.81855 & 1.00 & 0.10 & 0.70 & 2.20651 &   \\
 
9.00 & 0.10 & 0.10 & 24.2991 & 1.00 & 0.80 & 0.10 & -------- & 1.00 & 0.10 & 0.80 & 1.98268 &  \\
 
 10.0 & 0.10 & 0.10 & 26.9992 & 1.00 & 0.90& 0.10 & -------- &  1.00 & 0.10 & 0.90 & 1.50000 &   \\[0.2ex] 
 \hline \hline
\end{tabular}
\end{center}
\end{table}

\begin{table}[!h]
\begin{center}
\caption{\label{p1nspheres2} The critical orbit, denoted by \(r_{c_{in}}\), is illustrated for a range of values pertaining to mass \(M\), $Q$, and parameter \(l\).}
\begin{tabular}{c c c c ||| c c c c  ||| c c c c c } 
 \hline\hline
 $M$ & $l$ & $Q$ & $r_{c_{in}}$ & $M$ & $l$ & $Q$ & $r_{c_{in}}$ & $M$ & $l$ & $Q$ & $r_{c_{in}}$ &  \\ [0.2ex] 
 \hline 
 0.00 & 0.10 & 0.10 & -------- & 0.10 & 0.00 & 0.10 & 0.0066815 & 0.10 & 0.10 & 0.00 & 0.000000 &  \\ 

 1.00 & 0.10 & 0.10 & 0.0082557 & 0.10 & 0.05 & 0.10 & 0.0074061 & 0.10 & 0.10 & 0.05 & 0.002059 &  \\
 
 2.00 & 0.10 & 0.10 & 0.0041183 & 0.10 & 0.10 & 0.10 & 0.0082557 & 0.10 & 0.10 & 0.10 & 0.008255 & \\
 
 3.00 & 0.10 & 0.10 & 0.0027444 & 0.10 & 0.20 & 0.10 & 0.0104623 & 0.10 & 0.10 & 0.20 & 0.033333 &   \\
 
 4.00 & 0.10 & 0.10 & 0.0020580 & 0.10 & 0.30 & 0.10 & 0.0136947 & 0.10 & 0.10 & 0.30 & 0.076226 &  \\
 
 5.00 & 0.10 & 0.10 & 0.0016462 & 0.10 & 0.40 & 0.10 & 0.0187131 & 0.10 & 0.10 & 0.40 & 0.138825 & \\
 
 6.00 & 0.10 & 0.10 & 0.0013718 & 0.10 & 0.50 & 0.10 & 0.0271584 & 0.10 & 0.10 & 0.50 & 0.224414 &  \\
 
 7.00 & 0.10 & 0.10 & 0.0011758 & 0.10 & 0.60 & 0.10 & 0.0432236 & 0.10 & 0.10 & 0.60 & 0.338813 &   \\
 
 8.00 & 0.10 & 0.10 & 0.0010288 & 0.10 & 0.70 & 0.10 & 0.0814443 & 0.10 & 0.10 & 0.70 & 0.493489 &   \\
 
9.00 & 0.10 & 0.10 & 0.0009145 & 0.10 & 0.80 & 0.10 & -------- & 0.10 & 0.10 & 0.80 & 0.717325 &  \\
 
 10.0 & 0.10 & 0.10 & 0.0008230 & 0.10 & 0.90& 0.10 & -------- &  0.10 & 0.10 & 0.90 & 1.200000 &   \\[0.2ex] 
 \hline \hline
\end{tabular}
\end{center}
\end{table}

The examination of particle motion within Lorentz--violating frameworks has garnered significant attention due to its profound theoretical implications. A central aspect of this exploration involves the geodesic characteristics that arise when both an antisymmetric tensor field and an electric charge coexist in a black hole solution. This investigation holds remarkable importance in unraveling diverse astrophysical phenomena related to celestial entities, including the features of accretion disks and shadows.

Our specific focus revolves around a comprehensive examination of the behavior dictated by the geodesic equations. To accomplish this ambitious objective, we formulate the geodesic equations as follows:
\begin{equation}
\frac{\mathrm{d}^{2}x^{\mu}}{\mathrm{d}s^{2}} + \Gamma\indices{^\mu_\alpha_\beta}\frac{\mathrm{d}x^{\alpha}}{\mathrm{d}s}\frac{\mathrm{d}x^{\beta}}{\mathrm{d}s} = 0. \label{geogeo}
\end{equation}

Utilizing an arbitrary affine parameter denoted as $s$  and the notation that the symbol ``$\prime$'' represents the derivative with respect to it (namely, $\mathrm{d}/\mathrm{d}s$), our investigation leads to the formulation of four coupled partial differential equations, which are given below:

\ie
\begin{split}
&\frac{\mathrm{d}}{\mathrm{d}s}t' = -\frac{2 r' t' \left(Q^2-(l-1)^2 M r\right)}{r \left((l-1) r (2 (l-1) M+r)-Q^2\right)} ,
\end{split}
\fe
\ie
\begin{split}
&\frac{\mathrm{d}}{\mathrm{d}s}r' = \frac{\left(t'\right)^2 \left(Q^2-(l-1)^2 M r\right) \left(Q^2-(l-1) r (2 (l-1) M+r)\right)}{r^{5}(l-1)^4}-\frac{r^4 \left(r'\right)^2 \left((l-1)^2 M r-Q^2\right)}{r^{5}(l-1) r (2 (l-1) M+r)-Q^2}\\
& + \frac{r^4 \sin ^2(\theta ) \left(\varphi '\right)^2 \left(Q^2-(l-1) r (2 (l-1) M+r)\right)}{r^{5}(l-1)^2}-\frac{r^4 \left(\theta '\right)^2 \left((l-1) r (2 (l-1) M+r)-Q^2\right)}{r^{5}(l-1)^2},
\end{split}
\fe
\ie
\begin{split}
&\frac{\mathrm{d}}{\mathrm{d}s}\theta' = \sin (\theta ) \cos (\theta ) \left(\varphi '\right)^2-\frac{2 \theta ' r'}{r},  
\end{split}
\fe
and
\ie
\begin{split}
&\frac{\mathrm{d}}{\mathrm{d}s}\varphi' = -\frac{2 \varphi ' \left(r'+r \theta ' \cot (\theta )\right)}{r}\,.
\end{split}
\fe

\begin{figure}
    \centering
    \includegraphics[scale=0.4]{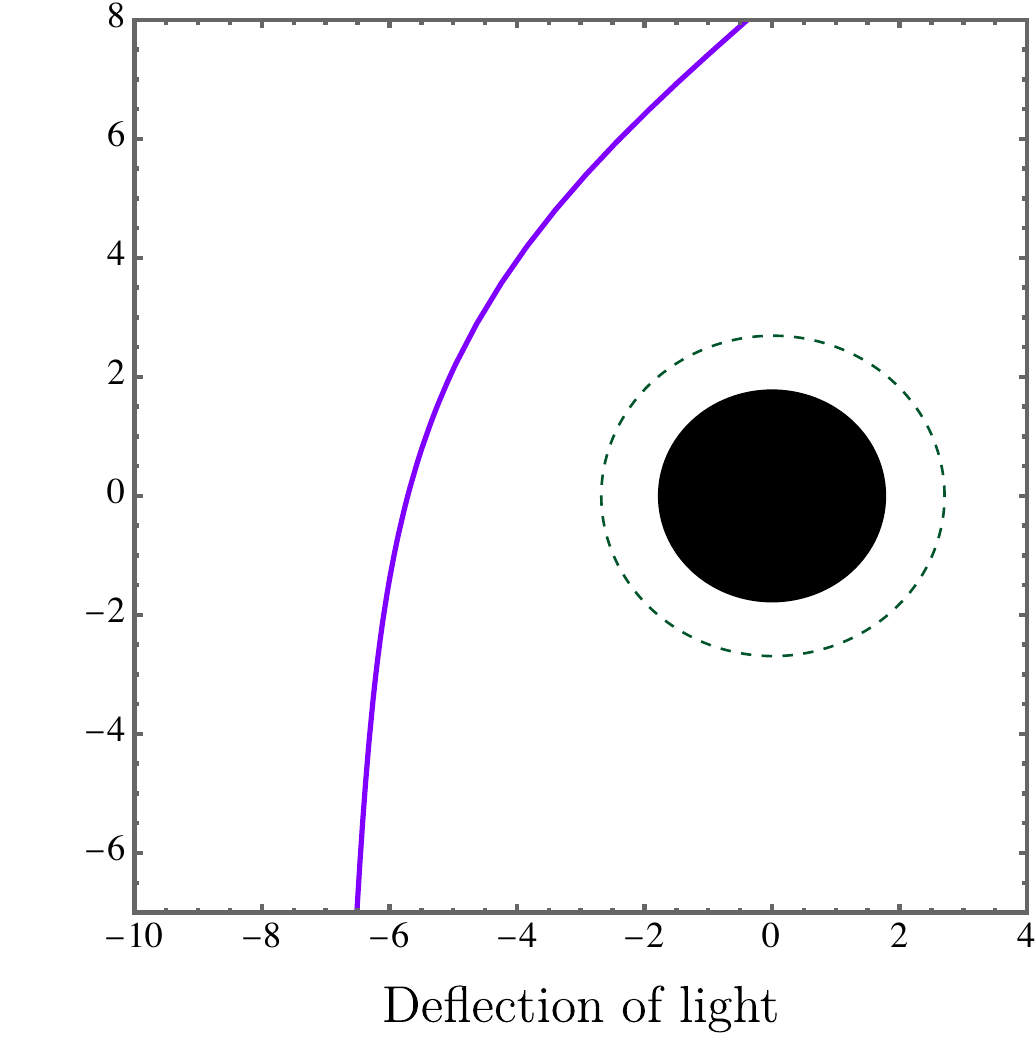}
    \caption{The presentation shows the deflection of light, with the light path depicted by the purple line. The photon sphere ($r_{c_{out}}$) is illustrated by dashed green lines, while the event horizon is represented by the black dot. The particular configuration of the system is regarded as following parameters: $Q=0.1$, $l=0.1$, and $M=1.0$.}
    \label{lightpath}
\end{figure}

\begin{figure}
    \centering
    \includegraphics[scale=0.5]{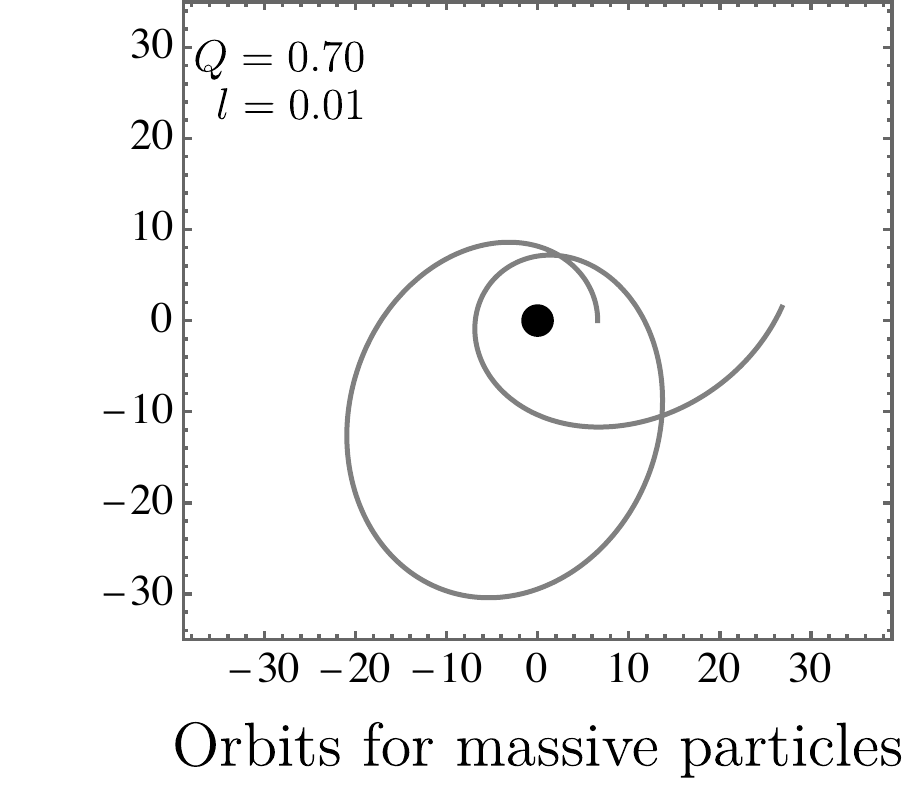}
    \caption{It is depicted one particular trajectory for massive particles when the system has the following configuration: $Q=0.7$, $l=0.01$, and $M=1.0$.}
    \label{massivetraj}
\end{figure}

It is important to mention that, since the spacetime under consideration is spherically symmetric, we focus on equatorial geodesics by setting $\theta = \pi/2$, where the energy and angular momentum along the $z$--direction remain conserved. To provide a more comprehensive understanding of the geodesic equation, we present Figs. \ref{lightpath} and \ref{massivetraj}. In Fig. \ref{lightpath}, we illustrate the deflection behavior of light for a specific system configuration, namely, $Q=0.1$, $l=0.1$, and $M=1.0$. Furthermore, in Fig. \ref{massivetraj}, we depict the trajectories of massive particles in the system characterized by $Q=0.7$, $l=0.01$, and $M=1.0$. Additionally, to represent the light trajectory, we have used the following numerical values:  $\text{ics} = \{t(0) = 0,\, r(0) = 8.15,\, \theta(0) = \pi/2,\, \varphi(0) = \pi/2\}$
and  
$\text{ivs} = \{\dot{r}(0) = -0.15,\, \dot{\theta}(0) = 0,\, \dot{\varphi}(0) = 0.02\}$.
For the massive particle modes, we have considered:  
$\text{ics} = \{t(0) = 0,\, r(0) = 6.6,\, \theta(0) = \pi/2,\, \varphi(0) = 0\}$  
and  $\text{ivs} = \{\dot{r}(0) = 0,\, \dot{\theta}(0) = 0,\, \dot{\varphi}(0) = 0.088\}$.


\subsection{Quasinormal frequencies}

In the ringdown phase, a notable feature comes to light: the presence of quasinormal modes \cite{94,95,96,97,98,99,100,101,102,103,104,105,106,107}. These modes exhibit distinctive oscillation patterns that correlate the influence of the system's initial state. Acting as unique signatures, they arise from the intrinsic vibrations of spacetime and persist consistently regardless of the specific starting conditions. In contrast to normal modes associated with closed systems, quasinormal modes bear relevance to open systems, resulting in a gradual energy dissipation through the emission of gravitational waves. Mathematically, these modes manifest as poles in the complex Green's function, offering a deeper understanding of the system's dynamic behavior during the ringdown phase.

Determining the frequencies of quasinormal modes necessitates solving the wave equation within a system governed by the background metric \(g_{\mu\nu}\). However, obtaining analytical solutions for these modes proves to be a formidable challenge. Various methods have been proposed in the scientific literature to tackle this task, with the WKB (Wentzel--Kramers--Brillouin) method emerging as one of the most prevalent. This technique draws significant influence from the pioneering work of Will and Iyer \cite{iyer1987black,iyer1987black1}. Subsequent advancements, including extending the method to the sixth order, have been contributed by Konoplya \cite{konoplya2003quasinormal}.

This section undertakes an examination of the propagation of a test electromagnetic field within the backdrop of a massive black hole. To accomplish this objective, we revisit the wave equations that govern a test electromagnetic field, expressed as follows:
\ie
\frac{1}{\sqrt{-g}}\partial_{\nu}\left[ \sqrt{-g} g^{\alpha \mu}g^{\sigma \nu} \left(A_{\sigma,  \alpha} -A_{\alpha,  \sigma}\right) \right]=0,  
\fe
where, the four--potential, denoted as \(A_{\mu}\), can be represented through an expansion in 4--dimensional vector spherical harmonics, taking the following form:
\begin{small}
\begin{align}\notag
& A_{\mu }\left( t,  r,  \theta, \phi \right)  \nonumber \\
&=\sum_{l, m} 
\begin{bmatrix} 
f(t,  r)Y_{l m}\left( \theta, \phi \right) \\
h(t,  r)Y_{l m}\left( \theta, \phi \right) \\
\frac{a(t,  r)}{\sin \left( \theta \right) }\partial _{\phi }Y_{l
m}\left( \theta, \phi \right) + k(t,  r)\partial _{\theta }Y_{l m}\left( \theta, \phi \right)\\
-a\left( t,  r\right) \sin \left( \theta \right) \partial _{\theta }Y_{lm}\left( \theta, \phi \right)+ k(t,  r)\partial _{\varphi }Y_{l m}\left( \theta, \phi \right)
\end{bmatrix}.%
\end{align}%
\end{small}
In this context, \(Y_{l m}(\theta, \phi)\) denotes the spherical harmonics utilized in the expansion. The first expression on the right side reflects a parity factor \((-1)^{l +1}\) and is recognized as the axial sector, whereas the second term possesses a parity factor \((-1)^l\), identified as the polar sector. By directly inserting this expansion into the Maxwell equations, we derive a second--order differential equation governing the radial component
\ie
\frac{\partial^{2} \psi}{\partial r^{*2}} - \left[  \omega^{2} - V_{eff}(r^{*})\right]\psi = 0.
\fe

The potential \(V_{eff}\) is commonly referred to as the \textit{Regge--Wheeler} potential, or simply the effective potential, encapsulating essential characteristics of the black hole's geometric structure. To facilitate our analysis, we introduce the tortoise coordinate \( r^{*} \), an extended coordinate system covering the entire spacetime and asymptotically approaching \( r^{*} \rightarrow \pm \infty \). The determination of this coordinate involves the equation \(\mathrm{d} r^{*} = \sqrt{[1/f(r)^{2}]}\mathrm{d}r\), which is elucidated as follows:
\ie
\begin{split}
r^{*}=  -(l-1)^2
&\left[ \frac{\left(2 (l-1)^3 M^2+Q^2\right) \tan ^{-1}\left(\frac{\sqrt{l-1} ((l-1) M+r)}{\sqrt{-(l-1)^3 M^2-Q^2}}\right)}{(l-1)^{3/2} \sqrt{-(l-1)^3 M^2-Q^2}} \right. \\
& \left. -M \ln \left((l-1) r (2 (l-1) M+r)-Q^2\right)+\frac{r}{l-1}\right].
\end{split}
\fe

The function $\psi(r^{*})$ is a complex composition that incorporates $a(t,  r)$, $f(t,  r)$, $h(t,  r)$, and $k(t,  r)$. Nevertheless, the specific functional dependencies vary depending on the parity. In the axial domain, the expression of the mode is stated as:
\ie
a(t,  r)=\psi \left( r^{*}\right),
\fe
and, for the polar sector, we obtain
\ie
\psi \left( r^{*}\right) =\frac{r^{2}}{%
\ell (\ell +1)}\left[ \partial _{t}h(t,  r)-\partial _{r}f(t,  r)\right].
\fe

By employing algebraic manipulation, we can express the effective potential in the following reformulated manner
\ie
V_{eff} = f(r) \left(\frac{\ell (\ell+1)}{r^2}\right). \label{ijasdijasd}
\fe

In the recent scientific discourse, a wave of investigations has explored quasinormal modes across various theoretical landscapes, spanning non--commutativity \cite{heidari2023gravitational,heidari2023exploring,hamil2024non}, bumblebee gravity \cite{hassanabadi2023gravitational}, regular black holes \cite{aa2023analysis,aa2023implications,a1jusufi2023dark,a2jusufi2024charged}, and other intriguing contexts. In our pursuit of computing these quasinormal modes, we employ the WKB approach as our chosen methodology. Our principal objective is to uncover stationary solutions for the system under consideration.

The WKB method, pioneered by Schutz and Will \cite{schutz1985black}, stands as an invaluable tool for unraveling quasinormal modes, particularly in the intricate realm of particle scattering near black holes. This method has witnessed refinements over time, marked by substantial contributions from Konoplya \cite{konoplya2003quasinormal, konoplya2004quasinormal}. It is essential to emphasize that the effectiveness of this technique is contingent upon the potential adopting a barrier--like shape that eventually levels off to constant values as \( r^{*} \to \pm \infty \). By aligning the terms in the power series solution with the turning points of the peak potential, precise computations of quasinormal modes become achievable. Under these stipulations, the formula developed by Konoplya is articulated as follows:
\ie
\frac{i(\omega^{2}_{n}-V_{0})}{\sqrt{-2 V^{''}_{0}}} - \sum^{6}_{j=2} \Lambda_{j} = n + \frac{1}{2}.
\fe
At its core, Konoplya's framework for determining quasinormal frequencies encompasses multiple key elements. Specifically, the term \( V^{''}_{0} \) represents the second derivative of the potential, evaluated at its apex \( r_{0} \). Moreover, the constants \( \Lambda_{j} \) are intricately influenced by both the effective potential and its derivatives at this pinnacle. Notably, recent strides in the field have ushered in a 13th--order WKB approximation, pioneered by Matyjasek and Opala \cite{matyjasek2017quasinormal}. This advancement significantly augments the precision of quasinormal frequency calculations.

It is crucial to emphasize that the quasinormal frequencies are characterized by a negative imaginary component. This distinctive attribute indicates that these modes undergo exponential decay over time, thereby embodying the mechanism of energy dissipation through waves. This characteristic aligns seamlessly with the existing scholarly literature that has explored perturbations involving scalar, electromagnetic, and gravitational fields within a spherically symmetric configuration \cite{konoplya2011quasinormal,berti2009quasinormal,chen2023quasinormal}.

In Tables \ref{laa11} and \ref{laa22}, we present the dynamics of quasinormal modes for varying values of $l$. These modes evolve in response to changes in two pivotal parameters, namely, $Q$ and $l$, utilizing the sixth--order WKB approximation. Strikingly, a uniform pattern emerges across all scenarios: with the increase of $Q$ and $l$, there is a noteworthy augmentation in the damping of these frequencies.

\begin{table}[!h]
\begin{center}
\caption{\label{laa11}By applying the sixth--order WKB approximation, we elucidate the quasinormal frequencies associated with diverse values of \( Q \) and \( l \), with a specific focus on cases where \( M=1 \) and $\ell=1$.}
\begin{tabular}{c| c | c | c} 
 \hline\hline\hline 
 \!\!\!\!\! $Q$ \,\,\,\,\,\,\,\,  $l$  & $\omega_{0}$ & $\omega_{1}$ & $\omega_{2}$  \\ [0.2ex] 
 \hline 
 0.00, \, 0.10 & 0.287540 - 0.113893$i$ & 0.243969 - 0.363696$i$  & 0.195309 - 0.658760$i$  \\ 
 
 0.10, \, 0.10  & 0.288306 - 0.114003$i$ & 0.244851 - 0.363967$i$ & 0.196349 - 0.659009$i$ \\
 
 0.20, \, 0.10  & 0.290657 - 0.114331$i$ & 0.247568 - 0.364757$i$  &  0.199563 - 0.659672$i$    \\
 
 0.30, \, 0.10  & 0.294757 - 0.114864$i$ & 0.252335 - 0.366006$i$  &  0.205221 - 0.660582$i$  \\
 
 0.40, \, 0.10  & 0.300926 - 0.115575$i$ & 0.259561 - 0.367579$i$  & 0.213830 - 0.661408$i$  \\
 
 0.50, \, 0.10  & 0.309720 - 0.116387$i$ & 0.269950 - 0.369181$i$ & 0.226213 - 0.661547$i$   \\
   [0.2ex] 
 \hline \hline \hline 
  \!\!\!\!\! $Q$ \,\,\,\,\,\,\,\, $l$ & $\omega_{0}$ & $\omega_{1}$ & $\omega_{2}$  \\ [0.2ex] 
 \hline 
 0.10, \, 0.00 & 0.248666 - 0.092700$i$ & 0.214835 - 0.294276$i$ & 0.174620 - 0.530213$i$  \\ 
 
 0.10, \, 0.10  & 0.288306 - 0.114003$i$ & 0.244851 - 0.363967$i$ & 0.196349 - 0.659009$i$ \\
 
 0.10, \, 0.20  & 0.339709 - 0.143613$i$ & 0.282338 - 0.461674$i$ &  0.223358 - 0.840482$i$  \\
 
 0.10, \, 0.30  & 0.408428 - 0.186513$i$  & 0.330058 - 0.604716$i$ &  0.258118 - 1.107510$i$  \\
 
 0.10, \, 0.40  & 0.503966 - 0.252132$i$ & 0.392207 - 0.826224$i$  & 0.305642 - 1.522770$i$   \\
 
 0.10, \, 0.50  & 0.644045 - 0.360294$i$ & 0.475697 - 1.196450$i$  & 0.378797 - 2.217680$i$   \\
   [0.2ex] 
 \hline \hline \hline 
\end{tabular}
\end{center}
\end{table}

\begin{table}[!h]
\begin{center}
\caption{\label{laa22}Utilizing the sixth--order WKB approximation, we present the quasinormal frequencies corresponding to various values of \( Q \) and $l$, particularly when $M=1$ and \( \ell=2 \).}
\begin{tabular}{c| c | c | c} 
 \hline\hline\hline 
  \!\!\!\!\! $Q$ \,\,\,\,\,\,\,\,  $l$   & $\omega_{0}$ & $\omega_{1}$ & $\omega_{2}$  \\ [0.2ex] 
 \hline 
 0.00, \, 0.10 &  0.534143 - 0.117132$i$ & 0.506847 - 0.359187$i$  & 0.461547 - 0.621860$i$ \\ 
 
 0.10, \, 0.10  & 0.53543 - 0.117227$i$ & 0.508207 - 0.359449$i$ & 0.463030 - 0.622225$i$ \\
 
 0.20, \, 0.10  & 0.539375 - 0.117507$i$ & 0.512376 - 0.360223$i$  &  0.467583 - 0.623290$i$   \\
 
 0.30, \, 0.10  & 0.546235 - 0.117959$i$ & 0.519643 - 0.361456$i$  &  0.475535 - 0.624946$i$   \\
 
 0.40, \, 0.10  & 0.556513 - 0.118546$i$ & 0.530564 - 0.363021$i$ & 0.487539 - 0.626921$i$  \\
 
 0.50, \, 0.10  & 0.571081 - 0.119183$i$ & 0.546109 - 0.364629$i$  & 0.504704 - 0.628636$i$ \\
   [0.2ex] 
 \hline \hline \hline 
  \!\!\!\!\! $Q$ \,\,\,\,\,\,\,\,  $l$  & $\omega_{0}$ & $\omega_{1}$ & $\omega_{2}$  \\ [0.2ex] 
 \hline 
 0.10, \, 0.00 & 0.458393 - 0.095066$i$ & 0.437374 - 0.290883$i$ & 0.401816 - 0.501948$i$ \\ 
 
 0.10, \, 0.10  & 0.535430 - 0.117227$i$  & 0.508207 - 0.359449$i$  & 0.463030 - 0.622225$i$ \\
 
 0.10, \, 0.20  & 0.636878 - 0.148155$i$  & 0.600559 - 0.455477$i$  &  0.541721 - 0.791419$i$   \\
 
 0.10, \, 0.30  & 0.775250 - 0.193183$i$ & 0.724963 - 0.595882$i$  &  0.645984 - 1.040070$i$   \\
 
 0.10, \, 0.40  & 0.972911 - 0.262428$i$ & 0.899878 - 0.812968$i$ & 0.789840 - 1.426820$i$ \\
 
  0.10, \, 0.50  & 1.274010 - 0.377102$i$ & 1.161020 - 1.174980$i$ & 1.000480 - 2.076090$i$ \\
   [0.2ex] 
 \hline \hline \hline 
\end{tabular}
\end{center}
\end{table}


{\subsection{Time--domain solution}

Investigating vectorial perturbations in the time domain is essential for understanding the influence of the quasinormal spectrum on time-dependent scattering processes. The complexity of the effective potential requires a precise methodology to capture its effects accurately. To this end, the characteristic integration technique proposed by Gundlach et al. \cite{Gundlach:1993tp} is employed, providing a reliable framework for examining quasinormal modes in dynamic scattering scenarios and their relevance to black hole physics.

The method described in Refs. \cite{Baruah:2023rhd, Bolokhov:2024ixe,Guo:2023nkd,Yang:2024rms,Gundlach:1993tp, Skvortsova:2024wly,Shao:2023qlt} relies on the use of light-cone coordinates, defined by $u = t - r^{*}$ and $v = t + r^{*}$. These coordinates simplify the structure of the wave equation, allowing for a more efficient analysis. Under this framework, the wave equation can be rewritten as
\ie
\left(  4 \frac{\partial^{2}}{\partial u \partial v} + V(u,v)\right) \bar{\psi} (u,v) = 0 \label{timedomain}.
\fe

The expression can be integrated effectively by employing a discretization scheme that combines a standard finite--difference method with numerical techniques
\ie
\bar{\psi}(N) = -\bar{\psi}(S) + \bar{\psi}(W) + \bar{\psi}(E) - \frac{h^{2}}{8}V(S)[\bar{\psi}(W) + \bar{\psi}(E)] + \mathcal{O}(h^{4}),
\fe
The points are defined as $S = (u, v), \quad W = (u + h, v), \quad E = (u, v + h), \quad N = (u + h, v + h),$ where $h$ represents the grid spacing factor. The null surfaces described by $u = u_{0}$ and $v = v_{0}$ are crucial, as they provide the reference points for initializing the system. In this work, the initial data on the null surface $u = u_{0}$ is specified by a Gaussian profile centered at $v = v_{c}$ with a width parameter $\sigma$
\ie
\bar{\psi}(u=u_{0},v) = A e^{-(v-v_{0})^{2}}/2\sigma^{2}, \,\,\,\,\,\, \bar{\psi}(u,v_{0}) = \bar{\psi}_{0}.
\fe
A constant initial condition $\bar{\psi}(u, v_{0}) = \bar{\psi}_{0}$ is specified at $v = v_{0}$, where $\bar{\psi}_{0}$ is set to zero without loss of generality. The integration progresses along surfaces of constant $u$ as $v$ increases, following the prescribed null data. This analysis considers a scalar test field with the choice $M = 1$ for simplicity. The initial data is modeled by a Gaussian profile centered at $v = -40$, with a width $\sigma = 1$ and $\bar{\psi}_{0} = 0$. The computational grid extends over $u \in [120, 200]$ and $v \in [120, 200]$.

In Figs. \ref{wavefunction1}--\ref{lnwavefunction2}, we show the behavior $\bar{\psi}$ as well as $\ln |\bar{\psi}|$ for the case where the cosmological constant vanishes.

\begin{figure}
    \centering
    \includegraphics[scale=0.4]{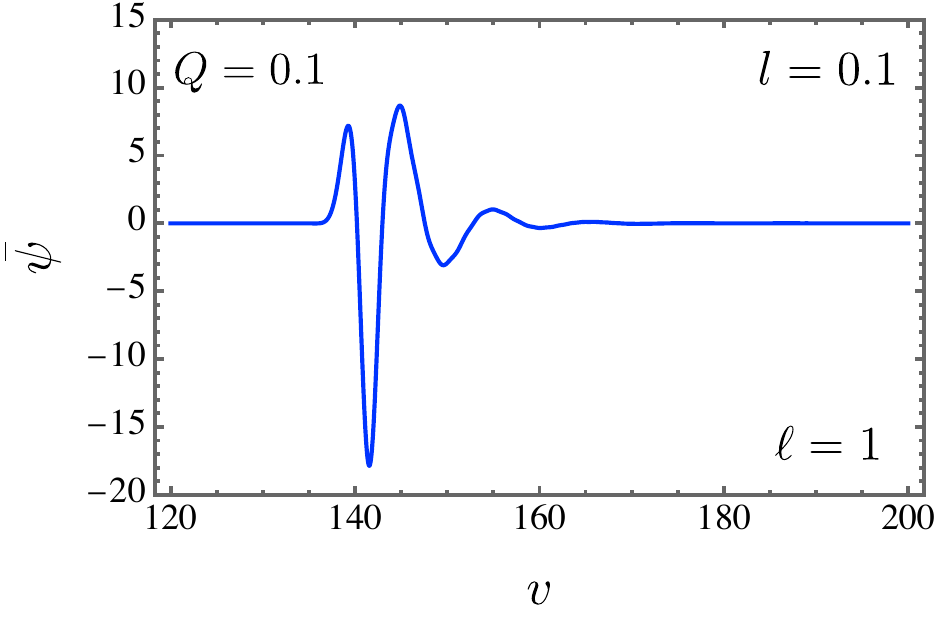}
    \includegraphics[scale=0.4]{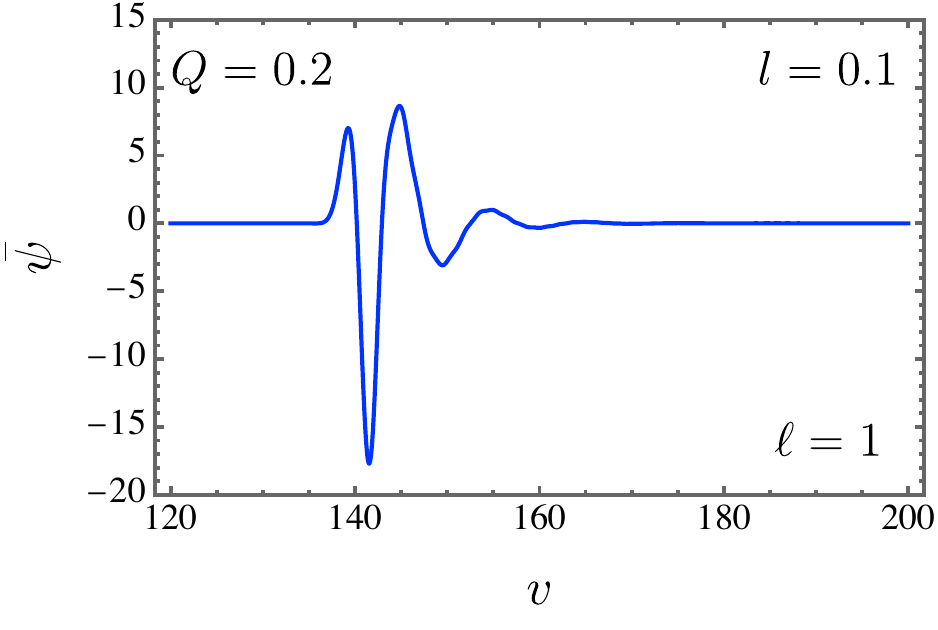}
    \includegraphics[scale=0.4]{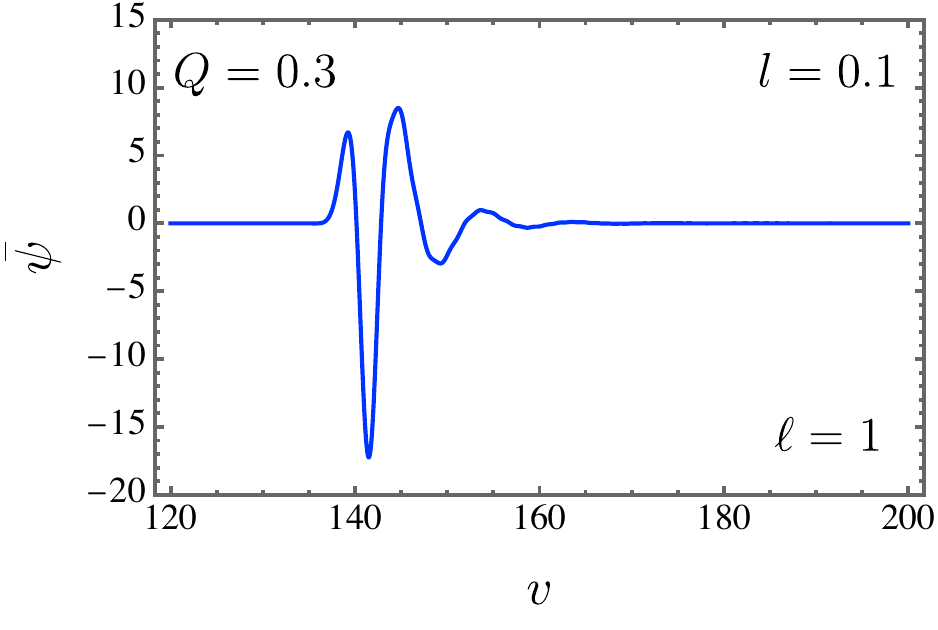}
    \includegraphics[scale=0.4]{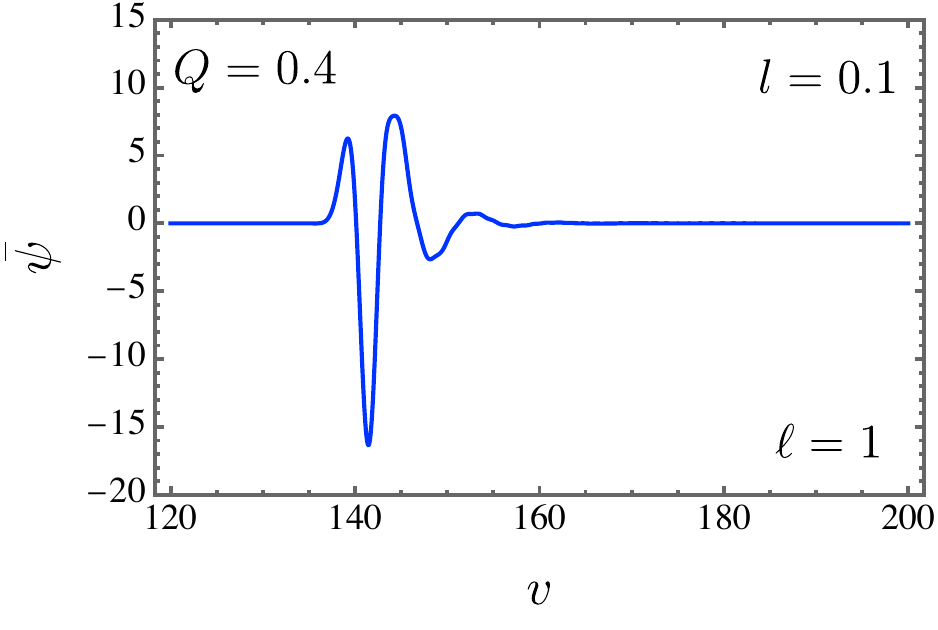}
    \caption{The wave function $\bar{\psi}$ is analyzed for different values of $v$, considering various values of $Q$ while keeping the parameters fixed as $l = 0.1$, $\ell = 1$, and $M = 0.5$.}
    \label{wavefunction1}
\end{figure}

\begin{figure}
    \centering
   \includegraphics[scale=0.4]{wavefunctionQ01.pdf}
   \includegraphics[scale=0.4]{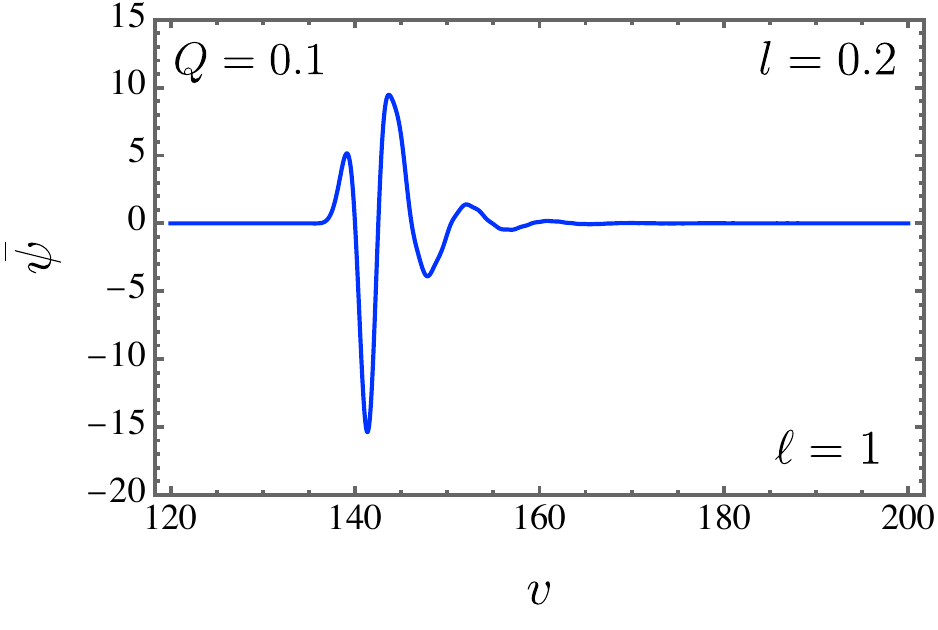}
   \includegraphics[scale=0.4]{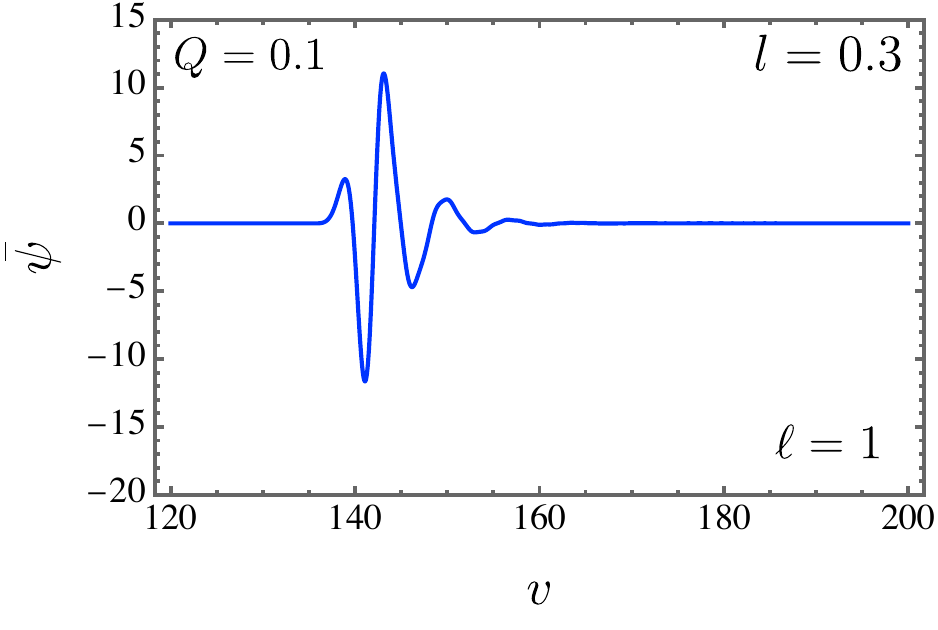}
   \includegraphics[scale=0.4]{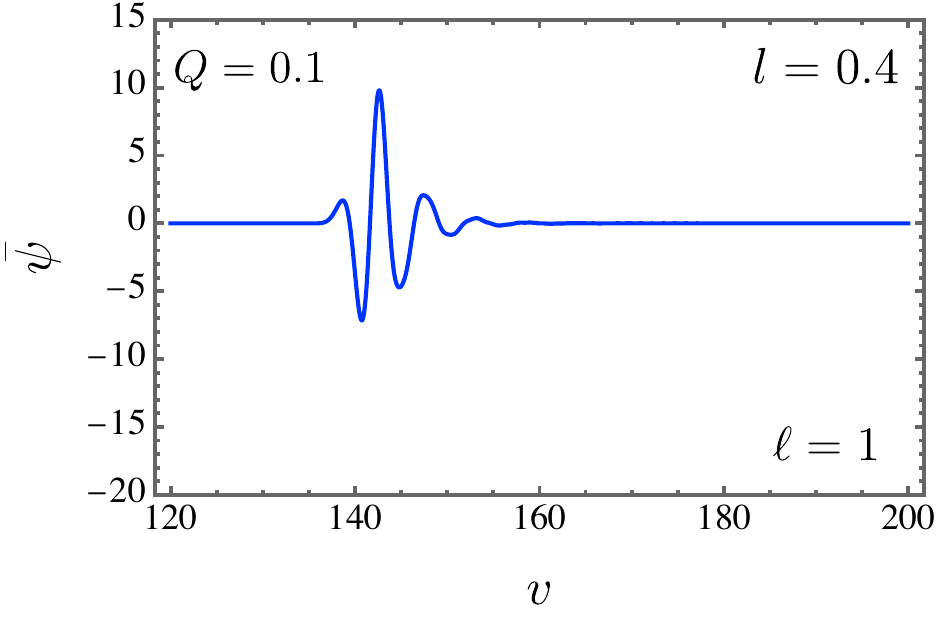}
    \caption{The wave function $\bar{\psi}$ is analyzed for different values of $v$, considering various values of $l$ while keeping the parameters fixed as $Q = 0.1$, $\ell = 1$, and $M = 0.5$.}
    \label{wavefunction2}
\end{figure}

\begin{figure}
    \centering
    \includegraphics[scale=0.4]{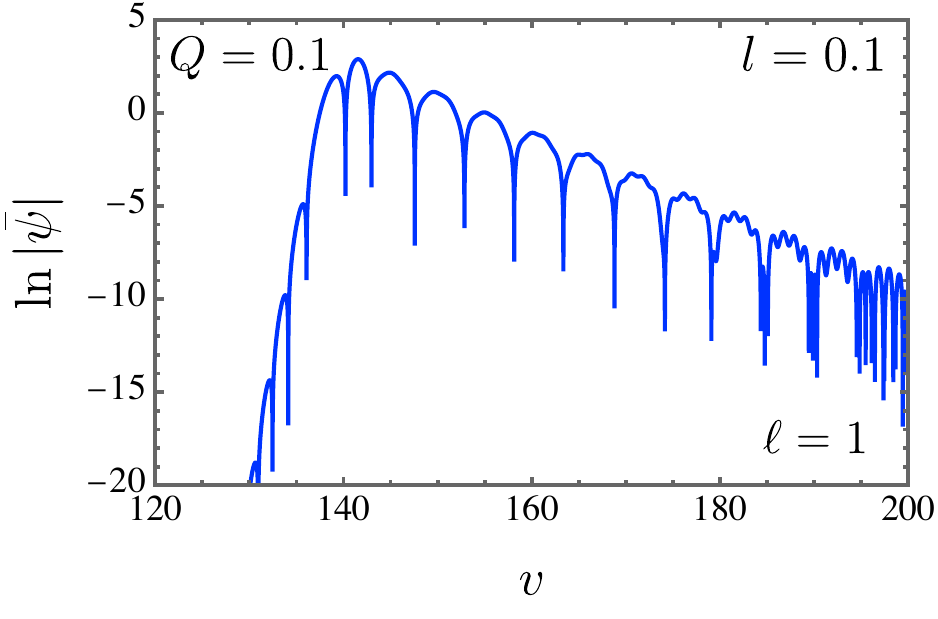}
    \includegraphics[scale=0.4]{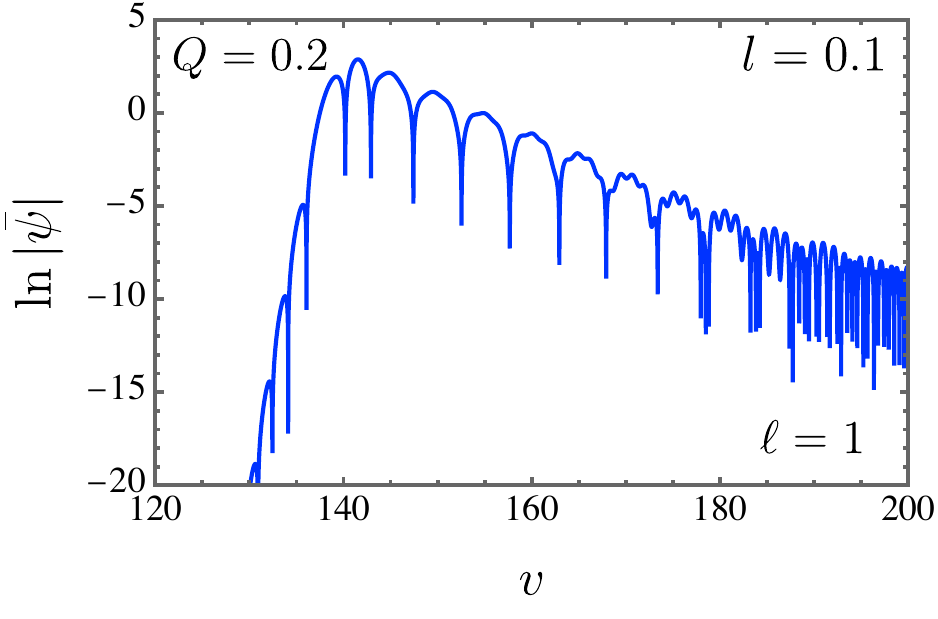}
    \includegraphics[scale=0.4]{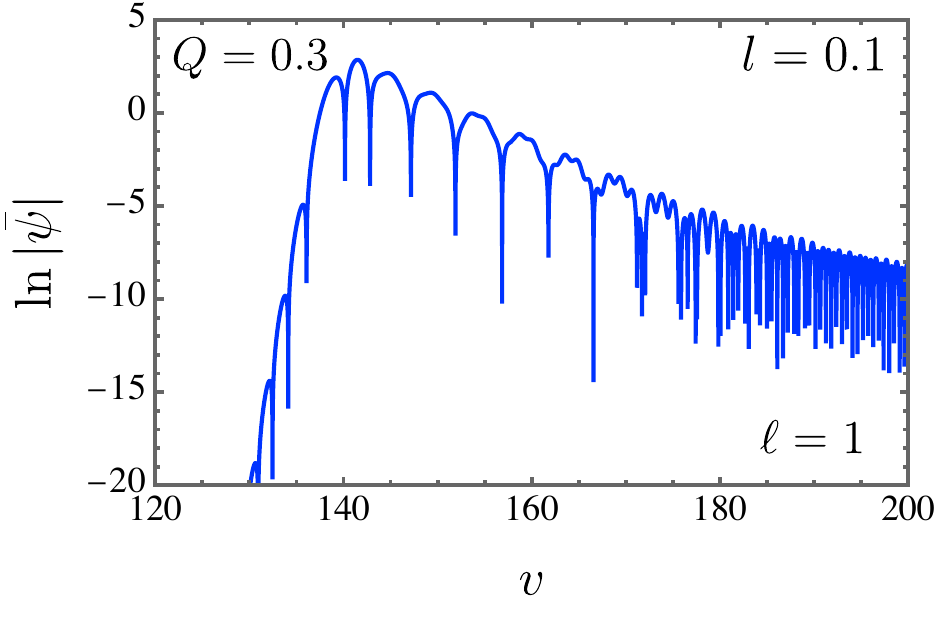}
    \includegraphics[scale=0.4]{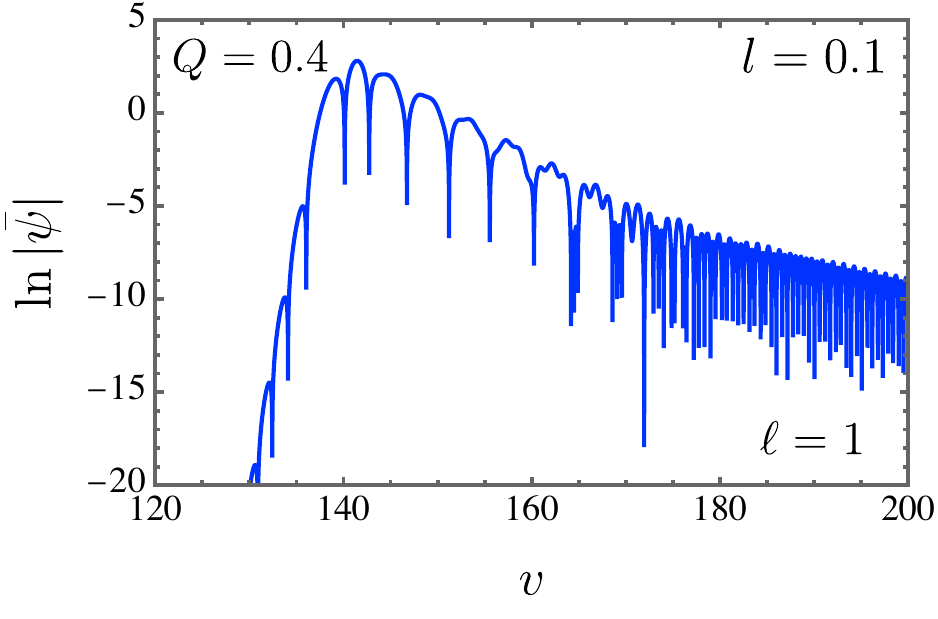}
    \caption{The wave function $\ln|\bar{\psi}|$ is analyzed for different values of $v$, considering various values of $Q$ while keeping the parameters fixed as $l = 0.1$, $\ell = 1$, and $M = 0.5$.}
    \label{lnwavefunction1}
\end{figure}

\begin{figure}
    \centering
    \includegraphics[scale=0.4]{lnwavefunctionQ01.pdf}
    \includegraphics[scale=0.4]{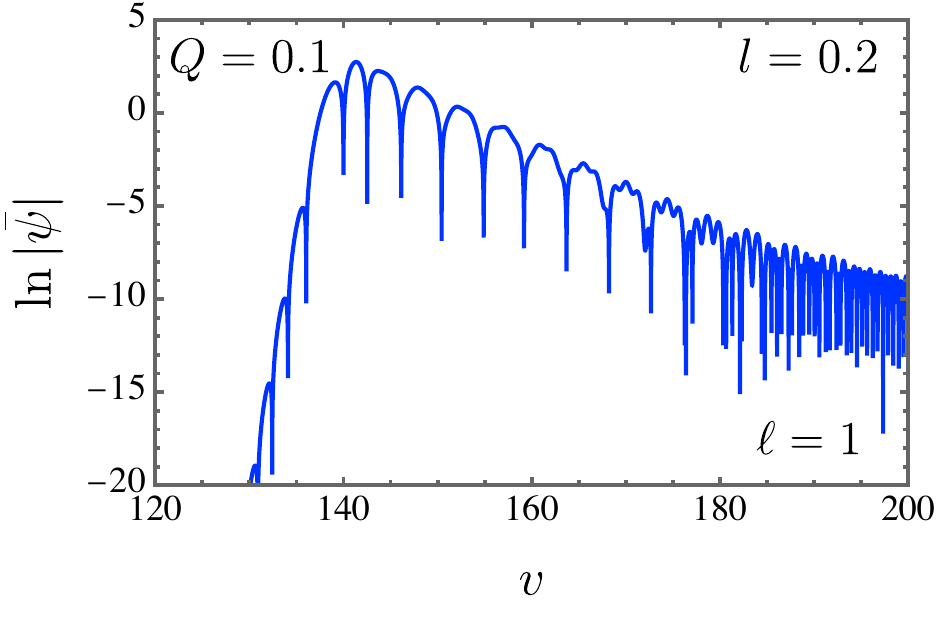}
    \includegraphics[scale=0.4]{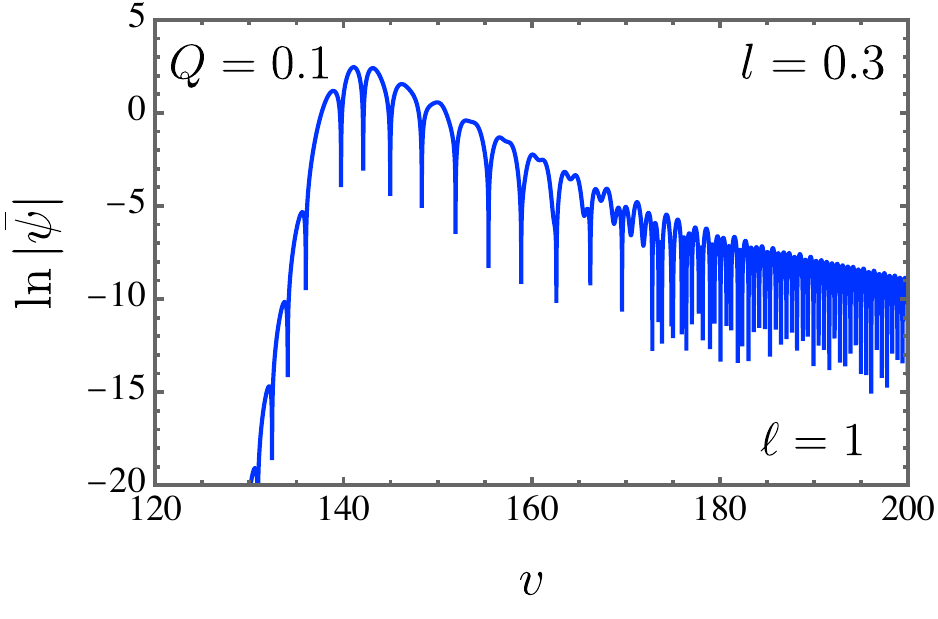}
    \includegraphics[scale=0.4]{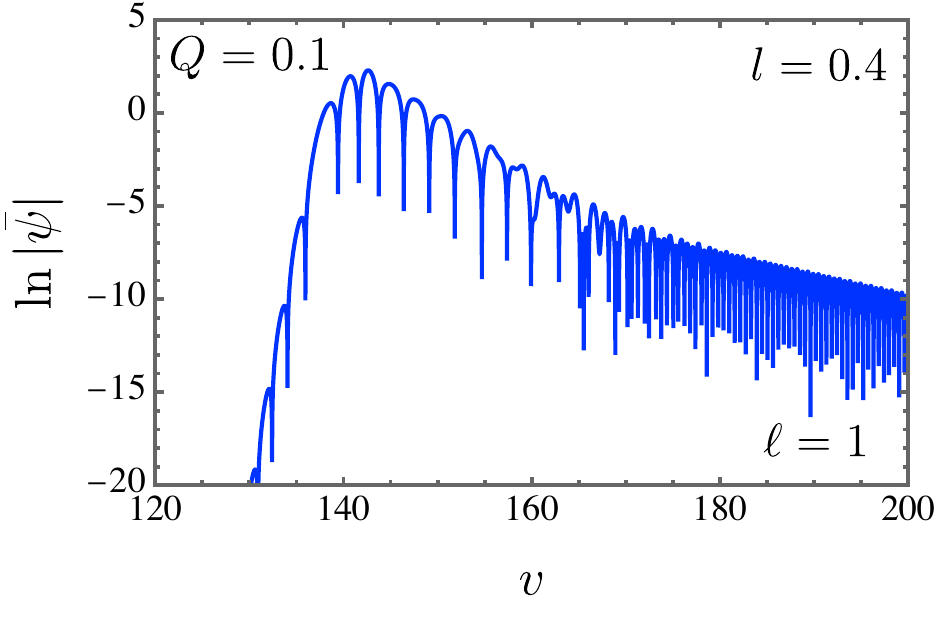}
    \caption{ The wave function $\ln|\bar{\psi}|$ is analyzed for different values of $v$, considering various values of $l$ while keeping the parameters fixed as $Q = 0.1$, $\ell = 1$, and $M = 0.5$.}
    \label{lnwavefunction2}
\end{figure}


\bigskip

\bigskip

\subsection{Temperature, evaporation and remnant mass}

In order to address the main contents of this subsection, the Hawking temperature must be calculated. Taking the advantage of working on spherically symmetric spacetime, we can write
\begin{equation}\label{Temp}
T_h = \frac{1}{4\pi}
\left. {\frac{\mathrm{d}g_{tt}(r)}{\mathrm{d}r}} \right|_{r = {r_{+}}} = \frac{1}{2\pi} \left( \frac{M}{r_{+}^2}-\frac{Q^2}{(l-1)^{2} r_{+}^3}\right).
\end{equation}
The horizon radius can be found by considering the lapse function to be zero as $F(r_+)=0$ and the mass $M$ will be written as a function of $r_+$
\begin{equation}\label{mass1}
	M= \frac{(1-l){r_+}^2+Q^2}{2 (1-l)^2 r_+}.
\end{equation}
By substituting Eq. (\ref{mass1}) in Eq. (\ref{Temp}), we arrive at the following equation

\begin{equation}
	T_h=\frac{(l-1) {r_+}^2-Q^2}{4 \pi  (1-l)^2 {r_+}^3}.
\end{equation}

\begin{figure}[h!]
	\centering
	\includegraphics[scale=0.45]{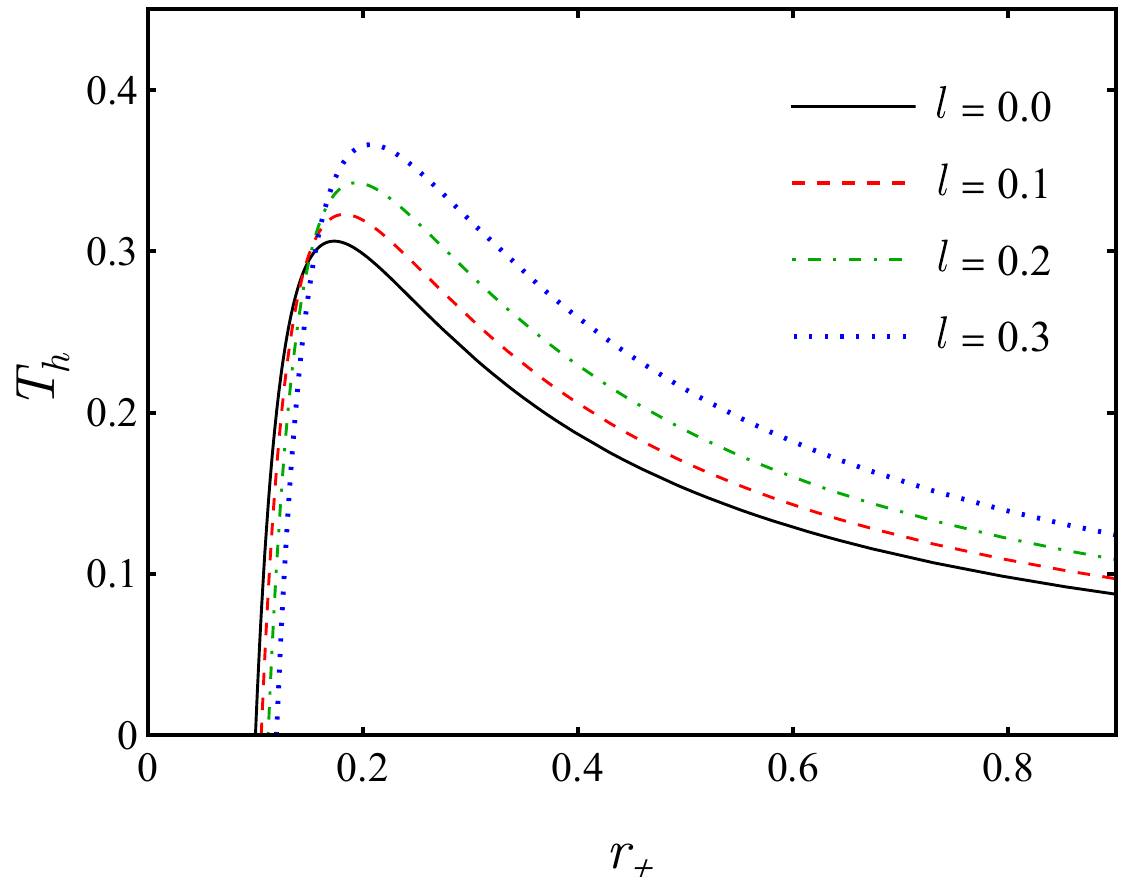}
	\caption{The temperature as a
		function of event horizon for fixed $Q = 0.1$, and different values of $l$.}
	\label{fig:ThL}
\end{figure}

In Fig. \ref{fig:ThL}, the Hawking temperature versus $r_{+}$ is shown for different values of $l$ when $Q = 0.1$. Notice that, in our case, the final stage of black hole evaporation ($T_{h} \to 0$) leads to a remnant mass, i.e., $M_{rem}$, written below
\ie
\label{renmantmass111}
M_{rem} = \frac{Q}{\sqrt{(1-l)^{3}}}.
\fe
As we can straightforwardly see from above expression, only two parameters play the role in modifying $M_{rem}$, i.e., $Q$ and $l$. In this sense, for a better comprehension of the reader, regarding the behavior of such a mass, we display Fig. \ref{remnantmass}. In addition, we also present Tab. \ref{remnanttable1} for a better visualization of the magnitude of $M_{rem}$ for different values of $Q$ and $l$. In addition, we can realize that there exists an increment of the remnant mass magnitude with higher values of $l$ and $Q$ as well. 

Another remarkable feature worthy to be investigated is the black hole lifetime. To do so, we write
\ie
\frac{\mathrm{d}M}{\mathrm{d}\tau} = - \alpha \sigma a T_{h}^{4},
\fe
where $a$ is the radiation constant, $\sigma$ is the cross section area and $\alpha$ is the greybody factor. Taking into account the geometric optics approximation, $\sigma$ turns out to be the photon capture cross section
\ie
\sigma = \pi \left. \left( \frac{g_{\varphi\varphi}}{g_{tt}} \right)  \right|_{r = {r_{out}}}= - \frac{\pi\left(\sqrt{9 (l-1)^4 M^2+8 (l-1) Q^2}-3 (l-1)^2 M\right)^4}{8 (l-1)^2 \left(-(l-1) M \sqrt{9 (l-1)^4 M^2+8 (l-1) Q^2}+3 (l-1)^3 M^2+2 Q^2\right)},
\fe
in a such way that
\ie
\begin{split}
\frac{\mathrm{d}M}{\mathrm{d}\tau} & = -\Upsilon
\frac{\pi\left(\sqrt{9 (l-1)^4 M^2+8 (l-1) Q^2}-3 (l-1)^2 M\right)^4}{8 (l-1)^2 \left(-(l-1) M \sqrt{9 (l-1)^4 M^2+8 (l-1) Q^2}+3 (l-1)^3 M^2+2 Q^2\right)} \\
& \times \left( \frac{(1-l)^3 \left(\sqrt{M^2-\frac{Q^2}{(1-l)^3}}+M\right)^2-Q^2}{4 \pi  (1-l)^5 \left(\sqrt{M^2-\frac{Q^2}{(1-l)^3}}+M\right)^3} \right)^{4},
\end{split}
\fe
where $\Upsilon = a \alpha$. Therefore, we have to solve
\ie
\begin{split}
\int_{0}^{t_{\text{evap}}} \Upsilon \mathrm{d}\tau & = - \int_{M_{i}}^{M_{rem}} 
\left[ \frac{\pi\left(\sqrt{9 (l-1)^4 M^2+8 (l-1) Q^2}-3 (l-1)^2 M\right)^4}{8 (l-1)^2 \left(-(l-1) M \sqrt{9 (l-1)^4 M^2+8 (l-1) Q^2}+3 (l-1)^3 M^2+2 Q^2\right)} \right.\\
& \left.  \times \left( \frac{(1-l)^3 \left(\sqrt{M^2-\frac{Q^2}{(1-l)^3}}+M\right)^2-Q^2}{4 \pi  (1-l)^5 \left(\sqrt{M^2-\frac{Q^2}{(1-l)^3}}+M\right)^3} \right)^{4}\right]^{-1} \mathrm{d}M,
\end{split}
\fe
with $M_{i}$ being the initial mass configuration, and $t_{\text{evap}}$ being the time for the final stage of the evaporation process. It leads to
\ie
t_{\text{evap}} = \frac{1}{\pi\Upsilon} \left( 2.62893\times 10^7 \right).
\fe
Here, some comments are worth pointing out. Initially, we have performed above integral numerically, considering $M_{i}=0.5$ and the values of $M_{rem}$ displayed in Tab. \ref{remnanttable1} (for $Q=0.1$ and $l=0.1$). For the sake of completeness, let us visualize the reduction of the mass $M$ when the time $t$ goes in Fig \ref{remnantmasss1}. Here, a specific choice of $l$ and $Q$ is taken into account, i.e., $l=Q=0.1$.

\begin{figure}
    \centering
      \includegraphics[scale=0.45]{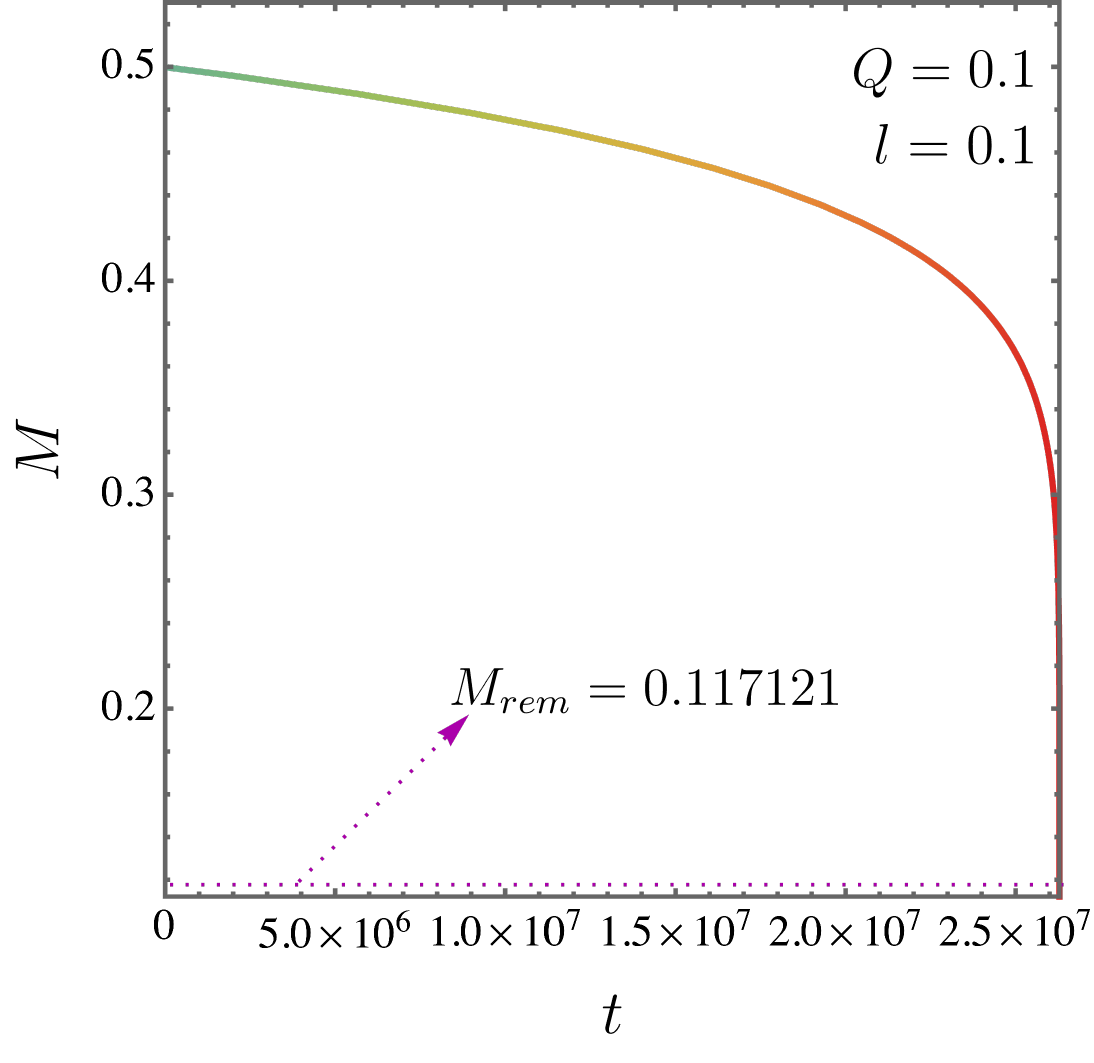}
    \caption{The mass reduction as a function of time for $l=Q=0.1$.}
    \label{remnantmasss1}
\end{figure}

\begin{figure}
    \centering
    \includegraphics[scale=0.35]{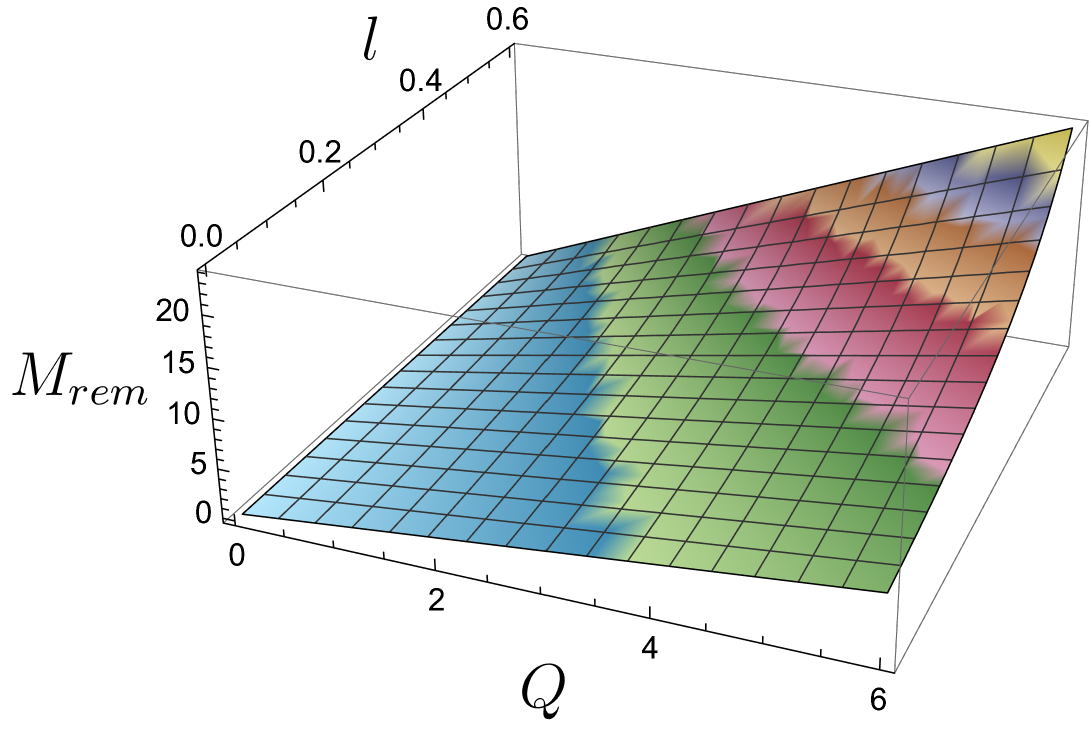}
     \includegraphics[scale=0.3]{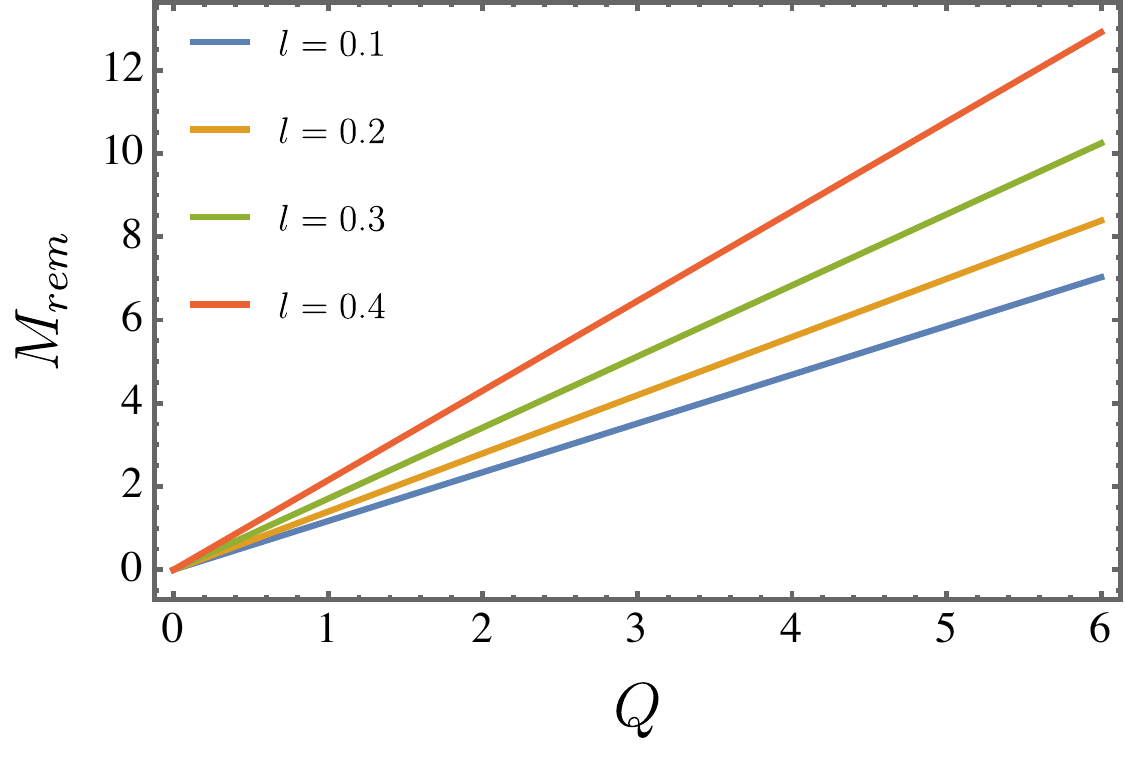}
      \includegraphics[scale=0.3]{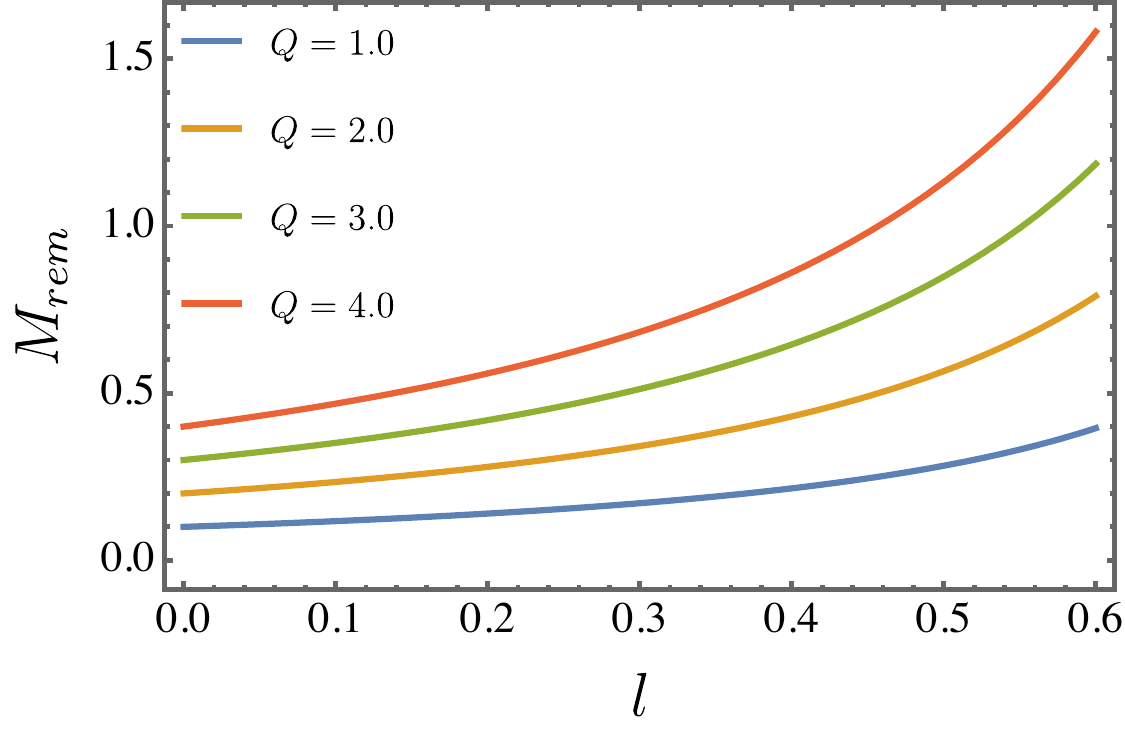}
    \caption{The remnant mass $M_{rem}$ for different configurations of $Q$ and $l$.}
    \label{remnantmass}
\end{figure}

\begin{table}[!h]
\begin{center}
\begin{tabular}{c c  c ||| c c c } 
 \hline\hline
 $Q$ & $l$ & $M_{rem}$ & $Q$ & $l$ & $M_{rem}$  \\ [0.2ex] 
 \hline 
  0.0 & 0.1 & 0.000000 & 1.0 & 0.0 & 1.00000  \\ 

  0.1 & 0.1 & 0.117121 & 1.0 & 0.1 & 1.17121  \\
 
  0.2 & 0.1 & 0.234243 & 1.0 & 0.2 & 1.39754  \\
 
  0.3 & 0.1 & 0.351364 & 1.0 & 0.3 & 1.70747  \\
 
  0.4 & 0.1 & 0.468486 & 1.0 & 0.4 & 2.15166  \\
 
  0.5 & 0.1 & 0.585607 & 1.0 & 0.5 & 2.82843  \\
 
  0.6 & 0.1 & 0.702728 & 1.0 & 0.6 & 3.95285  \\
 
  0.7 & 0.1 & 0.819850 & 1.0 & 0.7 & 6.08581   \\
 
 0.8 & 0.1 & 0.936971 & 1.0 & 0.8 & 11.1803    \\
 
 0.9 & 0.1 & 1.054090 & 1.0 & 0.9 & 31.6228    \\
 
 1.0 & 0.1 & 1.171210 & 1.0 & 1.0 & ---------    \\
 [0.2ex] 
 \hline \hline
\end{tabular}
\caption{\label{remnanttable1} The modification of the remnant mass $M_{rem}$ for different values of $Q$ and $l$.}
\end{center}
\end{table}


\subsection{Emission rate} \label{EMS}
Next, we are interested in finding the rate of energy emission by black holes, i.e., the Hawking radiation. It is known that the black hole shadow corresponds to its high energy absorption cross section for the observer at infinity. On the other hand, at high energy, it oscillates to a limiting constant value denoted as $\sigma_{lim}$, which, for a spherically symmetric black hole, reads \cite{wei2013observing}
\begin{equation}
	\sigma_{lim} \approx R_{sh}^{2},
\end{equation}
where $R_{sh}$ denotes the shadow radius. Now, applying this limiting constant value, the rate of the energy emission of a black hole is computed by \cite{wei2013observing,papnoi2022rotating,panah2020charged,sau2023shadow} 

\begin{equation}\label{emission}
	\frac{{{\mathrm{d}^2}E}}{{\mathrm{d}\omega \mathrm{d}t}} = \frac{{2{\pi ^2}\sigma_{lim} }}{{e^{{\omega }/T_h} - 1}}{\omega ^3},
\end{equation}
where $\omega$ denotes the frequency of the photon.

In Fig. \ref{fig:EmL}, the emission rate versus the frequency $\omega$ for various values of $l$ is plotted. We observe that when frequencies go to zero as well as to infinity the emission rate vanishes. It is obvious that, at a fixed $\omega$, the smaller values of $l$ lead to lower rate of the emission energy. In addition, it is also shown the behavior of the Hawking radiation by varying the charge $Q$.

\begin{figure}[h!]
    \centering
    \includegraphics[scale=0.39]{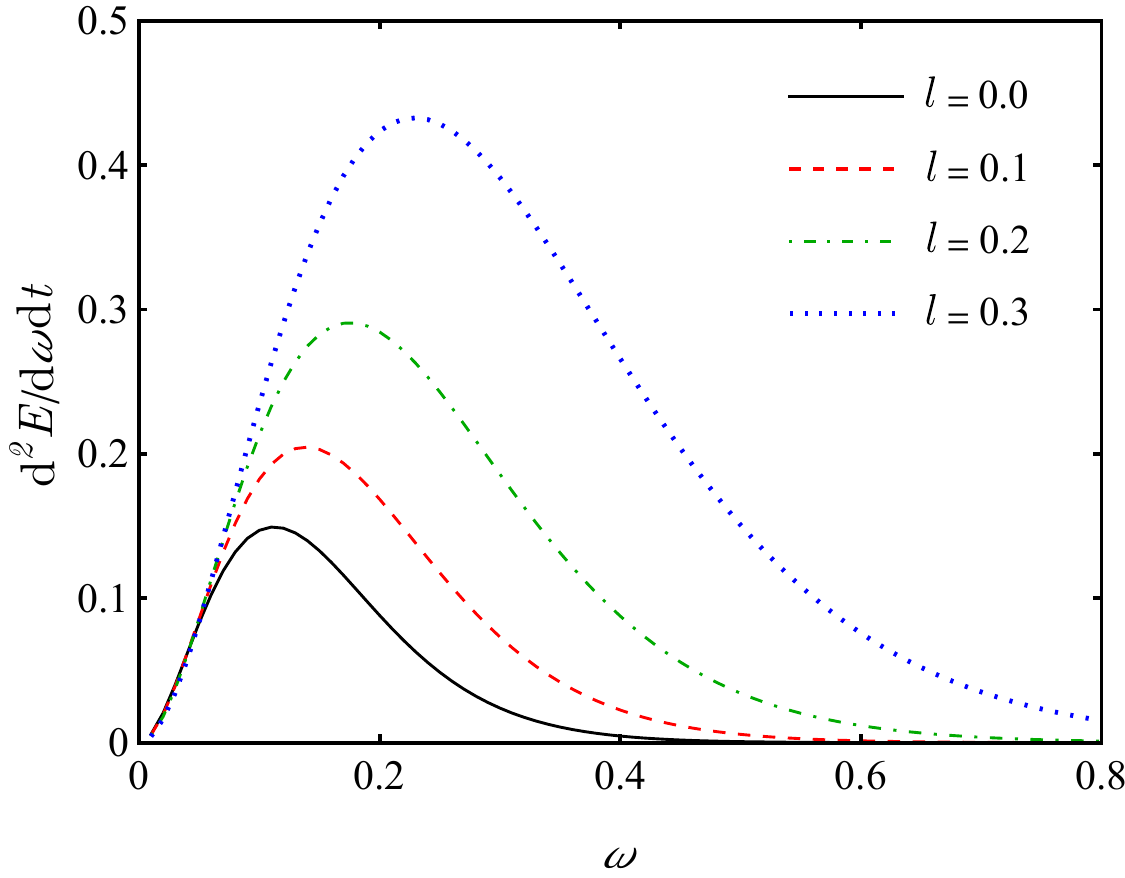}
    \includegraphics[scale=0.4]{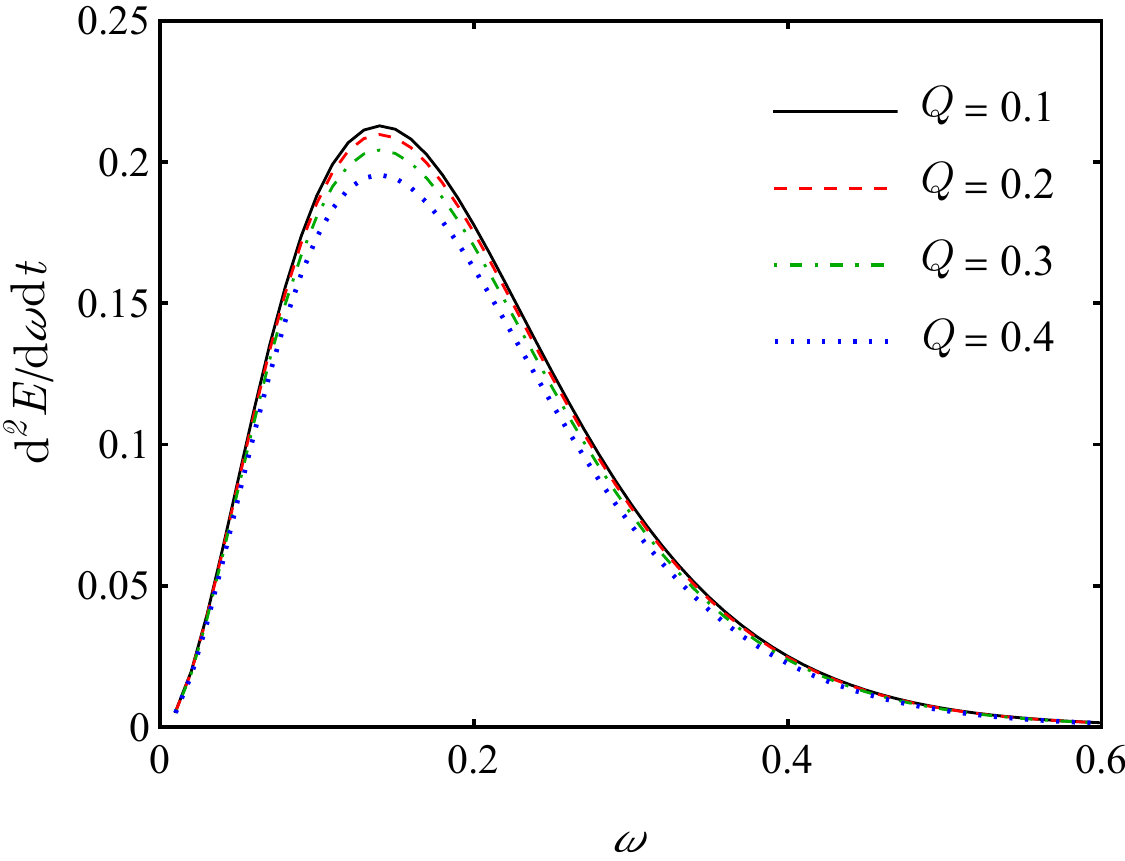}
    \caption{On the right hand, the emission rate, varying with frequency for $M = 1$, $Q = 0.1$ and $l = 0, 0.1, 0.2$ and $l = 0.3$. On the other hand, the Hawking radiation, concerning different values of $Q$, $M = 1$, and $l = 0.1$ is shown.}
    \label{fig:EmL}
\end{figure}


\subsection{Deflection angle}
The gravitational lensing is of great importance to explore the characteristics of the space time and deflection angle plays a significant role in measuring the gravitational lensing observations \cite{kumaran2023shadow,he2023deflection,junior2024gravitational,soares2023holonomy,soares2023gravitational}. 
In this section, the behavior of the deflection angle is investigated by analyzing the impact of the antisymmetric tensor field. Following Ref. \cite{weinberg1972gravitation}, the angle of deflected light can be computed from the following equation 
\begin{equation}\label{phi}
	\Delta \phi = 2\int_{{r_{min }}}^\infty  {\frac{\mathrm{d}r}{{r\sqrt {{{(\frac{r}{{{r_{min }}}})}^2}{F}({r_{min }}) - {F(r)}} }} - \pi } ,
\end{equation}
where $r_{min}$ denotes the minimum distance of the light ray through its pass near the black hole which is related to the impact parameter $b$ as follows 
\begin{equation}\label{rmin}
	b = \frac{{{r_{min }}}}{{\sqrt {F(r_{min})} }}.
\end{equation}
Regarding it, the minimum radius can be found based on Eq. \ref{rmin}. Then according to Eq. \ref{phi}, the deflection angle will be calculated.
In Table.\ref{deflection}, the effect of parameter $l$ on the deflection angle is shown. For each condition, the horizon radius $r_{+}$, the critical orbit $r_{out}$, the minimum radius $r_{min}$ and the deflection angle $\Delta\phi$ are computed for a light ray with impact parameter $b = 10$. In addition, according to the Table. \ref{deflection}, the deflection angle decreases when there is a bigger Lorentz--violating parameter.

\begin{table}[!h]
	\centering
	\caption{\label{deflection}The horizon radius $r_{+}$, the critical orbit $r_{out}$, the minimum radius $r_{min}$ and the deflection angle $\Delta\phi$ of a light ray with impact parameter $b = 10$ near an spacetime of a black hole with $M = 1$, $Q = 0.01$ and $l = 0$, $0.1$, $0.2$, $0.3$, $0.4$ and $0.5$.}
	\begin{tabular}{|c|c|c|c|c|}
		\hline
		$l$ & $r_{+}$ & $r_{out}$ & $r_{min}$ & $\Delta\phi$ \\ \hline
		0.0 & 1.99995 & 2.99993 &  8.78886 & 4.64437 \\ \hline
		0.1 & 1.79994 & 2.69992  & 9.48859 & 3.92897 \\ \hline
		0.2 & 1.59992 & 2.3999 &  10.2728 & 3.29044 \\ \hline
		0.3 & 1.3999 & 2.09986 &  11.1788 & 2.69819 \\ \hline
		0.4 & 1.19986 & 1.79981 & 12.2620 & 2.13066 \\ \hline
		0.5 & 0.9998 & 1.49973 &  13.6128 & 1.56986 \\ \hline
	\end{tabular}
	
\end{table}


\subsection{The correspondence of QNMs and shadows}
In Ref.\cite{cardoso2009geodesic}, it is discussed that the real part of the QNMs is
connected to the angular velocity of the last circular, null geodesic in the eikonal limit. Then, Ref. \cite{stefanov2010connection,wei2011relationship} argued about a connection between QNMs and strong lensing in eikonal limit and regarding these investigations, the following relation between the real part of the QNMs $(\omega_R)$ and the shadow radius has been introduced by Ref. \cite{jusufi2020connection,liu2020shadow}
\begin{equation}\label{eikon1}
	{\omega _R} = \mathop {\lim\limits_{\ell \gg 1}  \frac{\ell}{{{R_{sh}}}}},
\end{equation}
which it has been improved by more accuracy in Ref. \cite{cuadros2020analytical} with the following form

\begin{equation}\label{eikon2}
	{\omega _R} = \mathop {\lim\limits_{\ell \gg 1}  \frac{\ell+\frac{1}{2}}{{{R_{sh}}}}}.
\end{equation}

Applying the 6th order WKB method, the real part of the quasinormal frequencies for eikonal limits are computed, as seen in Fig. \ref{EikonalL}. Then, the real part of QNMs (divided by $\ell+\frac{1}{2}$) is plotted with respect to the multipole values $1 < \ell < 50$. Moreover, the inverse of the shadow radius is represented by dashed lines for $M = 1$, $Q = 0.1$, and various values of $l$. According to the plots, there is a convergence of quasinormal frequency values towards the inverse of the shadow radius. This means that in a charged black hole with an antisymmetric tensor, the correspondence of shadow radius and the real part of the quasinormal frequency, in the eikonal limit is preserved as expected from Eq. (\ref{eikon2}). Besides, it is worth mentioning that the correspondence between quasinormal modes and shadows has also been extended to axisymmetric spacetimes \cite{jusufi2020connection,yang2021relating,pedrotti2024see}.

\begin{figure}[h!]
    \centering
    \includegraphics[scale=0.45]{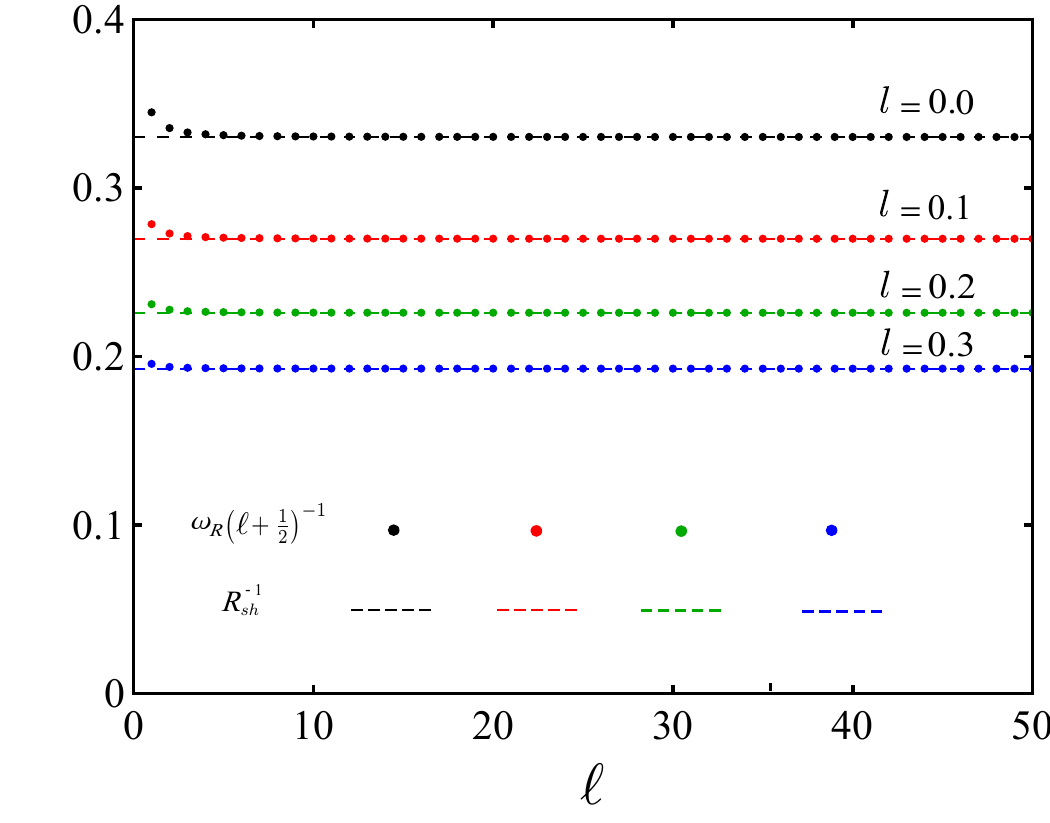}
    \caption{The dot points represent the real part of QNMs varying with $\ell$ and the dashed line is the value of inverse shadow radius. Each color corresponds to a different $l$.}
    \label{EikonalL}
\end{figure}


\section{The case where $\Lambda$ is nonzero}

When faced with the presence of the cosmological constant, the challenge lies in discovering a solution that simultaneously satisfies all equations of motion under the assumption \(V'(X) = 0\), where \(X \equiv B_{\mu\nu}B^{\mu\nu} + b^2\). Unfortunately, such a solution has proven elusive. Drawing inspiration from the methodology utilized in Bumblebee gravity \cite{maluf2021black}, we choose an alternate route by introducing a linear potential in the form \(V = \lambda X\), where \(\lambda\) signifies a Lagrange multiplier field \cite{bluhm2008spontaneous}. This approach relaxes the vacuum condition to \(V'(X) = \lambda\). The equation of motion for the Lagrange multiplier \(\lambda\) is determined by \(X = 0\), ensuring that \(b_{\mu\nu}\) embodies the vacuum configuration. The on--shell value of \(\lambda\) is subsequently determined by the vacuum field equations. It is crucial to note that the off--shell \(\lambda\) must share the same sign as \(X\) to maintain the positivity of the potential \(V\). Even in the presence of the cosmological constant, we can still deduce the same relationships from the field equations
\ie
\label{eee}
G(r) = F^{-1}(r),
\fe
and
\ie
\Phi(r) = \frac{Q}{(1-l)r}.
\fe
After some algebraic manipulations, we derive
\ie
\label{ddd}
F(r) = \frac{1}{1-l} - \frac{2M}{r} + \frac{1+l-2(3-l)\eta b^{2}Q^{2}}{r^{2}(1-l)^{4}} - \frac{(1-3l)\Lambda + (1-l) b^{2} \lambda r^{2}}{3(1-l)^{2}}.
\fe
When we substitute Eq. (\ref{eee}) in Eq. (\ref{ddd}), we verify that those solutions are consistent if and only if the conditions below are satisfied
\ie
\eta = \frac{l}{2 b^{2}},
\fe
and
\ie
\lambda = \frac{2l\Lambda}{(1-l)b^{2}}.
\fe
It is evident that the theory supports a Reissner--Nordström--(Anti--de Sitter) (RN--(A)dS)--like black hole solution with a non--vanishing cosmological constant if only \(\eta \neq 0\) and \(\lambda \neq 0\). Consequently, the metric function \(F(r)\) simplifies to{\footnote{For a comprehensive explanation of the full derivation of this solution, we kindly refer the readers to the original reference \cite{duan2023electrically}.}
\ie
F(r) = \frac{1}{1-l} - \frac{2M}{r} + \frac{Q^{2}}{(1-l)^{2}r^{2}} - \frac{\Lambda r^{2}}{3(1-l)},
\fe
in a such way that we can write the line element as
\ie
\begin{split}
\label{casc}
\mathrm{d}s^{2} = & - \left( \frac{1}{1-l} - \frac{2M}{r} + \frac{Q^{2}}{(1-l)^{2}r^{2}} - \frac{\Lambda r^{2}}{3(1-l)}         \right) \mathrm{d}t^{2} + \left(  \frac{\mathrm{d}r^{2}}{\frac{1}{1-l} - \frac{2M}{r} + \frac{Q^{2}}{(1-l)^{2} r^{2}} - \frac{\Lambda r^{2}}{3(1-l)}}   \right) \\
& + r^{2} \mathrm{d}\theta^{2} + r^{2} \sin^{2} \theta \mathrm{d}\phi^{2}.
\end{split}
\fe

When the cosmological constant is absent, the solution seamlessly transforms into the RN--like configuration as shown in the previous case studied in this manuscript. Likewise, in the scenario where the electric charge becomes null, the metric evolves into the Schwarzschild--(Anti--de Sitter) (Schwarzschild--(A)dS) form \cite{yang2023static}. Furthermore, a degeneration unfolds leading to the RN--(A)dS metric when the Lorentz--violating parameter $l$ is deliberately set to zero.

As the radial coordinate $r$ extends towards infinity, the metric functions converge towards \(F(r) = \frac{1}{G(r)} \rightarrow -\frac{\Lambda r^2}{3(1-l)}\). This behavior is indicative of the asymptotic traits characteristic of (Anti--)de Sitter ((A)dS) spacetime, illustrating the spacetime's convergence towards (A)dS in the infinite limit. In the specific scenario where \(V'(X)\) deviates from zero, the effective cosmological constant is denoted as \(\Lambda_{\text{eff}} \equiv  \frac{\Lambda}{1-l}   \) with the influential factor \(1-l\), in contrast to the conventional cosmological constant \(\Lambda\). Moreover, the interaction between the electromagnetic field and the Kalb--Ramond field introduces an intriguing transformation. The conventional bare electric charge \(Q\) gives way to an effective charge \(Q_{\text{eff}} = \frac{Q}{1-l}\) in the electrostatic potential \(\Phi(r)\). The horizons to Eq. (\ref{casc}), can be represented as follows
\ie
\begin{split}
\Tilde{r}_{1} = & + \frac{1}{2} \sqrt{\frac{3}{\Lambda }+\frac{\gamma }{\Lambda  (l-1) \rho }+\frac{\rho }{3 \sqrt[3]{2} \Lambda  (l-1)}-\frac{l-1}{\Lambda  l-\Lambda }} \\
& +\frac{1}{2} \sqrt{\frac{3}{\Lambda }-\frac{\gamma }{\Lambda  (l-1) \rho }-\frac{\rho }{3 \sqrt[3]{2} \Lambda  (l-1)}+\frac{l-1}{\Lambda  l-\Lambda }+\frac{12 (l M-M)}{\Lambda  \sqrt{\frac{3}{\Lambda }+\frac{\gamma }{\Lambda  (l-1) \rho }+\frac{\rho }{3 \sqrt[3]{2} \Lambda  (l-1)}-\frac{l-1}{\Lambda  l-\Lambda }}}}.
\end{split}
\fe
\ie
\begin{split}
\Tilde{r}_{2} = & + \frac{1}{2} \sqrt{\frac{3}{\Lambda }+\frac{\gamma }{\Lambda  (l-1) \rho }+\frac{\rho }{3 \sqrt[3]{2} \Lambda  (l-1)}-\frac{l-1}{\Lambda  l-\Lambda }} \\
& -\frac{1}{2} \sqrt{\frac{3}{\Lambda }-\frac{\gamma }{\Lambda  (l-1) \rho }-\frac{\rho }{3 \sqrt[3]{2} \Lambda  (l-1)}+\frac{l-1}{\Lambda  l-\Lambda }+\frac{12 (l M-M)}{\Lambda  \sqrt{\frac{3}{\Lambda }+\frac{\gamma }{\Lambda  (l-1) \rho }+\frac{\rho }{3 \sqrt[3]{2} \Lambda  (l-1)}-\frac{l-1}{\Lambda  l-\Lambda }}}},
\end{split}
\fe
\ie
\begin{split}
\Tilde{r}_{3} = & -\frac{1}{2} \sqrt{\frac{3}{\Lambda }+\frac{\gamma }{\Lambda  (l-1) \rho }+\frac{\rho }{3 \sqrt[3]{2} \Lambda  (l-1)}-\frac{l-1}{\Lambda  l-\Lambda }} \\
& -\frac{1}{2} \sqrt{\frac{3}{\Lambda }-\frac{\gamma }{\Lambda  (l-1) \rho }-\frac{\rho }{3 \sqrt[3]{2} \Lambda  (l-1)}+\frac{l-1}{\Lambda  l-\Lambda }-\frac{12 (l M-M)}{\Lambda  \sqrt{\frac{3}{\Lambda }+\frac{\gamma }{\Lambda  (l-1) \rho }+\frac{\rho }{3 \sqrt[3]{2} \Lambda  (l-1)}-\frac{l-1}{\Lambda  l-\Lambda }}}},
\end{split}
\fe
\ie
\begin{split}
\Tilde{r}_{4} = & -\frac{1}{2} \sqrt{\frac{3}{\Lambda }+\frac{\gamma }{\Lambda  (l-1) \rho }+\frac{\rho }{3 \sqrt[3]{2} \Lambda  (l-1)}-\frac{l-1}{\Lambda  l-\Lambda }} \\
& +\frac{1}{2} \sqrt{\frac{3}{\Lambda }-\frac{\gamma }{\Lambda  (l-1) \rho }-\frac{\rho }{3 \sqrt[3]{2} \Lambda  (l-1)}+\frac{l-1}{\Lambda  l-\Lambda }-\frac{12 (l M-M)}{\Lambda  \sqrt{\frac{3}{\Lambda }+\frac{\gamma }{\Lambda  (l-1) \rho }+\frac{\rho }{3 \sqrt[3]{2} \Lambda  (l-1)}-\frac{l-1}{\Lambda  l-\Lambda }}}},
\end{split}
\fe
where 
\ie
\gamma =3 \sqrt[3]{2} \left(l^2+4 \Lambda  l Q^2-2 l-4 \Lambda  Q^2+1\right),
\fe
and
\ie
\begin{split}
\rho = &\left\{972 (\Lambda  l-\Lambda ) \left(l^2 M-2 l M+M\right)^2 \right. \\
& \left. + \left[\left(972 (\Lambda  l-\Lambda ) \left(l^2 M-2 l M+M\right)^2+648 (l-1) Q^2 (\Lambda  l-\Lambda )-54 (l-1)^3\right)^2 \right.\right. \\
& \left.\left. - 4 \left(36 Q^2 (\Lambda  l-\Lambda )+9 (l-1)^2\right)^3\right]^{1/2}+648 (l-1) Q^2 (\Lambda  l-\Lambda )-54 (l-1)^3\right\}^{1/3}.
\end{split}
\fe

Nevertheless, it is worth mentioning that only two of them turn out to be acceptable physical solutions, i.e., $\Tilde{r}_{1}$ and $\Tilde{r}_{2}$. Actually, $\Tilde{r}_{1}$ and $\Tilde{r}_{2}$ represent the event and the Cauchy horizons respectively. To a better clarification, we present Figs. \ref{event3} and \ref{event4}. In Fig. \ref{event3}, we exhibit the event horizon, represented as $\Tilde{r}_{1}$, is depicted for three distinct configurations: the first involves an increase in electric charge (top left), the second explores variations in the parameter \(l\) (top right), and the third examines the growth in the magnitude of the cosmological constant (bottom plot). On the other hand, in Fig. \ref{event4}, we showcase the other event horizon denoted as $\Tilde{r}_{2}$, illustrated across two distinct configurations. The first scenario involves an augmentation in electric charge (top left), while the second investigates variations in the parameter \(l\) (top right). It is worth mentioning that we did not show the modification ascribed to $\Lambda$ to this case because it did not change substantially the present horizon as function of mass $M$.

\begin{figure}
    \centering
    \includegraphics[scale=0.42]{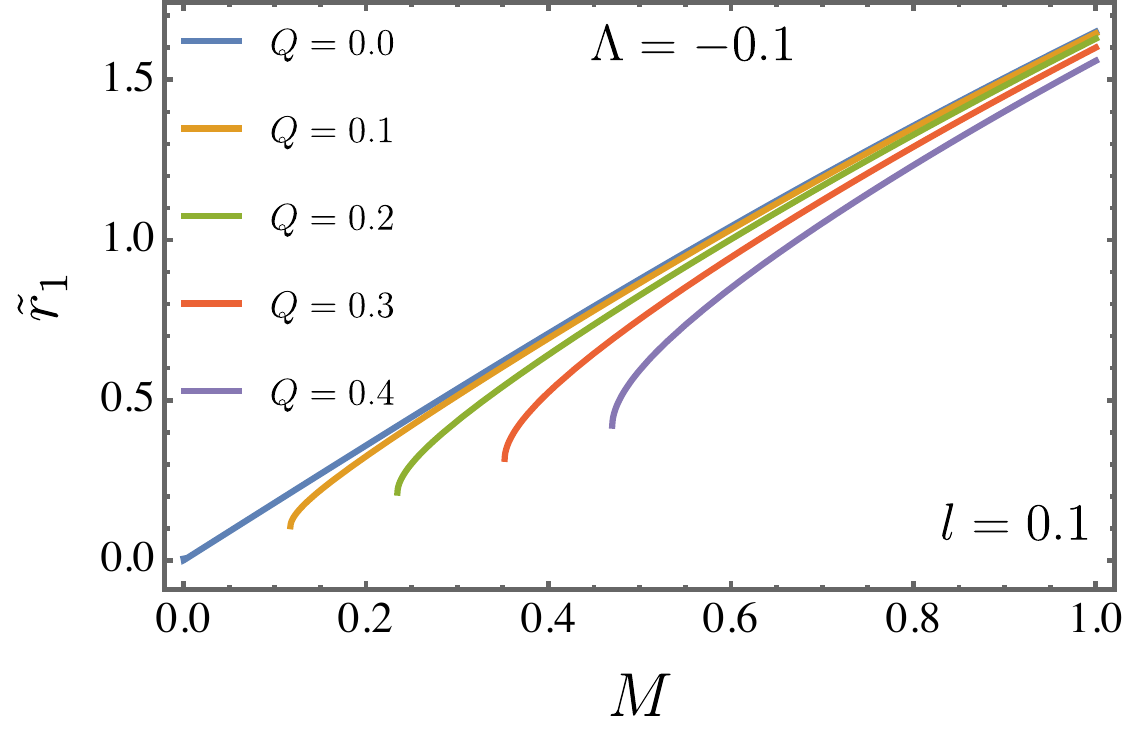}
    \includegraphics[scale=0.42]{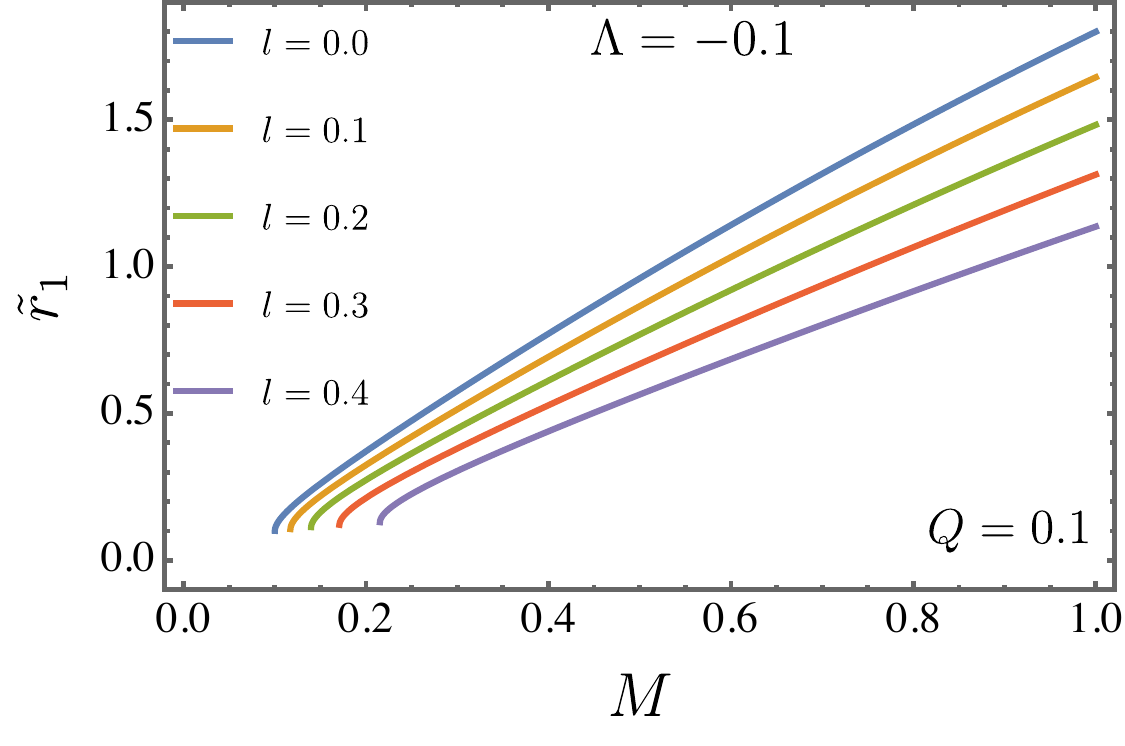}
    \includegraphics[scale=0.42]{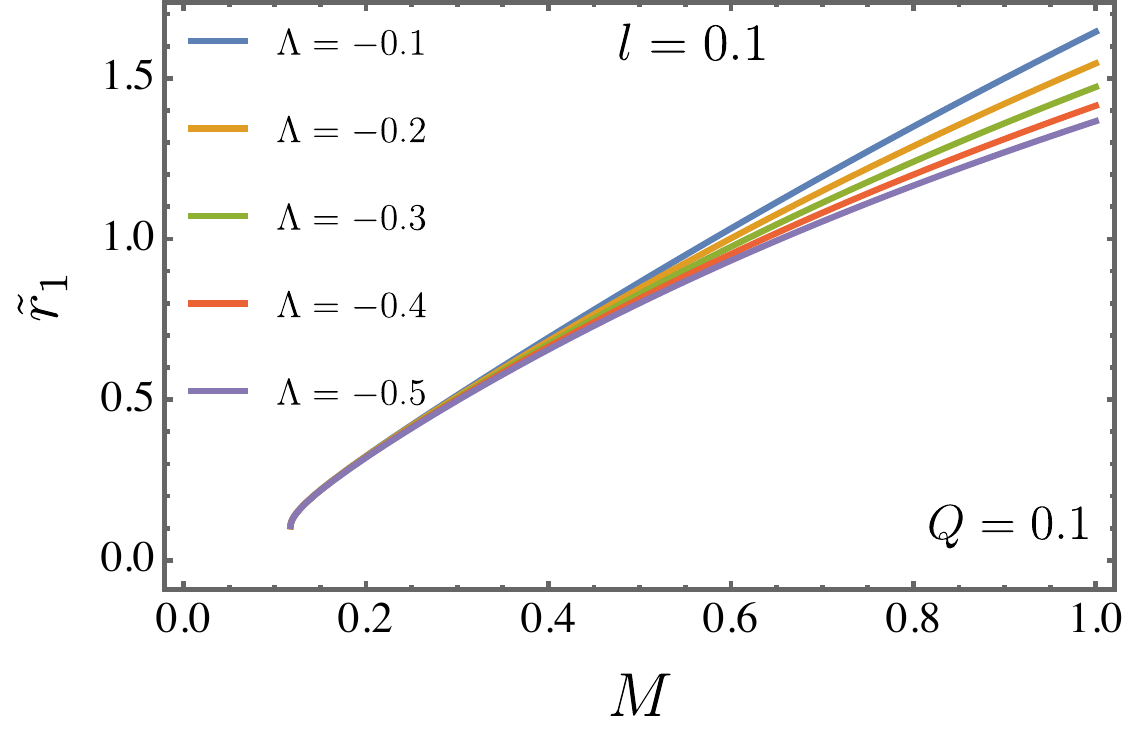}
    \caption{The event horizon is displayed, denoted as $\Tilde{r}_{1}$, for three distinct configurations: one where the electric charge increases (on the top left), another where the parameter $l$ is varied (on top the right), and when the magnitude of the cosmological constant grows (on the bottom plot).}
    \label{event3}
\end{figure}

\begin{figure}
    \centering
    \includegraphics[scale=0.42]{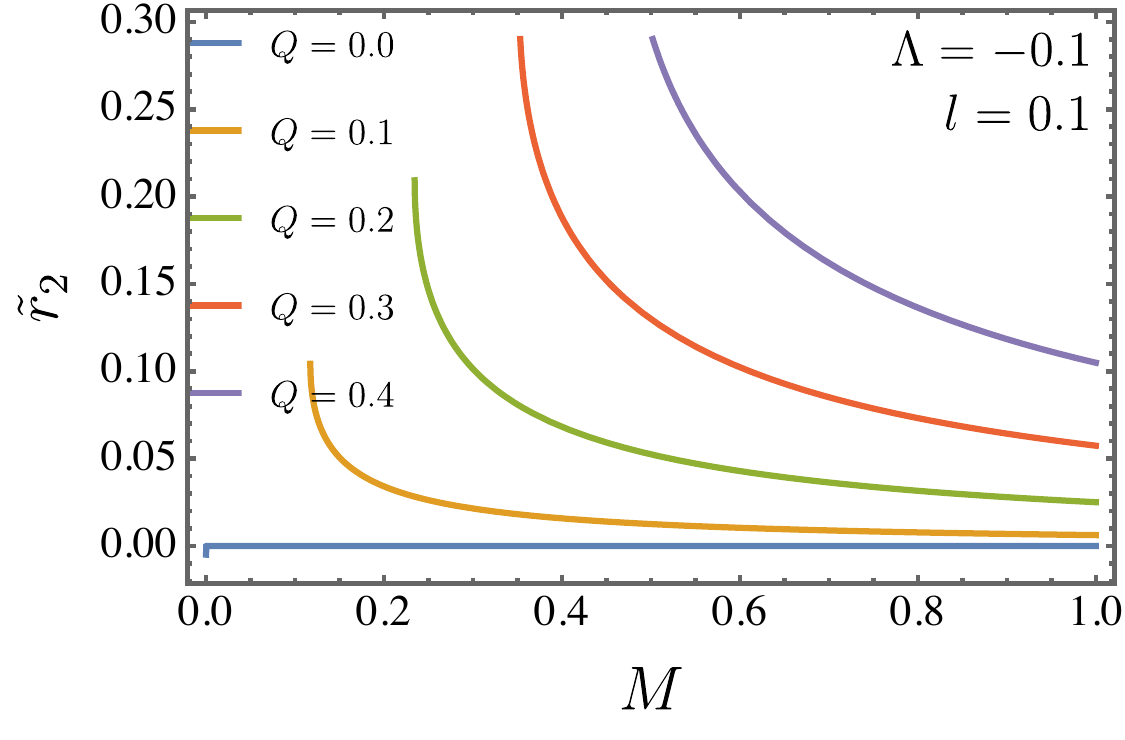}
    \includegraphics[scale=0.42]{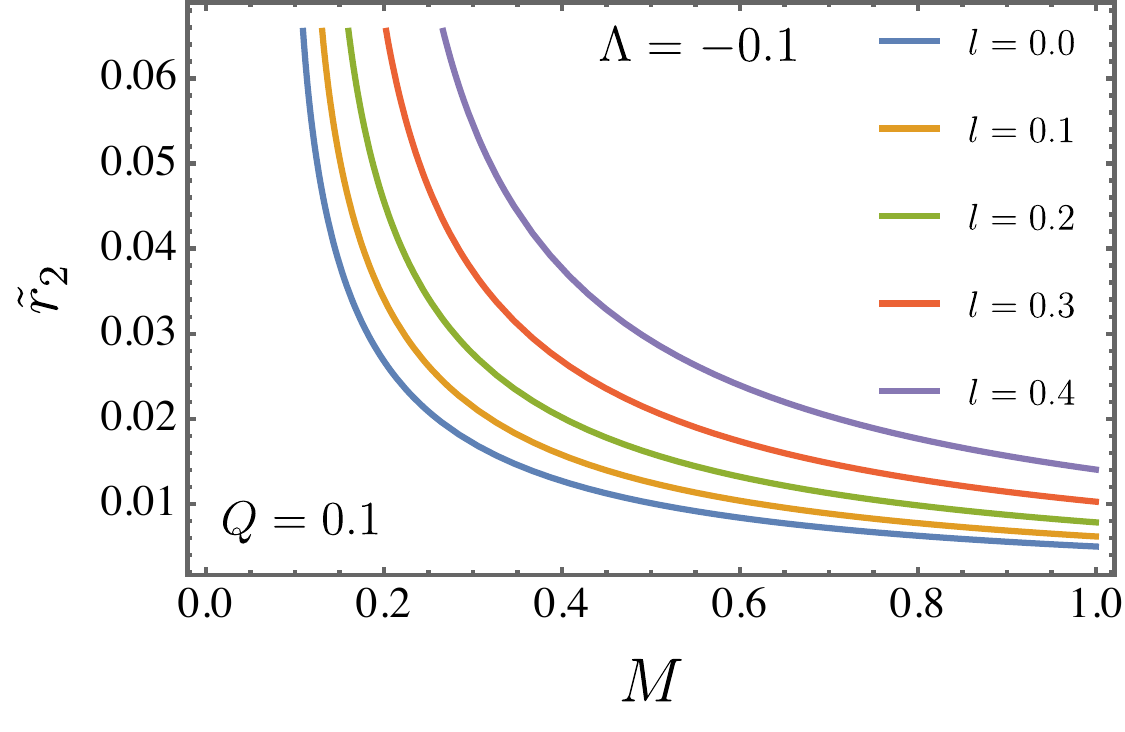}
    \caption{The event horizon is displayed, denoted as $\Tilde{r}_{2}$, for three distinct configurations: one where the electric charge increases (on the top left), another where the parameter $l$ is varied (on top the right), and when the magnitude of the cosmological constant grows (on the bottom plot).}
    \label{event4}
\end{figure}


\bigskip
\bigskip

\subsection{Critical orbits and geodesics}

As we did in the previous sections, we shall calculate the critical orbits as well as the geodesic trajectories. Remarkably, the cosmological constant did not affect the calculation of the photon sphere. However, the geodesic equation is written as

\ie
\frac{\mathrm{d}t^{\prime}}{\mathrm{d}s} = -\frac{2 r' t' \left((l-1) r \left(3 (l-1) M+\Lambda  r^3\right)-3 Q^2\right)}{r \left(3 Q^2-(l-1) r \left(6 (l-1) M-\Lambda  r^3+3 r\right)\right)}    ,
\fe

\ie
\begin{split}
\frac{\mathrm{d}r^{\prime}}{\mathrm{d}s} = & \frac{1}{9 r^{5}} \left[    \frac{9 r^4 \left(r'\right)^2 \left((l-1) r \left(3 (l-1) M+\Lambda  r^3\right)-3 Q^2\right)}{3 Q^2-(l-1) r \left(6 (l-1) M-\Lambda  r^3+3 r\right)} \right. \\ 
& \left. +\frac{\left(t'\right)^2 \left(3 Q^2-(l-1) r \left(6 (l-1) M-\Lambda  r^3+3 r\right)\right) \left(3 Q^2-(l-1) r \left(3 (l-1) M+\Lambda  r^3\right)\right)}{(l-1)^4} \right. \\
& \left.      -\frac{3 r^4 \left(\theta '\right)^2 \left((l-1) r \left(6 (l-1) M-\Lambda  r^3+3 r\right)-3 Q^2\right)}{(l-1)^2} \right. \\
& \left.   -\frac{3 r^4 \sin ^2(\theta ) \left(\varphi '\right)^2 \left((l-1) r \left(6 (l-1) M-\Lambda  r^3+3 r\right)-3 Q^2\right)}{(l-1)^2}   \right],
\end{split}
\fe

\ie
\frac{\mathrm{d}\theta^{\prime}}{\mathrm{d}s} = \sin (\theta ) \cos (\theta ) \left(\varphi '\right)^2-\frac{2 \theta ' r'}{r} ,
\fe

\ie
\frac{\mathrm{d}\phi^{\prime}}{\mathrm{d}s} = -\frac{2 \varphi ' \left(r'+r \theta ' \cot (\theta )\right)}{r}.
\fe

To a better illustration of the geodesic equations, we present Figs. \ref{lightpath2} and \ref{massivetraj2}. In particular, Fig. \ref{lightpath2} shows the light path, which is illustrated by the blue line. The dashed wine lines outlining the photon sphere, and the event horizon marked by the black dot. The particular configuration of the system is defined by the following parameters: $Q=0.1$, $l=0.1$, $M=1.0$, and $\Lambda=-1.0$. In addition, Fig. \ref{massivetraj2} represent various trajectories for massive particles. They are depicted under the configuration: $Q=0.1$, $l=0.01$, $M=1.0$, and $\Lambda=-0.1$. The left panel and the right panel differ in the mass values, namely, $M=0.3$ and $M=0.5$, respectively.

\begin{figure}
    \centering
    \includegraphics[scale=0.5]{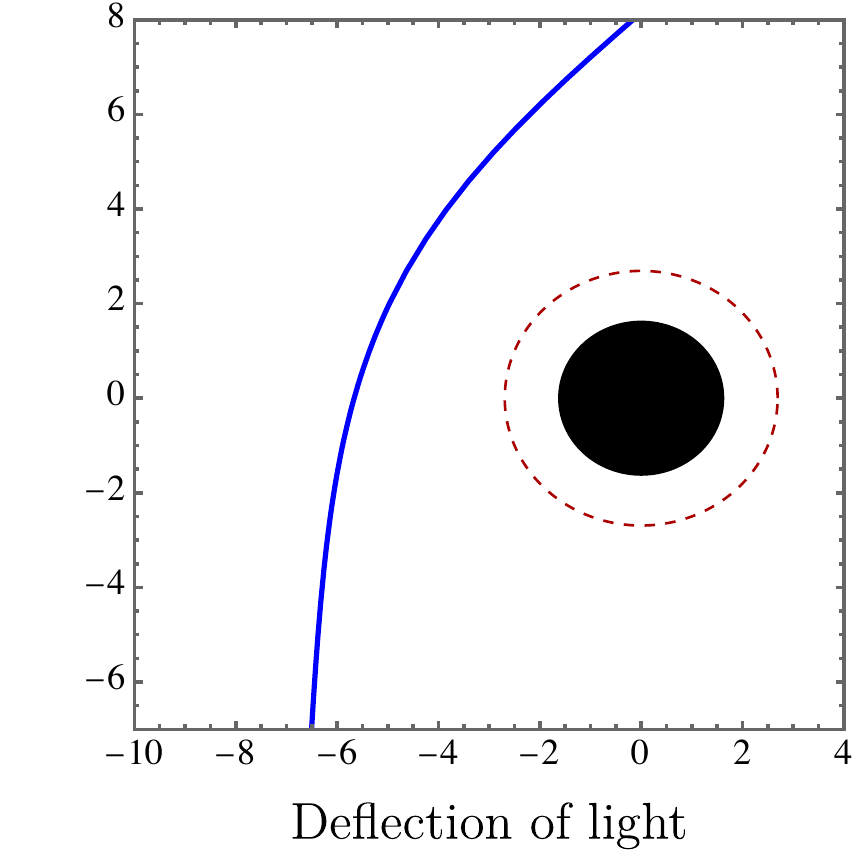}
    \caption{The trajectory of light, visually represented by the blue line. Dashed wine lines delineate the photon sphere ($r_{c_{out}}$), and the event horizon is denoted by the black dot. The specific configuration of the system is characterized by the following parameters: $Q=0.1$, $l=0.1$, $M=1.0$, and $\Lambda=-1.0$.}
    \label{lightpath2}
\end{figure}

\begin{figure}
    \centering
    \includegraphics[scale=0.36]{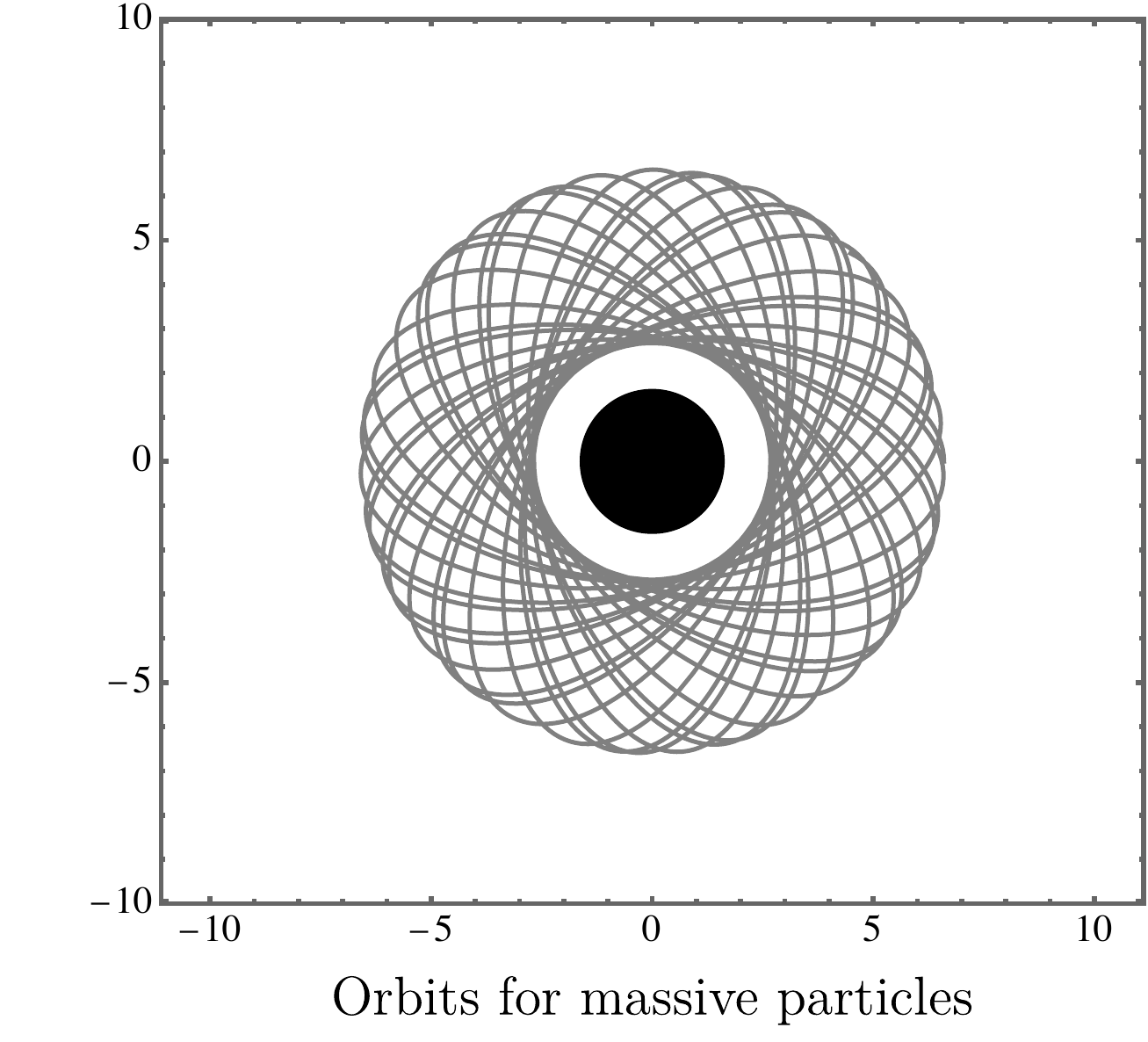}
     \includegraphics[scale=0.36]{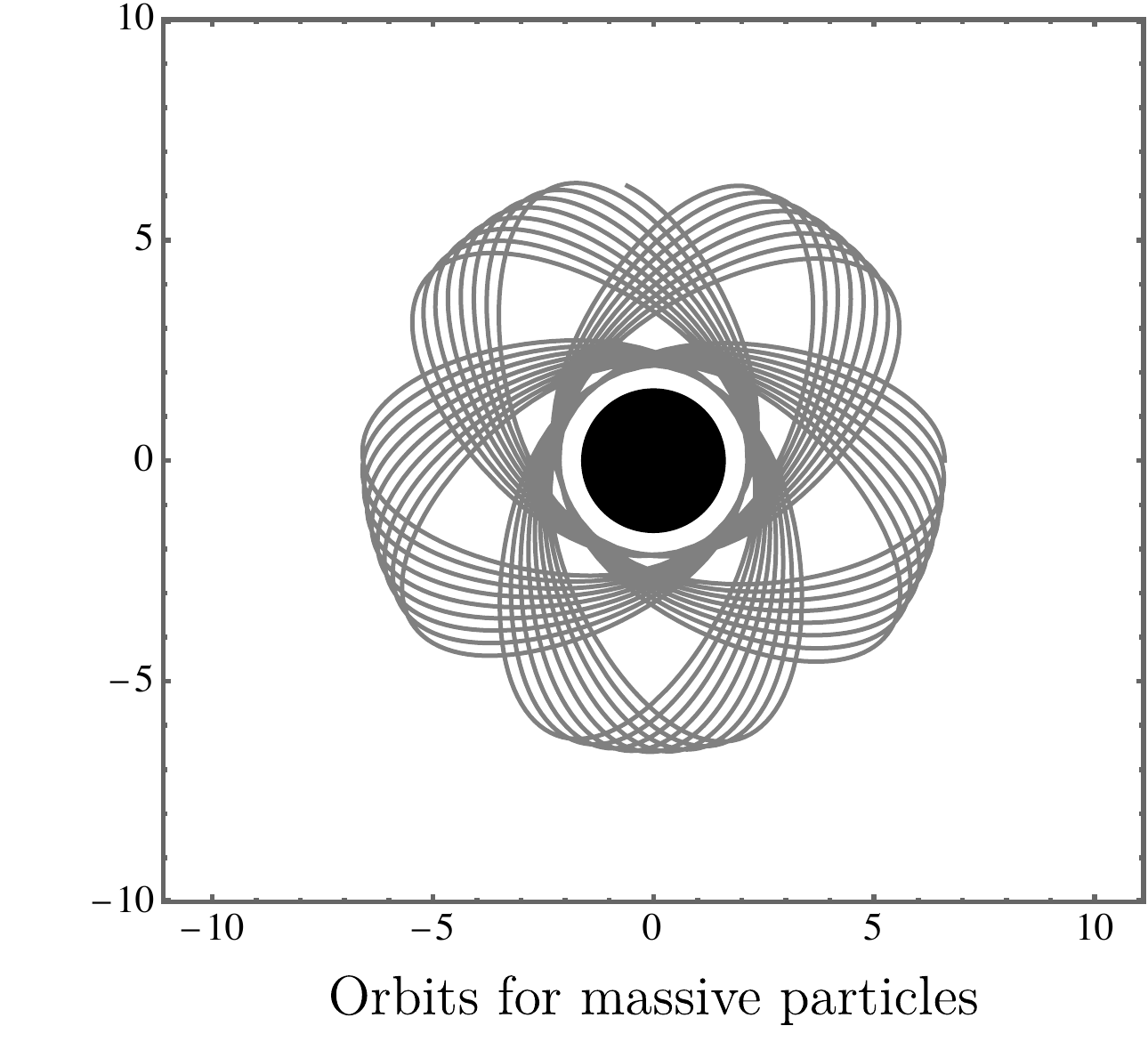}
    \caption{Some trajectories for massive particles are portrayed under the configuration: $Q=0.1$, $l=0.01$, $M=1.0$, and $\Lambda=-0.1$. The left panel and the right panel differ in the values of mass, i.e., $M=0.3$ and $M=0.5$ respectively.}
    \label{massivetraj2}
\end{figure}


\subsection{Quasinormal modes}

Since, all definitions were given in the previous sections, we present the tables for the quasinormal modes. Utilizing the sixth--order WKB approximation, we provide Tabs. \ref{laaLL11} and \ref{laaLL11}, which elucidate the quasinormal frequencies linked to varied values of \( Q \), \( l \), and $\Lambda$, with a particular emphasis on instances where \( M=1 \), considering \( \ell =1 \) and $\ell =2 $ respectively.

It is worth mentioning that, differently with what we have accomplished in the previous subsections, we are not going to address for the second case the correspondence between the quasinormal modes and shadows for a simple reason: although we have considered the eikonal limit for the quasinormal modes \cite{a1jusufi2023dark}, the results turn out to be identical to the first black hole studied here.

\begin{table}[!h]
\begin{center}
\caption{\label{laaLL11}By employing the sixth--order WKB approximation, we clarify the quasinormal frequencies corresponding to various values of \( Q \), \( l \), and $\Lambda$. Our specific emphasis is on scenarios where \( M=1 \) and \( \ell=1 \).}
\begin{tabular}{c| c | c | c} 
 \hline\hline\hline 
 \!\!\! $Q$ \,\,\,\,\,\,\,\,  $l$ \,\,\,\,\,\,\,\,\,\, $\Lambda$  & $\omega_{0}$ & $\omega_{1}$ & $\omega_{2}$  \\ [0.2ex] 
 \hline 
 0.00,    0.10,   -0.10 & 0.369093 - 0.148546$i$ & 0.214000 - 0.569097$i$  & --------- \\ 
 
 0.10,  0.10, -0.10  & 0.369813 - 0.148566$i$ & 0.216368 - 0.567078$i$ & --------- \\
 
 0.20, 0.10 , -0.10  & 0.372018 - 0.148616$i$  & 0.223526 - 0.561128$i$ &  ---------  \\
 
 0.30, 0.10, -0.10  & 0.375859 - 0.148665$i$ & 0.235638 - 0.551556$i$  &  ---------  \\
 
 0.40, 0.10, -0.10  & 0.381622 - 0.148643$i$ & 0.252982 - 0.538804$i$   &   0.000239514 - 1.48307$i$  \\
 
 0.50, 0.10, -0.10  & 0.389828 - 0.148397$i$ & 0.276010 - 0.523241$i$ & 0.038543900 - 1.36582$i$   \\
   [0.2ex] 
\hline\hline\hline 
 \!\!\! $Q$ \,\,\,\,\,\,\,\,  $l$ \,\,\,\,\,\,\,\,\,\, $\Lambda$  & $\omega_{0}$ & $\omega_{1}$ & $\omega_{2}$  \\ [0.2ex] 
 \hline 
 0.10,    0.00,   -0.10 & 0.333114 - 0.126616$i$ & 0.162163 - 0.527562$i$  & ---------  \\ 
 
 0.10,  0.10, -0.10  & 0.369813 - 0.148566$i$ & 0.216368 - 0.567078$i$ & --------- \\
 
 0.10, 0.20, -0.10  & 0.417218 - 0.178693$i$ & 0.271591 - 0.646176$i$ &  0.00769818 - 1.693030$i$   \\
 
 0.10, 0.30, -0.10  & 0.480765 - 0.221937$i$ & 0.330362 - 0.778284$i$  &  0.08378600 - 1.818360$i$  \\
 
 0.10, 0.40, -0.10  & 0.569741 - 0.287668$i$ & 0.397812 - 0.993552$i$  & 0.17485400 - 2.121130$i$  \\
 
 0.10, 0.50, -0.10  & 0.701528 - 0.395642$i$ & 0.482185 - 1.35994$i$ & 0.29075000 - 2.726480$i$  \\
   [0.2ex] 
\hline\hline\hline 
 \!\!\! $Q$ \,\,\,\,\,\,\,\,  $l$ \,\,\,\,\,\,\,\,\,\, $\Lambda$  & $\omega_{0}$ & $\omega_{1}$ & $\omega_{2}$  \\ [0.2ex] 
 \hline 
 0.10,    0.10,   0.00 & 0.288306 - 0.114003$i$ & 0.244851 - 0.363967$i$  & 0.196349 - 0.659009$i$  \\ 
 
 0.10,  0.10, -0.10  & 0.369813 - 0.148566$i$ & 0.216368 - 0.567078$i$ & --------- \\
 
 0.10, 0.10, -0.20  & 0.406356 - 0.177055$i$  & --------- &  ---------   \\
 
 0.10, 0.10, -0.30  & 0.249944 - 0.261463$i$ & ---------  &  ---------  \\
 
 0.10, 0.01, -0.40  & --------- & ---------  & ---------  \\
 
 0.10, 0.01, -0.50  & --------- & --------- & ---------  \\
   [0.2ex] 
   \hline\hline\hline 
\end{tabular}
\end{center}
\end{table}

\begin{table}[!h]
\begin{center}
\caption{\label{laaLL22}By employing the sixth--order WKB approximation, we clarify the quasinormal frequencies corresponding to various values of \( Q \), \( l \), and $\Lambda$. Our specific emphasis is on scenarios where \( M=1 \) and \( \ell=2 \).}
\begin{tabular}{c| c | c | c} 
 \hline\hline\hline 
 \!\!\! $Q$ \,\,\,\,\,\,\,\,  $l$ \,\,\,\,\,\,\,\,\,\, $\Lambda$  & $\omega_{0}$ & $\omega_{1}$ & $\omega_{2}$  \\ [0.2ex] 
 \hline 
 0.00,    0.10, -0.10 & 0.697526 - 0.153742$i$ & 0.636482 - 0.484706$i$  & 0.479412 - 0.916464$i$  \\ 
 
 0.10,  0.10, -0.10  & 0.698574 - 0.15372$i$ & 0.637958 - 0.484476$i$ & 0.482719 - 0.914274$i$ \\
 
 0.20, 0.10, -0.10  & 0.701788 - 0.153645$i$ & 0.642458 - 0.483753$i$ &  0.492679 - 0.907780$i$   \\
 
 0.30, 0.10, -0.10  & 0.707394 - 0.153482$i$ & 0.650219 - 0.482425$i$  &  0.509429 - 0.897173$i$  \\
 
 0.40, 0.01, -0.10  & 0.715827 - 0.153153$i$ & 0.661708 - 0.480241$i$  &  0.533250 - 0.882626$i$  \\
 
 0.50, 0.01, -0.10  & 0.727855 - 0.152500$i$ & 0.677764 - 0.476669$i$ &  0.564678 - 0.863959$i$  \\
   [0.2ex] 
    \hline\hline\hline 
 \!\!\! $Q$ \,\,\,\,\,\,\,\,  $l$ \,\,\,\,\,\,\,\,\,\, $\Lambda$  & $\omega_{0}$ & $\omega_{1}$ & $\omega_{2}$  \\ [0.2ex] 
 \hline 
 0.10,    0.00,   -0.10 & 0.626589 - 0.130756$i$ & 0.569746 - 0.412964$i$  & 0.393096 - 0.814531$i$  \\ 
 
 0.10,  0.10, -0.10  & 0.698574 - 0.15372$i$ & 0.637958 - 0.484476$i$ & 0.482719 - 0.914274$i$ \\
 
 0.10, 0.20, -0.10  & 0.793719 - 0.185388$i$ & 0.725356 - 0.583551$i$ &  0.579921 - 1.07268$i$   \\
 
 0.10, 0.30, -0.10  & 0.92427 - 0.231045$i$ & 0.842508 - 0.726995$i$  &  0.694379 - 1.31667$i$  \\
 
 0.10, 0.40, -0.10  & 1.11224 - 0.300732$i$ & 1.00777 - 0.946941$i$  & 0.841872 - 1.70228$i$  \\
 
 0.10, 0.50, -0.10  & 1.40126 - 0.415510$i$ & 1.25666 - 1.311240$i$ & 1.05134 - 2.351550$i$  \\
   [0.2ex] 
\hline\hline\hline 
 \!\!\! $Q$ \,\,\,\,\,\,\,\,  $l$ \,\,\,\,\,\,\,\,\,\, $\Lambda$  & $\omega_{0}$ & $\omega_{1}$ & $\omega_{2}$  \\ [0.2ex] 
 \hline 
 0.10,    0.10,   0.00 & 0.535430 - 0.117227$i$ & 0.508207 - 0.359449$i$  & 0.4630300 - 0.622225$i$  \\ 
 
 0.10,  0.10, -0.10  & 0.698574 - 0.153720$i$ & 0.637958 - 0.484476$i$ & 0.4827190 - 0.914274$i$  \\
 
 0.10, 0.10, -0.20  & 0.826290 - 0.183190$i$ & 0.618010 - 0.629579$i$  &  0.0184675 - 2.131830$i$   \\
 
 0.10, 0.10, -0.30  & 0.926378 - 0.207141$i$ & 0.217587 - 1.230770$i$  &  ---------  \\
 
 0.10, 0.01, -0.40  & 0.984458 - 0.224853$i$ & ---------  & ---------  \\
 
 0.10, 0.01, -0.50  & 0.953090 - 0.236589$i$ & --------- & ---------  \\
   [0.2ex] 
\hline\hline\hline 
\end{tabular}
\end{center}
\end{table}


\subsection{Time--domain solution}

Following the approach used in the initial analysis, where the cosmological constant was set to zero, we proceed similarly here. Figs. \ref{wavefunctionLambda} and \ref{lnwavefunctionLambda} display the behavior of $\bar{\psi}$ and $\ln |\bar{\psi}|$ for the case where the cosmological constant is nonzero.

\begin{figure}
    \centering
    \includegraphics[scale=0.4]{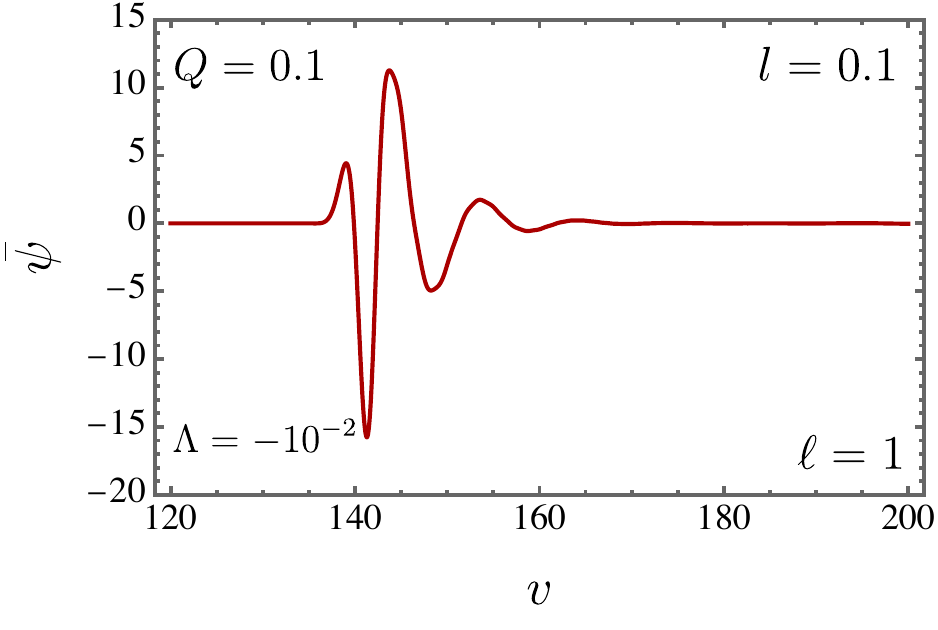}
    \includegraphics[scale=0.4]{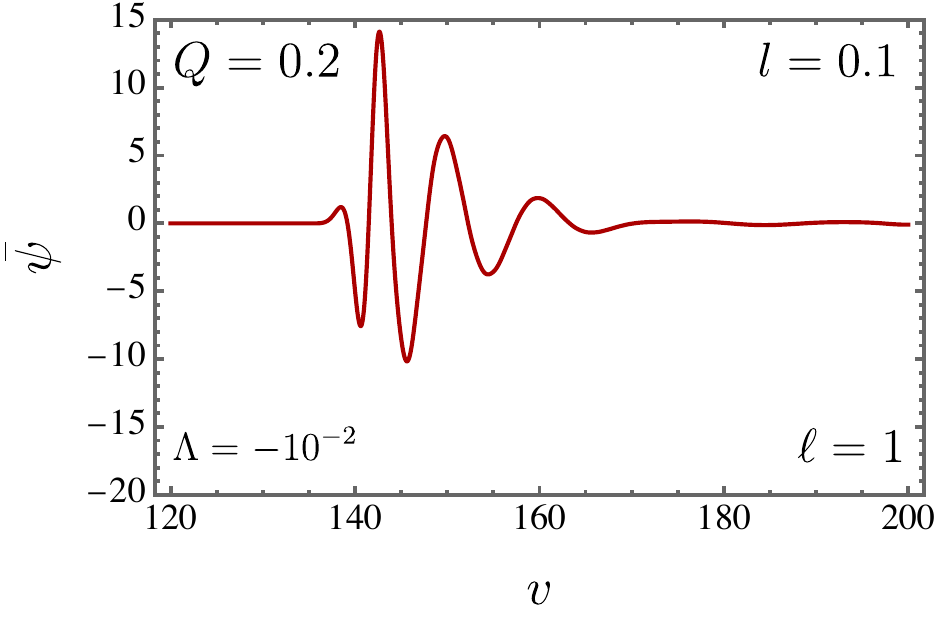}
    \includegraphics[scale=0.4]{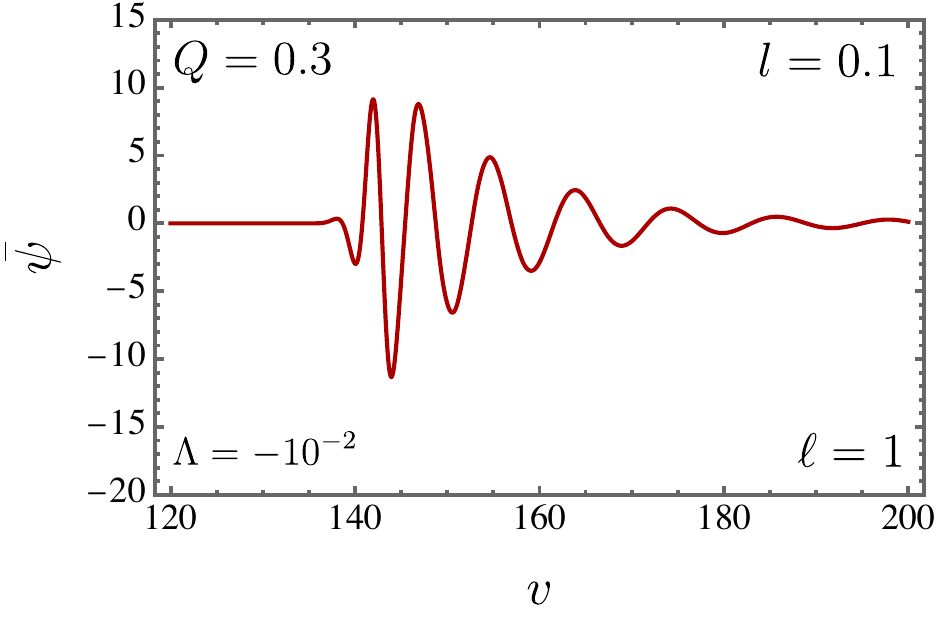}
    \includegraphics[scale=0.4]{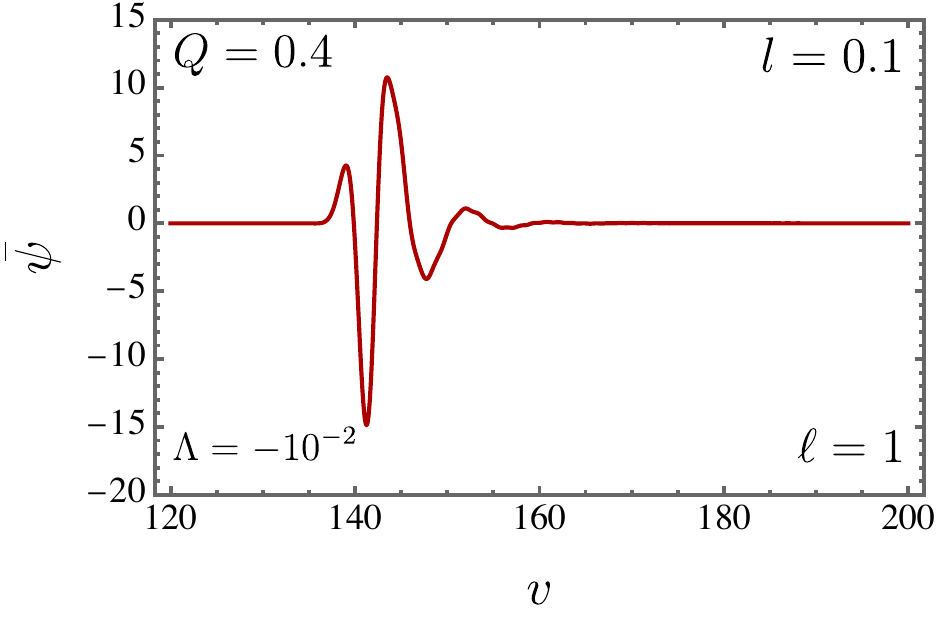}
    \caption{ The wave function $\bar{\psi}$ is analyzed for different values of $v$, considering various values of $Q$ while keeping the parameters fixed as $l = 0.1$, $\ell = 1$, $M = 0.5$ and $\Lambda = - 10^{-2}$.}
    \label{wavefunctionLambda}
\end{figure}

\begin{figure}
    \centering
    \includegraphics[scale=0.4]{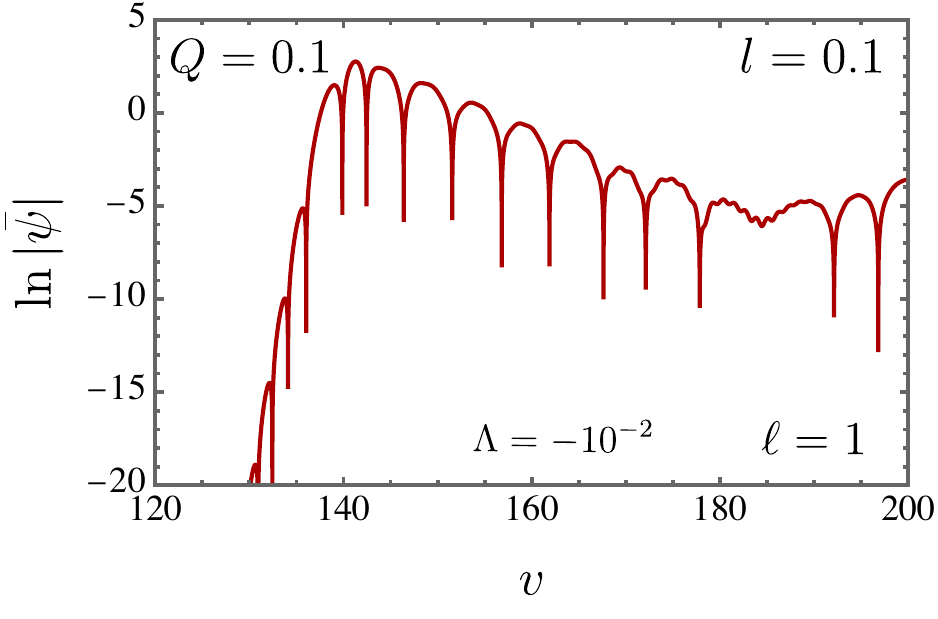}
    \includegraphics[scale=0.4]{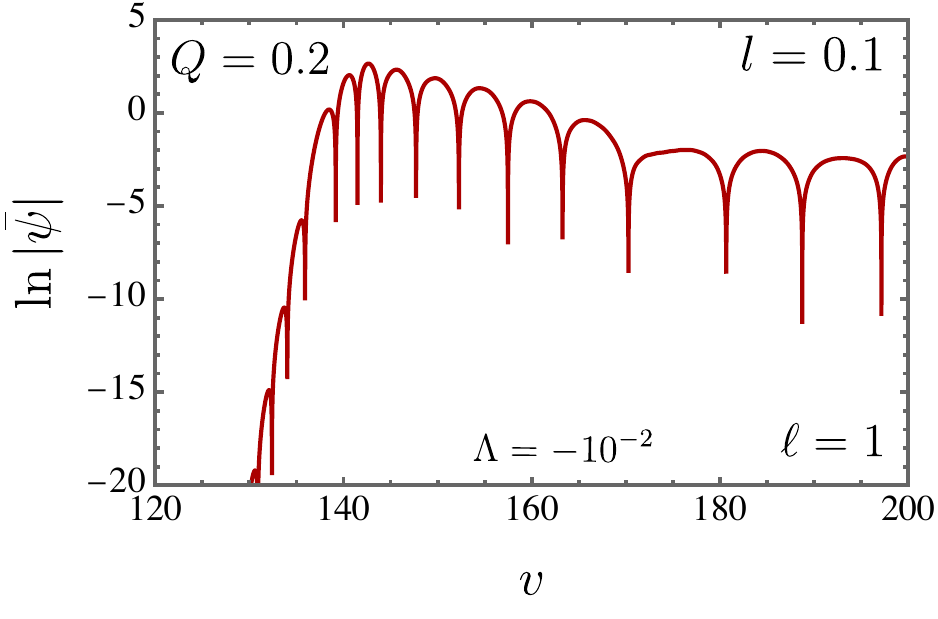}
    \includegraphics[scale=0.4]{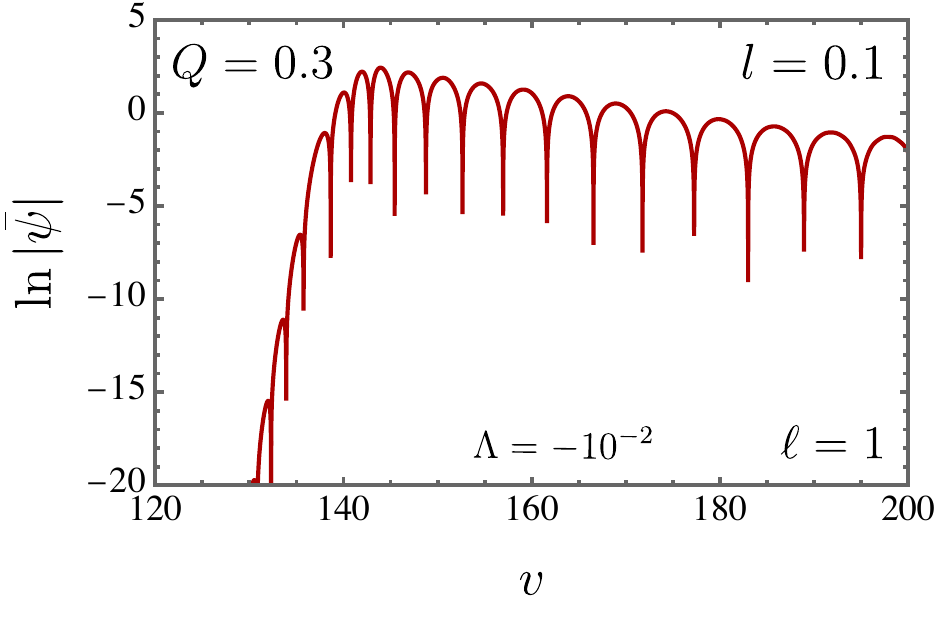}
    \includegraphics[scale=0.4]{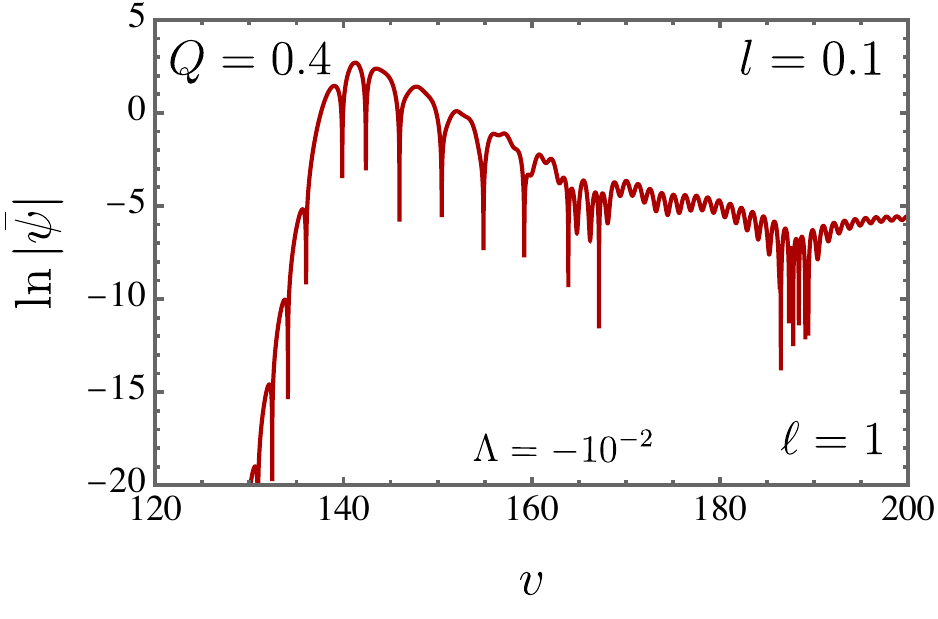}
    \caption{The wave function $\ln|\bar{\psi}|$ is analyzed for different values of $v$, considering various values of $Q$ while keeping the parameters fixed as $l = 0.1$, $\ell = 1$, $M = 0.5$, and $\Lambda = - 10^{-2}$.}
    \label{lnwavefunctionLambda}
\end{figure}


\subsection{Temperature and remnant mass}

In this section, by considering the presence of cosmological constant $\Lambda$, the Hawking temperature has the following form
\begin{equation}\label{Temp2}
	\Tilde{T}_h= \frac{1}{4\pi}
	\left. {\frac{\mathrm{d}g_{tt}(r)}{\mathrm{d}r}} \right|_{r = {\Tilde{r}_{1}}} = \frac{(l-1) \Tilde{r}_{1} \left(3 (l-1) M+\Lambda  \Tilde{r}_{1}^3\right)-3 Q^2}{6 \pi  (l-1)^2 \Tilde{r}_{1}^3}.
\end{equation}
Following the previous section, the mass is written related to the horizon radius as
\begin{equation}\label{mass2}
	\Tilde{M}=\frac{3 (1-l){\Tilde{r}_{1}}^2+3 Q^2-(1-l)\Lambda  {\Tilde{r}_{1}}^4}{6 (l-1)^2 {\Tilde{r}_{1}}}.
\end{equation}
Now, by substituting mass from Eq. (\ref{mass2}) in Eq. (\ref{Temp2}), the equation of Hawking temperature can be obtained
\begin{equation}
	\Tilde{T}_h=\frac{(1-l)  \left(1-\Lambda  {\Tilde{r}_{1}}^2\right){\Tilde{r}_{1}}^2-Q^2}{4 \pi  (1-l)^2 {\Tilde{r}_{1}}^3}.
\end{equation}

The Hawking temperature concerning event horizon radius is plotted for $Q = 0.1$. $\Lambda=-0.03$ and various values of $l$ in Fig. \ref{fig:TempLambda}. It is observed that when $l$ increases, $\Tilde{T}_{h} \to 0$ for larger $\Tilde{r}_{1}$. Furthermore, for a better investigation, the remnant mass is calculated for different $l$, and results are depicted in Tab. \ref{Remnant2}. Notice that the deviation of the latter Tab. in comparison with Tab. \ref{remnanttable1}, is so small. This result is rather predictable since the contributions ascribed to $\Lambda$ are expected to be small.

\begin{figure}
	\centering
	\includegraphics[scale=0.45]{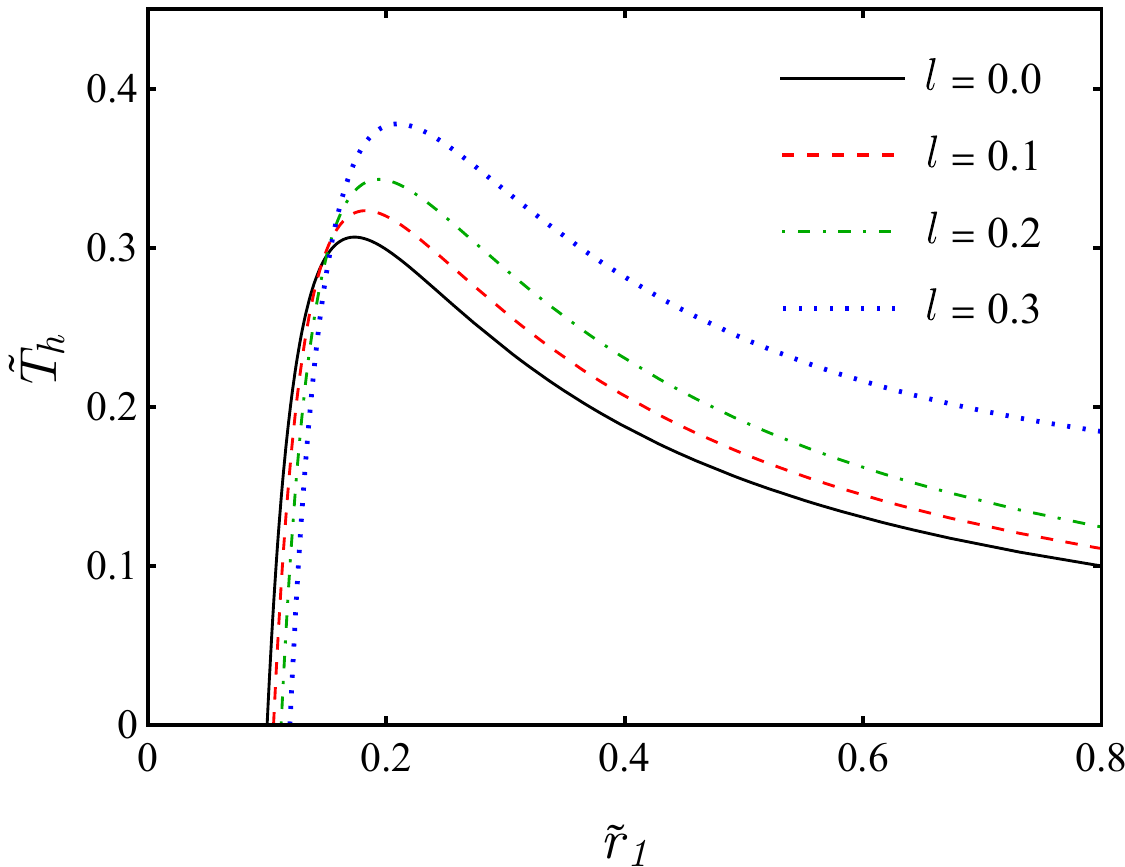}
	\caption{The temperature as a
		function of event horizon for fixed $Q = 0.1$, $\Lambda = -0.03$ and different values of $l$.}
	\label{fig:TempLambda}
\end{figure}

\begin{table}[!h]
\begin{center}
\begin{tabular}{c c c c } 
 \hline\hline
 $Q$ & $l$ & $\Lambda$ & $\Tilde{M}_{rem}$  \\ [0.2ex] 
 \hline 
  0.1 & 0.1 & -0.05 & 0.117132  \\ 

  0.1 & 0.1 & -0.10 & 0.117143   \\
 
  0.1 & 0.1 & -0.20 & 0.117165  \\
 
  0.1 & 0.1 & -0.30 & 0.117186   \\
 
  0.1 & 0.1 & -0.40 & 0.117208   \\
 
  0.1 & 0.1 &  -0.50 & 0.117229   \\
  
  0.1 & 0.1 & -0.60 & 0.117251  \\ 

  0.1 & 0.1 & -0.70 & 0.117272   \\
 
  0.1 & 0.1 & -0.80 & 0.117294  \\
 
  0.1 & 0.1 & -0.90 & 0.117315   \\
 
  0.1 & 0.1 & -1.00 & 0.117337   \\
 [0.2ex] 
 \hline \hline
\end{tabular}
\caption{\label{Remnant2} The modification of the remnant mass $\Tilde{M}_{rem}$ for different values $\Lambda$.}
\end{center}
\end{table}

Here, we proceed analogously to the previous subsections. In other words, we consider the final stage of black hole evaporation, i.e., $\Tilde{T}_{h} \to 0$, leading to a remnant mass as well, i.e., $\Tilde{M}_{rem}$
\ie
\label{ranmantmass222}
\Tilde{M}_{rem} = \frac{\sqrt{-\frac{\sqrt{1-\frac{4 \Lambda  Q^2}{1-l}}-1}{\Lambda }} \left(\sqrt{1-\frac{4 \Lambda  Q^2}{1-l}}+2\right)}{3 \sqrt{2} (1-l)}.
\fe
Considering above equation, one important aspect is worthy to be commended: if we regard the limit where $\lim\limits_{\Lambda \to 0} \Tilde{M}_{rem}$, the outcome result agrees with that one encountered in Eq. (\ref{renmantmass111}), as one can expected. It is clear that, from Eq. (\ref{ranmantmass222}), only two parameters modify $\Tilde{M}_{rem}$, i.e., $Q$, $l$, and $\Lambda$. Additionally, let us see the behavior of such a mass in Fig. \ref{remnantmass}.

Analogously, we have also accomplished the black hole lifetime to this case. In this sense, it reads
\ie
\nonumber
\frac{\mathrm{d}\Tilde{M}}{\mathrm{d}\tau} = - \alpha a \Tilde{\sigma}  \Tilde{T}_{h}^{4},
\fe
where 
\ie
\begin{split}
& \Tilde{\sigma} = \pi \left. \left( \frac{g_{\varphi\varphi}}{g_{tt}} \right)  \right|_{r = {r_{out}}}  = \frac{6 \pi Q^2 \left(-3 (l-1)^2 M\right)^2}{-9 (l-1)^4 M^2-3 (l-1)^2 M \left(\zeta +4 \Lambda  Q^2\right)+4 Q^2 (\zeta  \Lambda -3 l+3)},
\end{split}
\fe
with $\zeta =\sqrt{9 (l-1)^4 M^2+8 (l-1) Q^2}$, which leads to
\ie
\begin{split}
\frac{\mathrm{d}\Tilde{M}}{\mathrm{d}\tau} & = \Upsilon
\frac{6 \pi Q^2 \left(-3 (l-1)^2 M\right)^2}{9 (l-1)^4 M^2 + 3 (l-1)^2 M \left(\zeta +4 \Lambda  Q^2\right) - 4 Q^2 (\zeta  \Lambda -3 l+3)} \\
& \times \left(\frac{(1-l) \Tilde{r}_{1}^2 \left(1-\Lambda  \Tilde{r}_{1}^2\right)-Q^2}{4 \pi  (1-l)^2 \Tilde{r}_{1}^3} \right)^{4}.
\end{split}
\fe
Therefore, we have to solve
\ie
\begin{split}
\int_{0}^{\Tilde{t}_{\text{evap}}} \Upsilon \mathrm{d}\tau & =  \int_{\Tilde{M}_{i}}^{\Tilde{M}_{rem}} 
\left[ \frac{6 \pi Q^2 \left(-3 (l-1)^2 M\right)^2}{9 (l-1)^4 M^2 + 3 (l-1)^2 M \left(\zeta +4 \Lambda  Q^2\right) - 4 Q^2 (\zeta  \Lambda -3 l+3)} \right.\\
& \left.  \times \left(\frac{(1-l) \Tilde{r}_{1}^2 \left(1-\Lambda  \Tilde{r}_{1}^2\right)-Q^2}{4 \pi  (1-l)^2 \Tilde{r}_{1}^3} \right)^{4}\right]^{-1} \mathrm{d}\Tilde{M},
\end{split}
\fe
where $\Tilde{M}_{i}$ and $\Tilde{t}_{\text{evap}}$ are the initial mass configuration, and the time for final stage of the evaporation process for our black hole with cosmological constant. Thereby, we obtain
\ie
\Tilde{t}_{\text{evap}} = \frac{1}{\pi\Upsilon} \left( 2.20259\times 10^7 \right).
\fe
Some noteworthy comments arise at this point. Initially, we conducted the integral above numerically, with $\Tilde{M}_{i}=0.5$, while considering the values of $\Tilde{M}_{\text{rem}}$ outlined in Tab. \ref{Remnant2}, considering the variation of $\Lambda$ for fixed values of $Q=0.1$ and $l=0.1$. To ensure thoroughness, we illustrate the diminishing of mass $\Tilde{M}$ over time $\Tilde{t}$ in Fig. \ref{remnantmasss22}. Notably, this visualization pertains to a specific selection of parameters, namely $l=Q=0.1$, and $\Lambda = -0.1$. Notice that $\Tilde{t}_{\text{evap}}$ turns out to become larger than $\Tilde{t}_{\text{evap}}$, leading to the following radio difference: $t_{\text{evap}} = 1.19356 \,\Tilde{t}_{\text{evap}}$.

\begin{figure}
    \centering
      \includegraphics[scale=0.45]{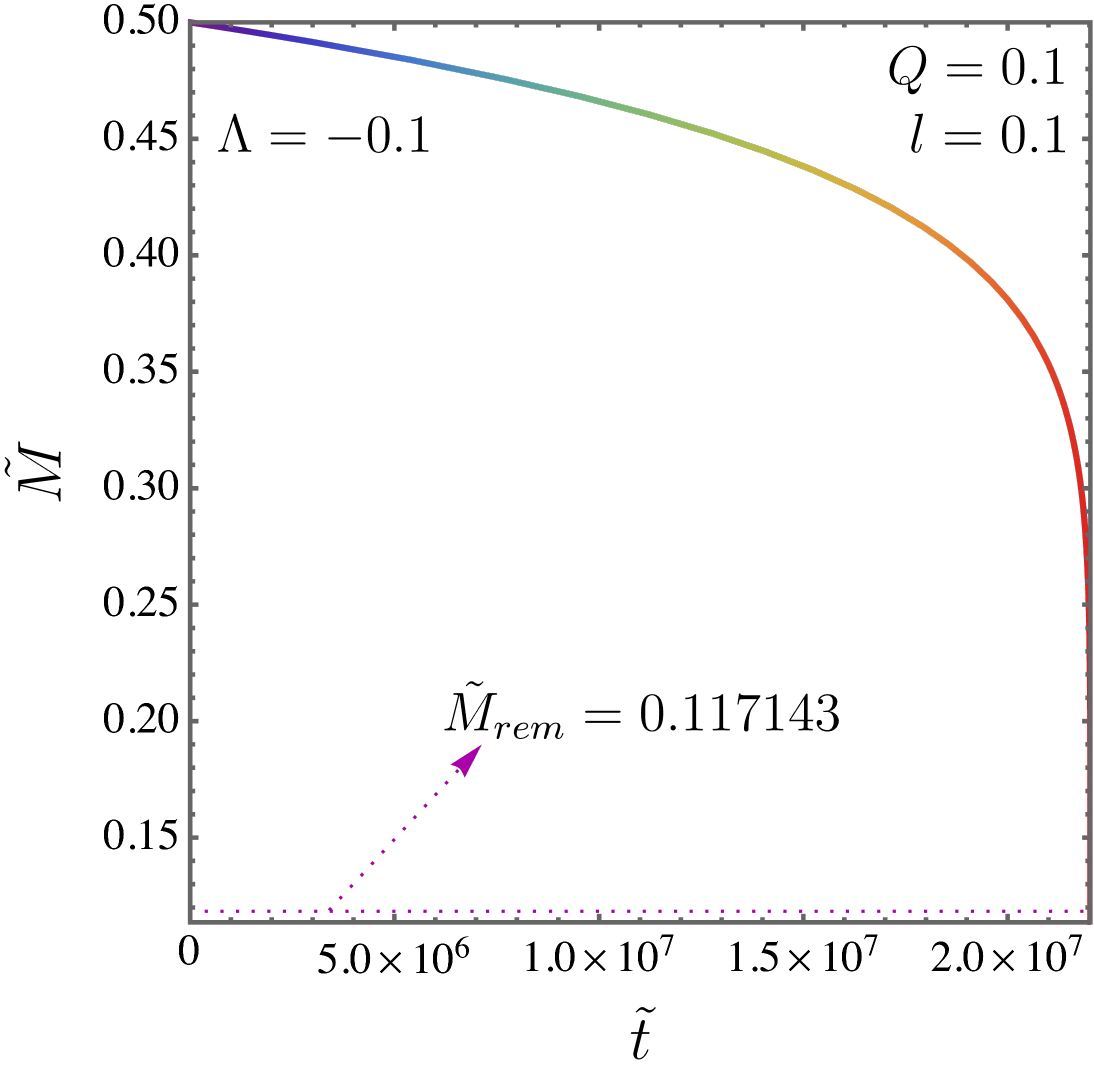}
    \caption{The mass reduction as a function of time $\Tilde{t}$ for $l=Q=0.1$ and $\Lambda=-0.1$.}
    \label{remnantmasss22}
\end{figure}

\begin{figure}
    \centering
    \includegraphics[scale=0.28]{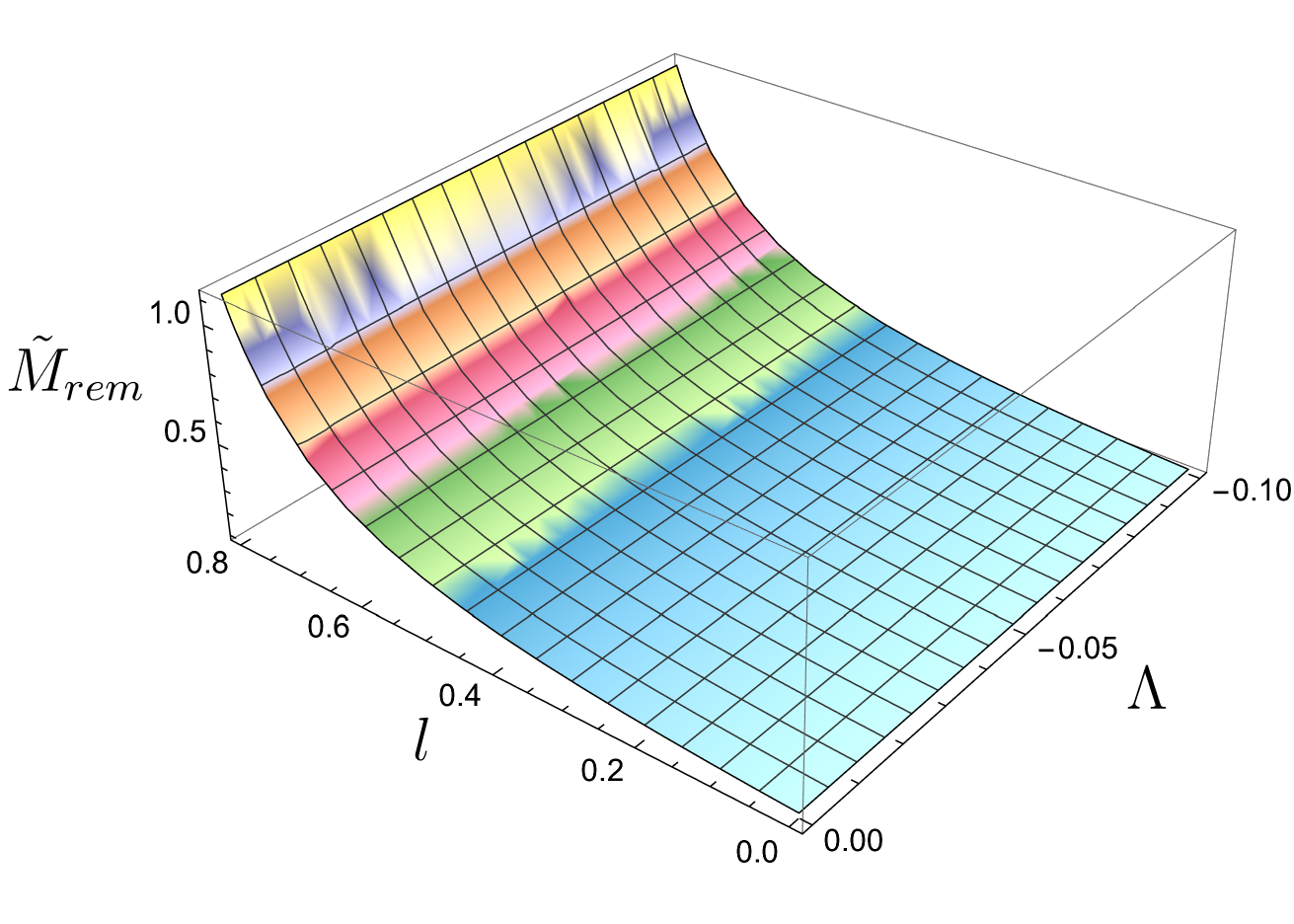}
     \includegraphics[scale=0.3]{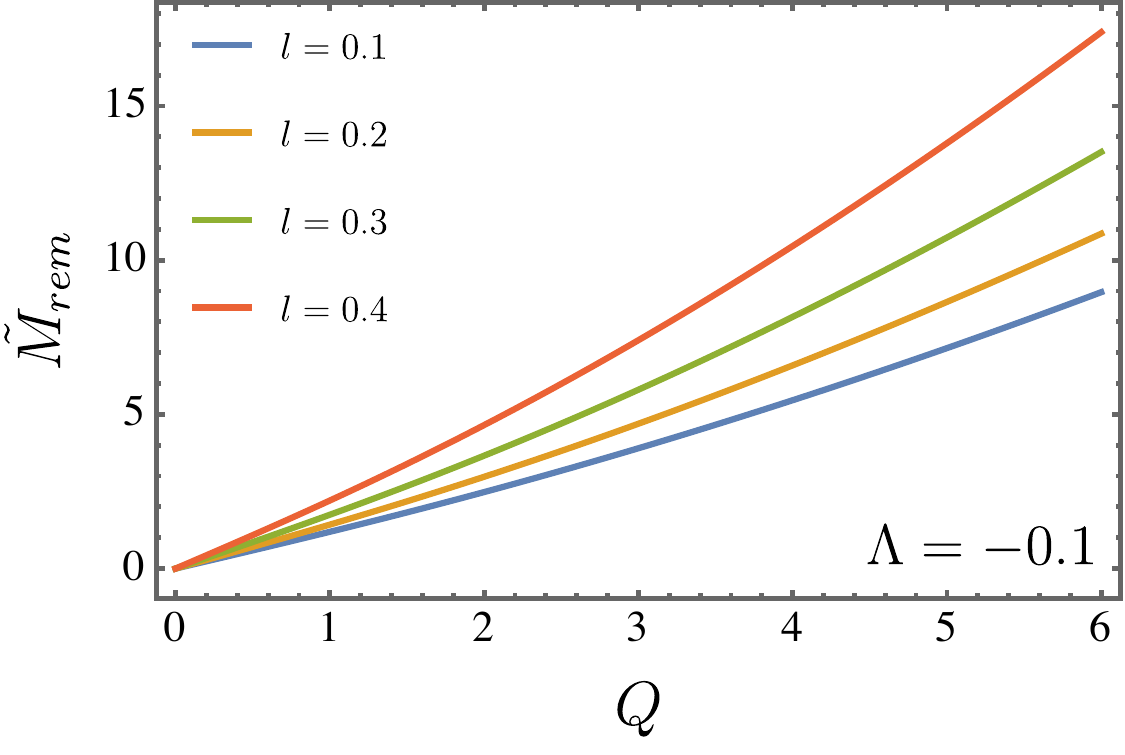}
      \includegraphics[scale=0.3]{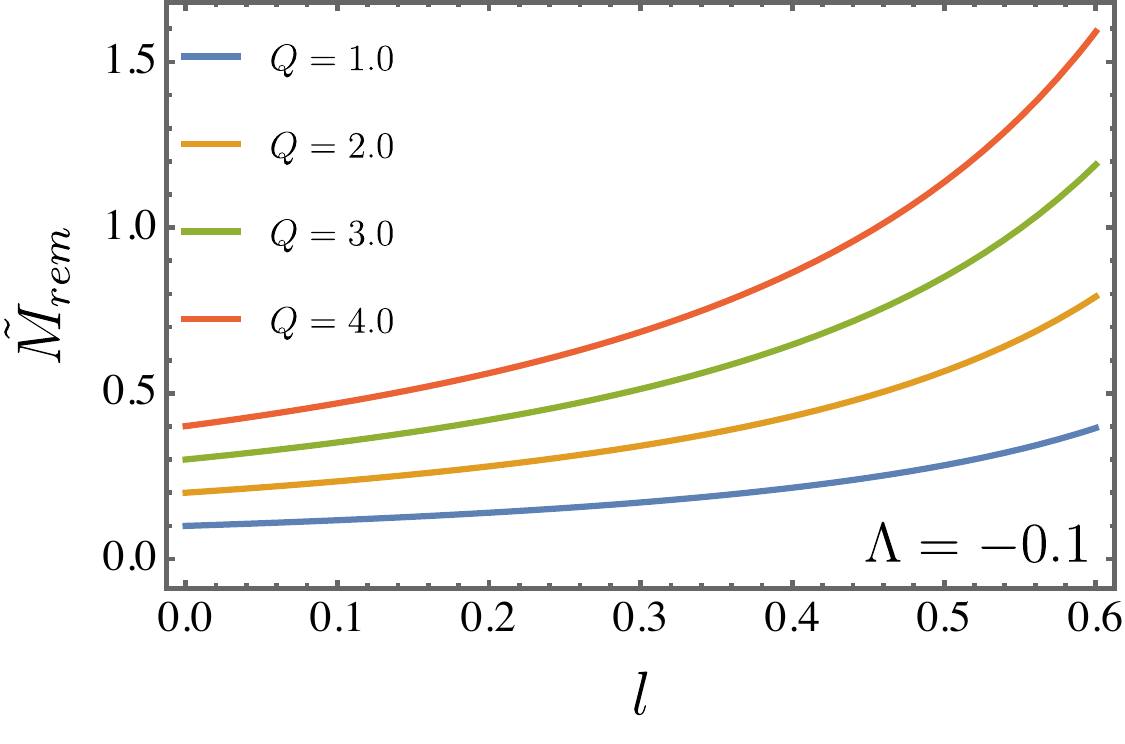}
        \includegraphics[scale=0.35]{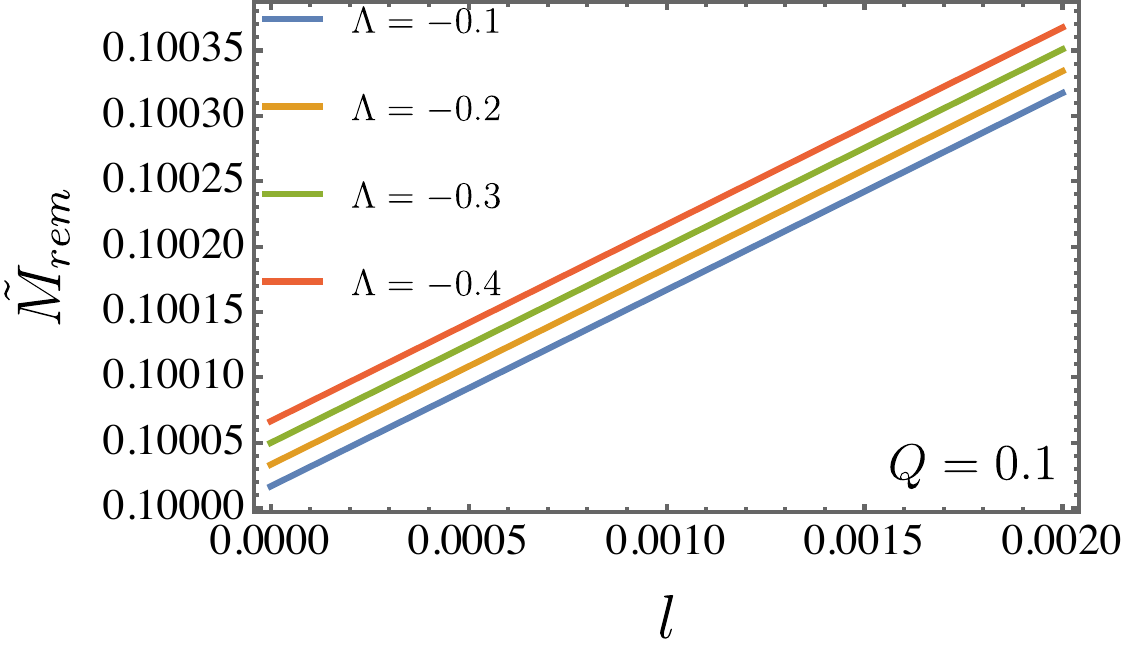}
    \caption{The remnant mass $\Tilde{M}_{rem}$ for different configurations of $Q$ and $l$.}
    \label{remnantmass}
\end{figure}


\subsection{Emission rate}

Following the previous sections, the emission rate is going to be addressed as well for the solution involving the cosmological constant. In this sense, we have
\begin{equation}\label{emission2}
\nonumber
	\frac{{{\mathrm{d}^2}E}}{{\mathrm{d}\omega \mathrm{d}t}} = \frac{{2{\pi ^3}R_{sh}^{2} }}{{e^{{\omega }/\Tilde{T}_h} - 1}}{\omega ^3}, 
\end{equation}
The result is investigated for $M = 1$, $Q = 0.1$, $\Lambda=-0.003$, and various values of $l$ and represented in Fig. \ref{fig:EmLambda}. Moreover, on the right panel, it exhibited the behavior of the Hawking radiation, considering different values of $Q$.
\begin{figure}
	\centering
	\includegraphics[scale=0.4]{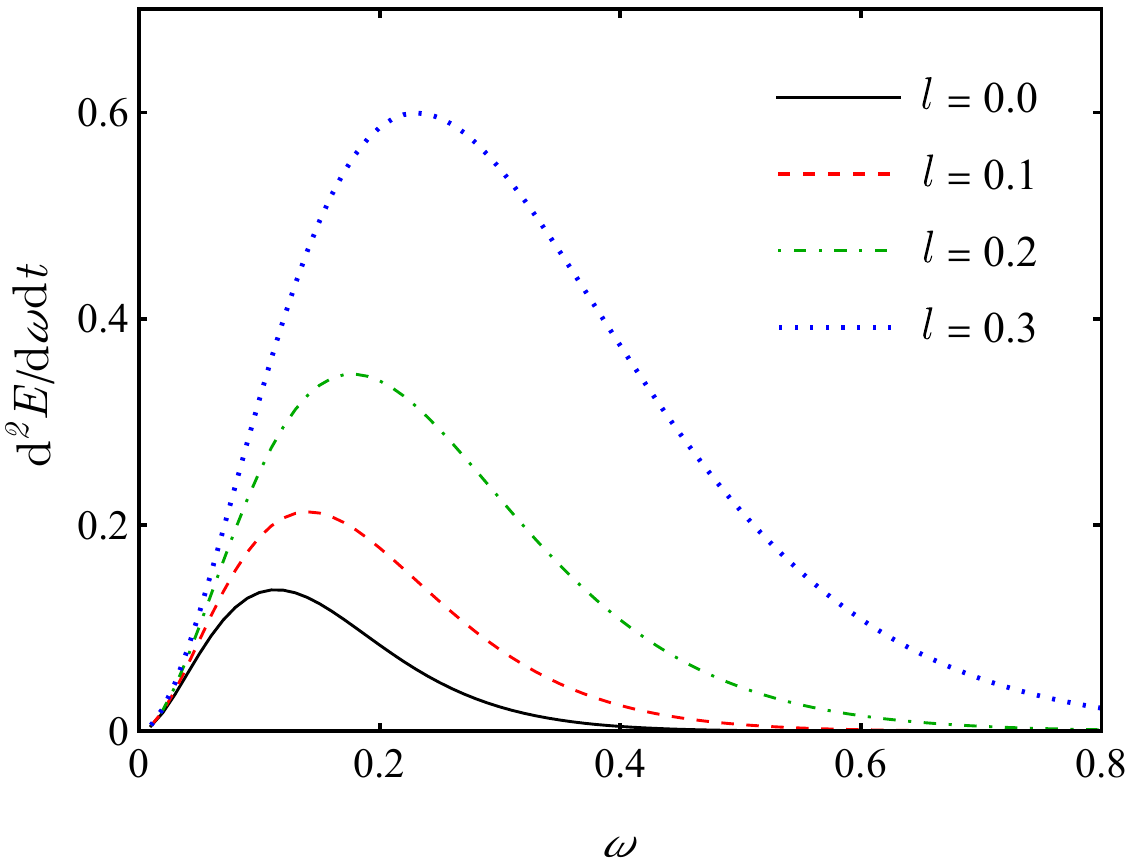}
 \includegraphics[scale=0.40]{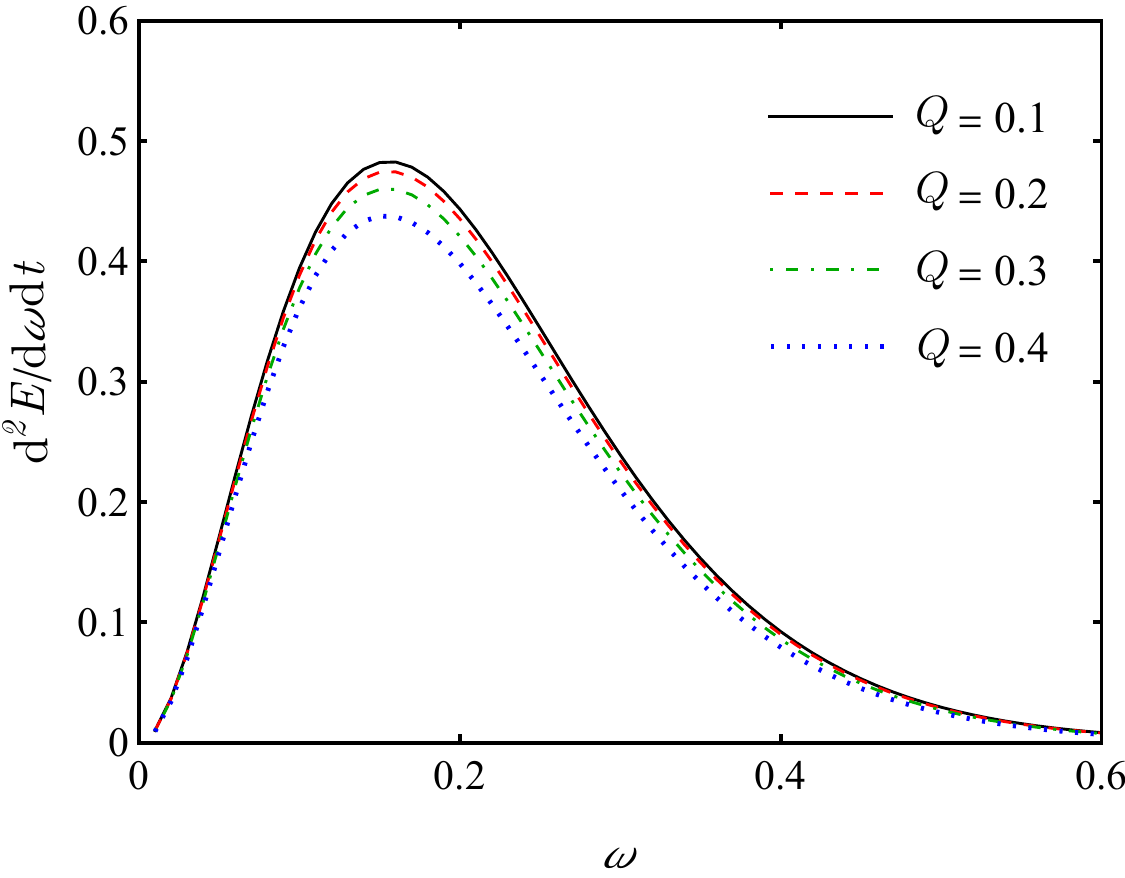}
	\caption{On the left, the rate of energy emission varies with frequency for $M = 1$, $Q = 0.1$, $\Lambda = -0.003$, $l = 0, 0.1, 0.2$, and $l = 0.3$. The observer position is considered to be $r_O = 10$. Also, on the right hand, it is displayed the behavior of the Hawking radiation, considering different values of $Q$ for $M = 1$, $l = 0.1$, $\Lambda = -0.003$ and $r_O = 10$.}
	\label{fig:EmLambda}
\end{figure}
It depicts the behavior of the Lorentz--violation parameter $l$
on the energy emission. As it is naturally to expect, the peaks of the energy emission rate increase and shift, depending on the values of $l$. Notice that the evaporation process would be slower when the effect of Lorentz--violation is higher.

Here, we have thoroughly investigated various properties of charged black holes, focusing on effects outside the event horizon. This naturally raises the question: for the black holes considered here, could information about their interior be obtained? Indeed, it seems plausible through at least two approaches. The entanglement (or its degradation) between the emitted particle and its infalling partner can encode details of the black hole's interior structure. If the Kalb--Ramond field or Lorentz--violating terms influence the interior, such effects could manifest as modifications of the quantum states, as discussed in Ref. \cite{Liu:2024wpa}. Additionally, in the tunneling picture commonly applied in effective field theories, involves particle creation near the event horizon. The correlation between the pairs suggests a connection between the emission spectrum and the interior dynamics, as energy conservation during tunneling might reflect aspects of the black hole's internal structure, as explored in Refs. \cite{Touati:2023cxy} and \cite{Calmet:2023gbw}. These an other ideas will be taken into account in the fourthcoming papers.


\section{Conclusion}

In this paper, we investigated the impact of anti--symmetric tensor effects, which induced Lorentz violation, on charged spherically symmetric black holes. Our study began with a comprehensive exposition of the model, establishing the necessary framework for deriving black hole solutions. We analyzed two distinct scenarios: one without a cosmological constant (the first case) and the other incorporating its presence (the second case).

Initially, for the first case, we addressed the horizons, critical orbits, and geodesics. We computed quasinormal modes, with a focused emphasis on vectorial perturbations, to gather information about the stability of the system. The increment of parameters $l$ and $Q$ was responsible for bringing about damped oscillating modes. We derived the Hawking temperature to perform the calculation of the remnant mass. Moreover, we estimated the lifetime of the black hole until it reached its final stage after the evaporation process, i.e., $t_{evap} = (1/\pi \Upsilon) 2.62893 \times 10^{7}$. Furthermore, we also explored the emission rate where, with the increase of $l$, larger intensities of emanating radiation were observed; the deflection angle turned out to be smaller when $l$ increased. In addition, we also investigated the correspondence between quasinormal modes and shadows.

Subsequently, for the second case, we analyzed the horizons, revealing the existence of just one physical event horizon, along with geodesics and critical orbits. Notably, $\Lambda$ contributed without modifying the photon sphere. The quasinormal modes were computed, indicating that the cosmological constant contributed to damped oscillations as well. For the parameters considered in this paper, some modes exhibited instabilities in this case. Furthermore, the time--domain solution was employed to investigate the system's perturbations, focusing on their evolution and oscillatory behavior.  Analogous to the first case, we derived the Hawking temperature to calculate the remnant mass. Using this quantity, we accurately estimated the lifetime until reaching its final stage after occurring the evaporation process, i.e., $\Tilde{t}_{evap} = (1/\pi \Upsilon) 2.20259 \times 10^{7}$. Compared to the first case, $\Tilde{t}_{\text{evap}}$ was shorter, leading to the following relation: $t_{\text{evap}} = 1.19356 \,\Tilde{t}_{\text{evap}}$. The investigation of the emission rate in this scenario was also accomplished.

From another perspective, to obtain direct or indirect information about the interior region of the black holes considered here, we aim to investigate the particle creation process, similar to Refs. \cite{Touati:2023cxy,Calmet:2023gbw,df1,df2,df3}, as well as the entanglement of quantum states, inspired by Ref. \cite{Liu:2024wpa}. These and related concepts are currently under development.


\section*{Acknowledgments}
\hspace{0.5cm}

A. A. Araújo Filho would like to thank Fundação de Apoio à Pesquisa do Estado da Paraíba (FAPESQ) and Conselho Nacional de Desenvolvimento Cientíıfico e Tecnológico (CNPq)  -- [150891/2023-7] for the financial support. Most of the calculations were performed by using the \textit{Mathematica} software. Also, H. Hassanabadi is grateful to excellent project FoS UHK 2023/2025-2026 for the financial support.


\bibliographystyle{ieeetr}
\bibliography{main}

\end{document}